\documentclass[aps,prd,preprint,groupedaddress,nofootinbib,showpacs,eqsecnum]{revtex4-1}

\usepackage{amsmath}    
\usepackage{graphicx}   
\usepackage{slashed}	
\usepackage{amssymb}    
\usepackage{array}

\newcommand   \<   {\langle}
\renewcommand \>   {\rangle}
\def\be#1\ee{\begin{align}#1\end{align}}
\newcommand \nn {\nonumber}

\def\subsubsubsection#1 {\noindent {\it #1}}

\def\mysubsubsection#1 {
\vspace{1.8em}
\textit{\small #1}
\vspace{.2em}
} 

\usepackage{hyperref}
\hypersetup{
    colorlinks=false,	
    linktoc=all,		
    linkcolor=black,	
}

\def\tree{{\rm tree}}
\def\fig#1{Fig.~\ref{#1}}
\def\eqn#1{Eq.~(\ref{#1})}
\def\eqns#1#2{Eqs.~(\ref{#1}) and~(\ref{#2})}
\def\sect#1{Sect.~\ref{#1}}

\def\newPiece#1{\{#1\}}
\def\spa#1.#2{\left\langle#1\,#2\right\rangle}
\def\spb#1.#2{\left[#1\,#2\right]}
\def\nn{\nonumber}
\def\eps{\epsilon}

\def\footnotelineskip{\baselineskip 13 pt}
\def\captionlineskip{\baselineskip 18 pt}
\newbox\charbox
\newbox\slabox
\def\s#1{{      
        \setbox\charbox=\hbox{$#1$}
        \setbox\slabox=\hbox{$/$}
        \dimen\charbox=\ht\slabox
        \advance\dimen\charbox by -\dp\slabox
        \advance\dimen\charbox by -\ht\charbox
        \advance\dimen\charbox by \dp\charbox
        \divide\dimen\charbox by 2
        \raise-\dimen\charbox\hbox to \wd\charbox{\hss/\hss}
        \llap{$#1$} }}

\begin{document}

\begin{flushleft}
 \hfill CERN-TH-2021-035
\end{flushleft}

\title{Gravitational Effective Field Theory Islands, \\
Low-Spin Dominance,  and the Four-Graviton Amplitude \\}

\author{Zvi Bern$^{a}$, Dimitrios Kosmopoulos$^a$ and Alexander Zhiboedov$^b$\\}

\affiliation{
$\null$
$^a$ Mani L. Bhaumik Institute for Theoretical Physics\\
UCLA Department of Physics and Astronomy\\
Los Angeles, CA 90095, USA\\
$^b$ CERN, Theoretical Physics Department\\
 CH-1211 Geneva 23, Switzerland
}

\begin{abstract}

We analyze constraints from perturbative unitarity and crossing on the
leading contributions of higher-dimension operators to the
four-graviton amplitude in four spacetime dimensions, including
constraints that follow from distinct helicity configurations.  We
focus on the leading-order effect due to exchange by massive degrees
of freedom which makes the amplitudes of interest infrared finite. In
particular, we place a bound on the coefficient of the $R^3$ operator
that corrects the graviton three-point amplitude in terms of the $R^4$
coefficient.  To test the constraints we obtain nontrivial effective
field-theory data by computing and taking the large-mass expansion of
the one-loop minimally-coupled four-graviton amplitude with massive
particles up to spin 2 circulating in the loop. Remarkably, we observe
that the leading EFT coefficients obtained from both string and
one-loop field-theory amplitudes lie in small islands.  The shape and
location of the islands can be derived from the dispersive
representation for the Wilson coefficients using crossing and assuming
that the lowest-spin spectral densities are the largest.  Our analysis
suggests that the Wilson coefficients of weakly-coupled gravitational
physical theories are much more constrained than indicated by bounds
arising from dispersive considerations of $2 \to 2$ scattering.  The
one-loop four-graviton amplitudes used to obtain the EFT data are
computed using modern amplitude methods, including generalized
unitarity, supersymmetric decompositions and the double copy.

\end{abstract}

\maketitle

\setcounter{tocdepth}{1} 
\tableofcontents

\newpage


\section{Introduction}
\label{GRMatterSection}

Remarkably, systematic bounds can be placed on possible corrections 
to Einstein gravity~\cite{causality,BiaHigherOrder, Tolley:2020gtv,
  Caron-Huot:2020cmc, Arkani-Hamed:2020blm, Sinha:2020win,Caron-Huot:2021rmr}.  Such
corrections naturally appear due to the presence of heavy particles in
the theory. To leading order in Newton's constant $G$, such particles
can be exchanged at tree-level, as in string theory, or at one-loop,
as in the case of matter minimally coupled to gravity. By expanding
such amplitudes at low energies and matching to a low-energy effective
field theory one finds an infinite series of higher-derivative
corrections to Einstein gravity. The coefficients in front of these
higher-derivative operators, or Wilson coefficients, satisfy various
bounds due to unitarity and causality of the underlying
amplitude~\cite{causality,BiaHigherOrder}. In this paper, we focus on
the leading corrections to Einstein
gravity.\footnote{\footnotelineskip In particular, higher-loop effects
  do not affect the discussion in this paper since by assumption
  gravity is weakly coupled and we are focusing on the leading-order
  effect.} A central question, which we investigate in this paper, is
to understand if there are principles that can greatly
restrict the values of physically allowed Wilson coefficients.

Consistency bounds on the Wilson coefficients received a lot of
attention recently in the context of $2 \to 2$ scattering, which is
also the subject of our paper. The basic tool to derive such bounds is
given by dispersion relations which express low-energy Wilson
coefficients as weighted sums of the discontinuity of the
amplitude. Unitarity constrains the form of the discontinuity of the
amplitude which can be further used to derive the bounds. The simplest
examples of this type constrain the sign of Wilson coefficients. More
interesting bounds arise when one accommodates constraints coming from
crossing symmetry. Including those leads to the two-sided bounds on
the Wilson coefficients \cite{Tolley:2020gtv, Caron-Huot:2020cmc,
  Arkani-Hamed:2020blm, Sinha:2020win}. In this way the ultraviolet
(UV) complete theories form bounded regions in the space of
couplings. 
Ref.~\cite{Arkani-Hamed:2020blm} also analyzed a few examples of physical EFTs
in the context of scattering of scalars and noted that they lie near the boundaries of the allowed region due to
the importance of low-spin contributions to the partial-wave
expansions (see Sect.~10.3 and Appendix~D of Ref.~\cite{Arkani-Hamed:2020blm}).
 
In the context of physical theories, especially gravitational ones, it is then natural to ask the
following question:
\begin{center}
{\it Is it possible that the Wilson coefficients of physical theories
  live in much smaller regions than the bounds coming from considerations of $2 \to 2$ scattering suggest?} 
\end{center}
By physical theories in this paper we mean perturbatively consistent
$S$-matrices that satisfy unitarity, causality, and crossing for any
$n \to m$ scattering processes. Constructing such $S$-matrices is far
beyond the scope of bootstrap methods that focus on $2 \to 2$
scattering, but such examples are provided to us by string theory and
matter minimally coupled to gravity.\footnote{\footnotelineskip
  Perturbatively consistent $S$-matrices occupy a somewhat
  intermediate position between fully non-perturbatively consistent
  quantum gravities (often referred to as landscape) and consistency
  of $2 \to 2$ scattering studied by bootstrap methods.} We can then
imagine that consistency of the full $S$-matrix is reflected back on
the $2\to 2$ scattering through more stringent constraints on Wilson
coefficients that one would naively have found by analyzing $2 \to 2 $
scattering. In this paper we present data extracted from
field-theory and string-theory $2 \to 2$ scattering amplitudes that
suggest that the above assertion is indeed true and we identify a principle behind it. 
 This principle is {\it low-spin dominance} (LSD), which, if
    fundamentally correct, might be traced back to the
    consistency of the full gravitational $S$-matrix, beyond $2
    \to 2$ scattering. However, demonstrating this is beyond the scope
    of the present paper. 

The universal nature of gravity together with the strict consistency
requirements that graviton scattering obeys make such an assertion
 plausible. It is a well-known fact that scattering of massless
spinning particles is very
constrained~\cite{Benincasa:2007xk}. In fact, massless particles of
spin larger than two do not admit a non-trivial
$S$-matrix~\cite{Weinberg}. Gravitons, being massless spin-two
particles, are thus expected to have an especially constrained
$S$-matrix, and the assertion made in the previous paragraph is thus
particularly plausible for graviton scattering which is the subject of
the present paper.

Firstly, we use the techniques of Ref.~\cite{Arkani-Hamed:2020blm} to
derive bounds between low-energy couplings of the same dimensionality
in gravitational scattering.\footnote{\footnotelineskip It would be
  very interesting to generalize our analysis to include bounds that
  relate couplings of different dimensionality along the lines of
  Refs.~\cite{Tolley:2020gtv,Caron-Huot:2020cmc,Caron-Huot:2021rmr}.}
We focus on the first few corrections to Einstein gravity. We then ask
where do the Wilson coefficients obtained from string theory and from
the low-energy limit of the one-loop minimally-coupled amplitudes land
in the space allowed by the general bounds.  Remarkably, in all cases
studied here we find that both the string and field-theory
coefficients land on a small {\it theory island}, which to a good
approximation is a thin line segment in the space of EFT
coefficients. (See, for example, Fig.~\ref{fig:a412} and
Fig.~\ref{fig:k6scaled} in
\sect{Sec:Bounds}).\footnote{\footnotelineskip In
  Ref.~\cite{Huang:2020nqy} the string-theory island was interpreted
  in terms of unitarity constraints coupled with world-sheet monodromy
  constraints.} The location of this island can be found by assuming
that lowest-spin partial waves dominate the dispersive representation
of the low-energy couplings, which is the LSD principle. See \sect{sec:nonsusymmetric} for the
precise mathematical formulation.  More generally, we show how one can
combine an assumed hierarchy among the spectral densities of various
spins with crossing symmetry to systematically derive stronger bounds
on the Wilson coefficients.  We impose crossing symmetry via the use
of null constraints~\cite{Tolley:2020gtv,Caron-Huot:2020cmc}.

The idea that LSD is a true property of physical theories can be
traced back to causality, or the statement that the amplitude cannot
grow too fast in the Regge limit. Otherwise we could have simply added
a tree-level exchange by a large-spin particle which would contribute
to a given spin partial wave. Due to causality we cannot do
  this (see e.g. Ref.~\cite{Camanho:2014apa}). The situation is particularly dramatic in gravity. In this
case the only particle that can be exchanged at tree-level in graviton
scattering without violating causality is the graviton itself.
Moreover, its self-coupling has to be the one of Einstein gravity Ref.~\cite{Camanho:2014apa,Meltzer:2017rtf,Belin:2019mnx,Simons-Duffin-Zhiboedov}
.  Alternatively, particles of all spins have
to be exchanged at tree level to preserve causality, which is the
mechanism realized in string theory.  It is important to emphasize
that at the level of $2 \to 2$ scattering LSD does not follow from
causality and we do not prove it in this paper, rather we use it as a
principle to organize the known data, and suggest that it may hold
more generally.  It would be interesting to understand if it follows
from considerations on the consistency of $n \to m$ graviton
scattering. 

Alternatively, it is also possible that our finding of LSD could be special for the models considered here and bears little significance for more general gravitational models. This possibility, which we cannot exclude, would be still very interesting. Indeed, as we demonstrate, any such violation is an indication for non-stringy, non-weakly-coupled-matter physics. For example, it would be very interesting to see if one can somehow violate LSD by making the matter sector strongly coupled, e.g. by considering large-$N$ QCD coupled to gravity \cite{Kaplan:2020tdz}.

Curiously, the phenomenon of LSD generates hierarchies
between different Wilson coefficients in the absence of any
symmetry. We call this phenomenon {\it hierarchy from unitarity}
and it is something that could have puzzled an unassuming low-energy
physicist. We find that  specific combinations of Wilson
coefficients whose dispersive representation does not involve the
lowest-spin partial waves can be much smaller than their counterparts
that do have them in their definitions.

Secondly, we apply the dispersive sum rules~\cite{Caron-Huot:2020adz,Caron-Huot:2020cmc} to
amplitudes with various helicity configurations of the external gravitons.\footnote{\footnotelineskip Flat space
  superconvergence considered in Ref.~\cite{Simons-Duffin-Zhiboedov} is a
  particular example of these more general sum rules.} We derive various bounds on the inelastic scattering (the one in which the final and initial state gravitons have different helicities) in terms of the elastic one (see e.g. Ref. \cite{Trott:2020ebl}). We also place a
precise bound on the $R^3$ coefficient in terms of the $R^4$
coefficient (see \eqref{eq:boundR3}). Such a bound translates the problem
of making the analysis of Ref.~\cite{Camanho:2014apa} quantitatively
precise to the problem of the bounding the leading $R^4$ contact
coefficient in terms of the gap of the theory. This has been recently
done in Ref.~\cite{Caron-Huot:2021rmr} in a similar perturbative setting
for $D=10$ maximal supergravity; see also Ref.~\cite{Guerrieri:2021ivu}
for the nonperturbative analysis of the same problem. It would, of course, be very
interesting to generalize these studies to more general cases of
graviton scattering.

In order to provide data for checking and understanding the derived constraints, we
first compute the one-loop four-graviton scattering amplitude
with the gravitons minimally coupled to massive matter up to
spin 2.  Amplitudes corresponding to the ones discussed here, but with
massless particles circulating in the loop were obtained a while ago
in Ref.~\cite{Dunbar:1994bn} and corresponding gauge-theory amplitudes with massive particles in
the loop were computed in Ref.~\cite{BernMorgan}.  
We use the same type of organization of the amplitude in
terms of supersymmetric multiplets as applied in the earlier 
calculations, since they naturally group contributions according to their
analytic properties.

To evaluate the amplitudes, we make use of standard tools including
the unitarity method~\cite{Unitarity} and the Bern-Carrasco-Johansson
(BCJ)~\cite{BCJ,BCJReview} double copy, which gives gravity integrands
in terms of corresponding gauge-theory ones.  We build on the
$D$-dimensional version of the unitarity method of
Ref.~\cite{BernMorgan} in order to fix the rational terms in the
amplitudes.  At four points gauge-theory tree-level amplitudes
automatically satisfy the duality between color and kinematics, so the
associated double-copy relations also hold automatically on the
unitarity cuts.  We use this to express the cuts of the gravity loop
integrands directly in terms of the corresponding gauge-theory ones.
By using the double copy our computation parallels the corresponding
gauge-theory one~\cite{BernMorgan} allowing us to import many of the
same steps into the gravitational amplitude calculations.

A complication with massive amplitudes is that there is a class of
terms that depend on the mass but do not have branch cuts in any
kinematic variable.  This makes their construction tricky in the
context of the unitarity method. Ref.~\cite{BrittoMirabella}
introduced an approach to this problem. Here we instead solve the problem
differently by  making use of a special property of the scattering
amplitudes under study that exploits their simple dependence
on the mass of the particle circulating in the loop.  
In our case (i.e. a single mass circulating in
the loop) we instead  use knowledge of the ultraviolet properties of the
amplitudes to fix all remaining functions in the amplitude not
determined by unitarity.   This procedure is greatly aided by arranging the
amplitude in terms of integrals that have no mass or spacetime 
dependence in their coefficients. To ensure the veracity of our amplitudes 
we perform a number of nontrivial checks on the mass dependence,
and infrared and ultraviolet properties.
Related to this, we also note a simple relation between ultraviolet
divergences of appropriate spacetime dimension shifts of the amplitudes
and the terms in the large-mass expansion in four dimensions (see \eqn{eq:UVIRequiv}).

We analyze our amplitudes in the large-mass limit and match to a
low-energy effective field theory.  In this way we systematically
obtain corrections to Einstein gravity due to the presence of a heavy
spinning particle. These corrections are organized in inverse powers of the particle's mass.  As already noted 
not only are our results for the Wilson coefficients fully consistent with
the general analysis of bounds on gravitational scattering, but are
restricted to small islands.

Since we focus on the leading effect due to heavy particles in the weakly
coupled setting neither IR divergences, nor logarithms due to the loops
of massless particles  make an appearance in our analysis. Taking
these into consideration is an important task which we leave for the future.

Our paper naturally consists
of two parts: In the first part we explain in detail the
construction of the one-loop massive amplitudes used to provide
theoretical data that we interpret in the second part in terms of
bounds on coefficients of gravitational EFTs.  Readers who are
interested in the EFT constraints can skip
\sect{AmplitudeConstructionSection} on the construction of the
one-loop amplitudes. Particularly important plots that illustrate the
theory islands and the concept of low-spin dominance in the
partial-wave expansion are given in
Figs.~\ref{fig:a412}-\ref{fig:k6scaled}.

In more detail, the sections are organized as follows: In
\sect{AmplitudeConstructionSection} we describe our construction of
the one-loop four-graviton amplitude with massive matter up to spin 2
in the loop.  In \sect{EFTMatchingSection} we compute graviton
scattering in a general low-energy effective theory.  By expanding our
amplitudes in the low-energy limit, we extract the Wilson coefficients
of the effective field theory.  In \sect{Sec:GravAmplitudes} we
describe the general properties of the gravitational amplitudes
stemming from unitarity and causality.  In \sect{Sec:Bounds} we derive
two-sided bounds on Wilson coefficients that follow from a single
helicity configuration that describes elastic scattering; comparing to known data from string theory and
our computed one-loop amplitude, we show that the results fall into
small islands. We trace the position of these islands using low-spin
dominance of partial waves. In \sect{Sec:BoundMultipleHelicity} we
obtain bounds that arise from considering multiple helicities. We bound the low-energy expansion coefficients of inelastic amplitudes in terms of elastic ones.  We also derive a bound for the coefficient of the $R^3$
operator in terms of the $R^4$ coefficient.  Finally, we provide our concluding remarks in
\sect{ConclusionSection}.  We include various appendices.  In
Appendix~\ref{app:MinimalCoupling} we describe in some detail our
definition of minimal coupling of gravity to a massive spinning
particle.  In Appendix~\ref{StringAppendix} we collect tree-level
graviton four-point amplitudes in various string theories.  In
Appendix~\ref{app:fullk6analysis} we present details on the derivation
of some low-energy bounds that are not listed in the main text of the
paper. In Appendix~\ref{app:Mstumodel} we analyze an amplitude
function with an accumulation point in the spectrum that partially violates
low-spin dominance, but show that the corresponding low-energy coefficients still land on the small islands.
Appendix~\ref{app:wignerd} collects the Wigner d-matrices used
throughout the paper. In Appendix~\ref{AmplitudesResultsAppendix} we
present our results for the one-loop amplitudes.  We give the expressions for one-loop
integrals in terms of which the amplitudes are expressed 
in Appendix~\ref{IntegralValuesAppendix}.  Finally, in
Appendix~\ref{HighOrdersAppendix} we expand these results to high
orders in the large-mass expansion.

\section{Construction of one-loop four-graviton scattering amplitudes}
\label{AmplitudeConstructionSection}

In this section we  describe the construction of the one-loop
four-graviton amplitudes with massive matter up to spin 2 in the loop. We collect the results in
Appendix~\ref{AmplitudesResultsAppendix}.  
We first briefly review the methods
used to obtain the amplitudes.
Then, following the generalized-unitarity
method we build the integrand-level generalized-unitarity cuts.
We describe a natural and efficient
organization of the unitarity cuts and the amplitudes motivated by supersymmetry.  This
organization also meshes well with the double-copy construction which
we use to obtain gravitational unitarity cuts from gauge-theory ones.
Having obtained the unitarity cuts we describe the necessary integral reduction and cut merging into the amplitudes.  This process fixes all but a few pieces of the amplitudes, which we obtain by exploiting the known ultraviolet properties of the amplitudes. 
After calculating the amplitudes, we comment on some interesting ultraviolet properties we observe.
Finally, we conclude this section by listing the consistency checks we performed on our calculation.

\subsection{Basic methods}
\label{ReviewSection}

\mysubsubsection{Spinor helicity}

We use the spinor-helicity method~\cite{SpinorHelicity} to describe
the external graviton states of amplitudes (for reviews see Refs.~\cite{Mangano:1990by}).
The natural quantities in this formalism are two component  Weyl spinors 
\begin{equation}
  (\lambda_i)_\alpha \equiv [u_+(k_i)]_\alpha \,, \hskip 1.3 cm  (\tilde \lambda_i)_\alpha \equiv [u_-(k_i)]_\alpha \,,
\end{equation}
which we write in a `bra'  and `ket' notation as
\begin{equation}
| k_i^+ \rangle = |\, i\,\rangle  = \lambda_i  \,, \hskip .8 cm 
| k_i^- \rangle = |\, i\, ]  = \tilde \lambda_i  \,, \hskip .8 cm
\langle k_i^-  | = \langle \, i \, |  = \lambda_i  \,, \hskip 1 cm 
\langle k_i^+ | = [\, i\, |  = \tilde \lambda_i  \,.
\end{equation}
where $k_i^\mu$ refers to the null momentum of the $i$-th external
particle, while the `$\pm$' superscript refers to the helicity of the
corresponding state. The spinor inner products are defined using the antisymmetric
tensors $\varepsilon^{\alpha\beta}$ and  $\varepsilon^{\dot \alpha \dot \beta}$,
\begin{equation}
\langle k_i^- | k_j^+ \rangle = \langle i j \rangle = \varepsilon^{\alpha\beta} 
(\lambda_i)_\alpha  (\lambda_j)_\alpha \,,
 \qquad
\langle k_i^+ | k_j^- \rangle = [ i j ] =  -\varepsilon^{\dot\alpha\dot \beta} 
(\tilde \lambda_i)_{\dot\alpha} (\tilde \lambda_i)_{\dot \alpha} \,.
\end{equation}
These spinor products are antisymmetric in their arguments and we choose a convention where they satisfy $\langle i j
\rangle [i j] =  2 k_i \cdot k_j$.

In order to construct amplitudes with external gravitons, our starting
point is the corresponding ones with external gluons. For calculations
involving external gluons  the helicity polarization
vectors are defined as
\begin{equation}
\varepsilon_\mu^+(k_i;q_i)=  \frac{\langle q_i| \gamma_\mu |i]}{\sqrt{2} \langle  i q_i \rangle} \,, 
\hskip 2 cm \varepsilon_\mu^-(k_i;q_i)=\frac{[ q_i| \gamma_\mu |i \rangle}{\sqrt{2} [ i q_i ]} \,,
\end{equation}
where $q_i$ are arbitrary null `reference momenta' which drop out of
the final gauge-invariant amplitudes. Note that we do not use a
shorthand notation for the spinors corresponding to the reference
momenta. The polarization tensors for gravitons are simply given in
terms of products of gluon polarization vectors,
\begin{equation}
\varepsilon_{\mu\nu}^{+}(k;q) = \varepsilon_{\mu}^{+}(k;q) \, \varepsilon_{\nu}^{+}(k;q) \,, \hskip 2 cm 
\varepsilon_{\mu\nu}^{-}(k;q) = \varepsilon_{\mu}^{-}(k;q) \, \varepsilon_{\nu}^{-}(k;q) \,, \hskip 2 cm 
\end{equation}
which automatically satisfy the graviton tracelessness condition, due
to the Fierz identity.
When these polarization vectors are contracted into external momenta
$k_i^\mu$ or loop momenta $\ell^\mu$ we define,
\begin{equation}
k_1 \cdot \varepsilon_2^+ = \frac{\langle q_2| \slashed{k}_1 |2]}{\sqrt{2} \langle q_2 2 \rangle} 
\equiv \frac{\langle q_2| 1 |2]}{\sqrt{2} \langle q_2 2 \rangle} \,,
\hskip 1.5 cm 
\ell \cdot \varepsilon_2^+ = \frac{\langle q_2| \slashed{\ell} |2]}{\sqrt{2} \langle q_2 2 \rangle} 
\equiv \frac{\langle q_2| \ell |2]}{\sqrt{2} \langle q_2 2 \rangle} \,,
\quad \text{etc.},
\end{equation}
where we also use the abbreviation $\varepsilon_{2}^+ \equiv
\varepsilon^+(k_2;q_2)$.

We note that, in general, for loop calculations some care is needed
when using dimensional regularization.  To take advantage of the
spinor-helicity formulation in a one-loop calculation we need to
choose an appropriate version of dimensional
regularization. Specifically, instead of taking the external
polarization tensors and momenta to be ($4-2\epsilon$)-dimensional as
in conventional dimensional regularization \cite{Collins}, we use the
so called four-dimensional helicity (FDH) scheme~\cite{FDHScheme,
  FDHExamples}  where both external and loop state counts are kept in
four dimensions and only the loop momentum is continued to
$4-2\epsilon$ dimensions.  Because the massive one-loop amplitudes
that we obtain here are neither ultraviolet nor infrared divergent,
the precise distinction between the different versions of dimensional
regularization drops out from the final results for the amplitudes.
We do, however, need to regularize intermediate steps because
individual loop integrals are ultraviolet divergent, with the
divergence canceling in final results.

\mysubsubsection{Generalized unitarity}

In order to construct the loop integrands we use the
generalized-unitarity method~\cite{Unitarity}.  This method
systematically builds complete loop-level integrands using as input
on-shell tree-level amplitudes.  A central advantage is that
simplifications and features of the latter are directly imported into
the former.  Reviews of the generalized-unitarity methods are found in
Refs.~\cite{UnitarityReview1, UnitarityReview2}.

In general, the task of computing an amplitude is to reduce it to a
linear combination of known scalar integrals.  Using standard
integral-reduction techniques (see e.g. Refs.~\cite{IntegralReduction,
  Smirnov:2019qkx}) any four-point one-loop amplitude can be written as
a linear combination of box, triangle, bubble and tadpole integrals,
\begin{equation}
\mathcal{M}_4^\text{1-loop} = \Bigl(d_{s,t} I_4(s,t) +  c_s I_3(s)  + b_s I_2(s) + \hbox {perms.} \Bigr)
     + b_0 I_2(0) + a_0 I_1\,,
\label{ReductionTarget}
\end{equation}
where the permutations run over distinct relabelings of the integrals.
At the four-point level there are a total of 11 coefficients.  These
coefficients depend on polarization vectors, momenta, masses and the
dimensional-regularization parameter $\epsilon$.  We define the basis
integrals appearing in \eqn{ReductionTarget} by
\begin{align}
& I_4(s,t)  =  \int \frac{d^{D} L}{(2 \pi)^{D}} \frac{ -i  (4 \pi)^{D/2} }{(L^2 -m^2)
((L + k_1)^2 - m^2)((L + k_1+k_2)^2 - m^2)((L - k_4)^2 - m^2)} \,, \nonumber \\
& I_3(s)  =  \int \frac{d^{D} L}{(2 \pi)^{D}} \frac{i (4 \pi)^{D/2}}{(L^2 -m^2)
((L + k_1)^2 - m^2)((L +k_1+k_2)^2 - m^2)} \,, \nonumber \\
& I_2(s)  = \int \frac{d^{D} L}{(2 \pi)^{D}} \frac{-i (4 \pi)^{D/2}} {(L^2 -m^2)
((L +k_1+k_2)^2 - m^2)} \,, \nonumber \\
& I_1  =  \int \frac{d^{D} L}{(2 \pi)^{D}} \frac{i (4 \pi)^{D/2}} {(L^2 -m^2)} \,, 
\label{4dMasters}
\end{align}
where $D = 4-2\epsilon$, $s = (k_1 + k_2)^2$, $t = (k_2
+ k_3)^2$ and $u = (k_1 + k_3)^2$.  We obtain the remaining integrals
in \eqn{ReductionTarget} by permuting the external legs.
The unitarity method efficiently targets the coefficients of the
integrals in \eqn{ReductionTarget}.  The integrals $I_2(0)$ and $I_1$
are respectively bubble on external leg and tadpole contributions, and
are independent of kinematic variables.  As we discuss below, because they lack 
dependence on kinematic variables, these
latter integrals require special treatment to determine their
coefficients.

\begin{figure}[t]
     \centering
     \includegraphics[width=\textwidth]{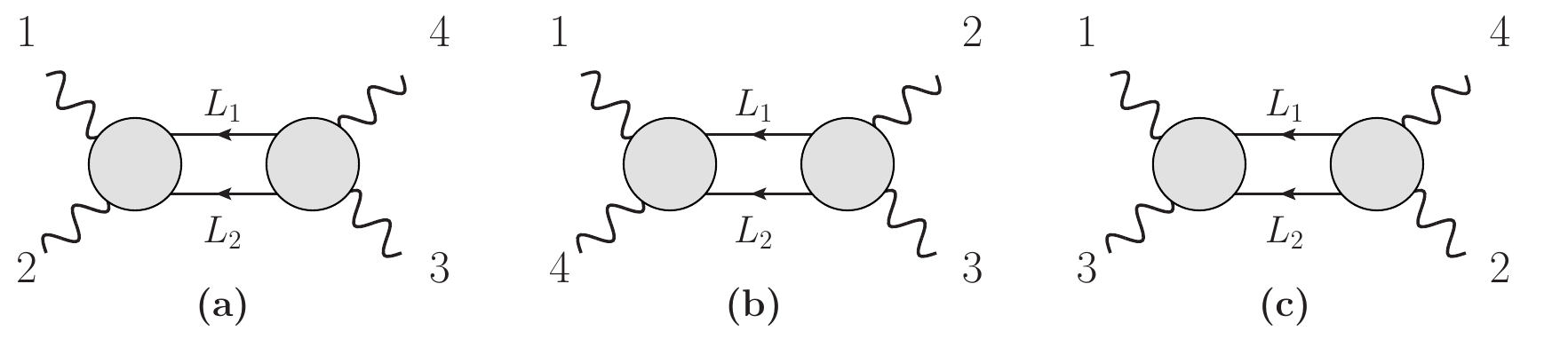}
     \caption{\captionlineskip The (a) $s$-, (b) $t$- and (c) $u$-channel two-particle
          cuts of a one-loop four-point amplitude. The exposed lines
          are all on shell and the blobs represent tree-level amplitudes. }
      \label{TwoParticleCutFigure}
\end{figure}

Traditionally, unitarity of the scattering matrix is implemented at
the integrated level via dispersion relations (see
e.g. Ref.~\cite{AnalyticSMatrixBook}). However, for our purposes, it is
much more convenient to use an integrand-level version of unitarity~\cite{Unitarity}.
This is based on the concept of a generalized-unitarity cut that reduces an integrand
to a sum of products of tree-level amplitudes.  For example, for the
$s$-channel cut displayed in \fig{TwoParticleCutFigure}(a),
\begin{equation}
i \mathcal{M}_4^\text{1-loop} \Big\rvert_{s\rm\hbox{-}cut} \!\!= \! \int\!\!
\frac{d^{4-2\epsilon} L}{(2\pi)^{4-2\epsilon}} \frac{1}{L_1^2-m^2}
\frac{1}{L_2^2-m^2} \sum_{\rm states} \! \mathcal{M}_4^\tree(1, 2, L_1,
L_2) \mathcal{M}_4^\tree(-L_1, -L_2, 3, 4)
\Big\rvert_{s\rm\hbox{-}cut} ,
\label{GenericCut}
\end{equation}
where $L_1 = L$ and $L_2 = -L-k_1-k_2$ represent the two cut legs, and
$\mathcal{M}_4^\tree$ denote the tree-level amplitudes.  The sum runs
over all intermediate physical states that contribute for a given set
of external states.
The three generalized-unitarity cuts of the one-loop four-point amplitude are
shown in \fig{TwoParticleCutFigure}.  In this figure
the exposed lines are all on shell and the blobs
represent on-shell tree-level amplitudes.

To obtain the full one-loop amplitude we must combine the unitarity
cuts.  One possibility is to carry this out prior to integration by
finding a single integrand with the correct unitarity cuts in all
channels~\cite{Bern:2004cz}. Some non-trivial examples where this
approach was implemented are high-loop computations in
super-Yang-Mills and supergravity (see
e.g. Refs.~\cite{SuperHighLoopExamples}).  On the other hand, in
high-multiplicity QCD calculations (see e.g. Ref.~\cite{BlackHat}) the
cuts are usually combined after reducing to a basis of integrals.  We
apply the latter approach here.  We do so by promoting each cut
propagator to a Feynman propagator, and each cut to a Feynman
integral. We then use FIRE6~\cite{Smirnov:2019qkx} to reduce each
Feynman integral to the scalar integrals appearing in
\eqn{ReductionTarget}. In each cut channel we only determine
coefficients of basis integrals with cuts in that channel. By
systematically evaluating each cut we determine all coefficients
except for those of integrals without kinematic dependence,
i.e. $I_2(0)$ and $I_1$.  In the case of gauge theory, the
corresponding coefficients are determined by imposing the known
ultraviolet behavior of the amplitudes~\cite{BernMorgan}.  Below, we
describe an analogous procedure for the case of gravitational
amplitudes.

\mysubsubsection{Double copy}

To efficiently obtain the unitarity cuts of the four-graviton amplitude,
we use the double-copy construction~\cite{Kawai:1985xq,BCJ} which expresses gravitational
scattering amplitudes directly in terms of gauge-theory ones. Here we
use the BCJ form of the double-copy relations~\cite{BCJ,BCJReview}, which
is more natural when organizing expressions in terms of diagrams. 

\begin{figure}[tb]
     \centering
     \includegraphics[width=0.8 \textwidth]{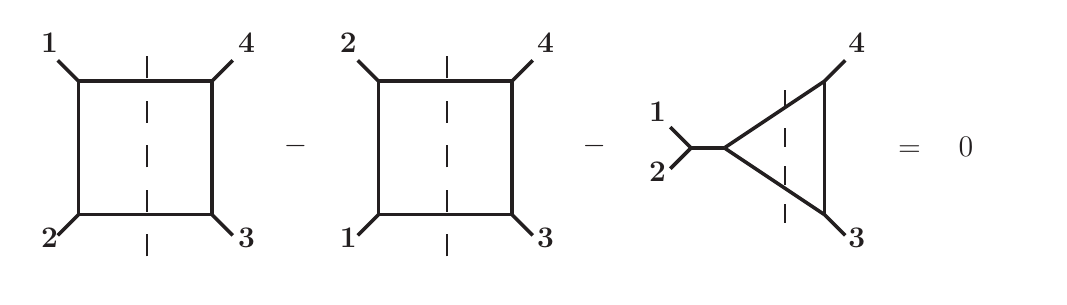}
     \caption{\captionlineskip Example of a color or numerator relation for the one-loop four-point amplitudes. 
       Here the diagram represent either color or kinematic numerators.  We use these relations 
       on the generalized-unitarity cuts, as indicated by the dashed lines. 
       The relations are effectively tree-level ones
       except that the state and color sums on the cut legs are carried out. }
     \label{colorGraphs}
\end{figure}

To apply the BCJ double copy, we start by writing a four-point
one-loop gauge-theory amplitude in the following form:
\begin{equation}
i \mathcal{A}_4^\text{1-loop} = g^4 \sum_i \int \frac{d^{4-2\epsilon} L}{(2\pi)^{4-2\epsilon}} \frac{1}{S_i} 
\frac{n_i c_i}{D_i} \,.
\label{GeneralGauge}
\end{equation}
The sum runs over all distinct four-point one-loop graphs with
trivalent vertices. We denote the gauge-theory coupling constant by $g$. We label each graph by an integer $i$. The $S_i$
are the symmetry factors of the graphs.  The color factor $c_i$ of
each graph is obtained by dressing each vertex by a structure constant
$\tilde{f}^{abc}$, since we take all particles to be in the adjoint
representation. Our normalization of the structure constants follows
that of Ref.~\cite{BCJ}. The denominator $D_i$ contains the propagators of
each graph. Finally, we capture all non-trivial kinematic dependence
by the numerator $n_i$.

The color factors obey color-algebra relations of the type
\begin{equation}
c_i - c_j - c_k = 0 \,,
\label{jacobiColor}
\end{equation}
where $i$, $j$ and $k$ are some graphs. These relations follow from the Jacobi identity obeyed by
the structure constants $\tilde{f}^{abc}$. For a 
representation obeying color-kinematics duality, the numerators satisfy the same Jacobi 
relations, i.e.
\begin{equation}
\tilde n_i - \tilde n_j - \tilde n_k = 0\,.
\label{jacobiNumerator}
\end{equation}
The tildes on the numerators signify that these numerators do
not have to be the same as the ones appearing in
\eqn{GeneralGauge}, but as noted in Ref.~\cite{BCJ} these can
be kinematic numerators from a different theory.  Given such a representation we may obtain 
the corresponding gravitational amplitude simply by replacing
the color factor with the corresponding kinematic numerator, 
\begin{equation}
c_i \rightarrow \tilde n_i \,,
\label{DoubleCopy}
\end{equation}
so that
\begin{equation}
i \mathcal{M}_4^\text{1-loop} = \left( \frac{\kappa}{2} \right)^4 \sum_i 
\int \frac{d^{4-2\epsilon} L}{(2\pi)^{4-2\epsilon}} \frac{1}{S_i} 
\frac{n_i \tilde n_i }{D_i} \,,
\end{equation}
where $\kappa$ is the gravitational coupling constant, which
is given in terms of Newton's constant by,
\begin{equation}
\kappa^2 = 32 \pi G\,.
\label{kappaDef}
\end{equation}
The matter content of the resulting gravitational theory is determined
by the choice of the numerators $n_i$ and $\tilde{n}_i$.  We use
this to control the type of particle circulating in
the loop.

\begin{figure}
     \centering
     \includegraphics[width=0.8 \textwidth]{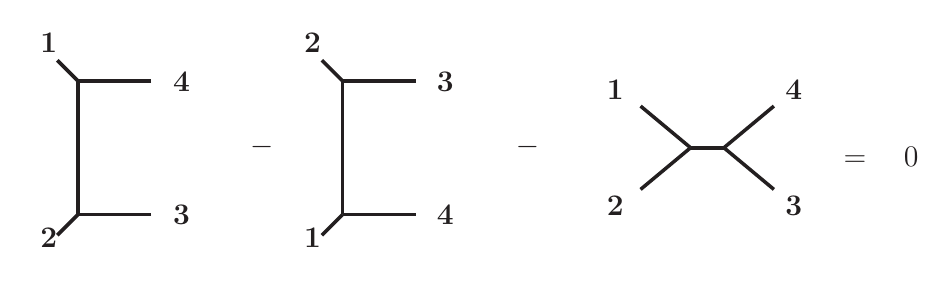}
     \caption{\captionlineskip The tree-level four-point color or numerator Jacobi identity.
         This can be used to set the numerator of one of the diagrams to zero.}
            \label{TreeFigure}
\end{figure}

While the BCJ double copy is usually formulated at the level of the full
integrand, since we extract the final answer directly from the cuts 
it is more convenient to use it at the level of generalized-unitarity
cuts. In Fig.~\ref{colorGraphs} we
depict an example of a color relation in terms of cut graphs. In this way we can ignore duality relations that contain diagrams without support on the given cut. For duality relations with support on the cut, the particles of each tree-level amplitude entering the cut remain on the same side of the cut for all three diagrams, as is the case in Fig.~\ref{colorGraphs}.
Effectively this amounts to using the duality relations for the two
tree-level amplitudes on each side of the cut (see e.g. Fig.~\ref{TreeFigure}).  The tree-level relations are sufficient 
to ensure that for the cut diagram, the double-copy replacement formula \eqref{DoubleCopy} holds.

We also take advantage of a property of four-point tree-level amplitudes
noted in the original BCJ paper~\cite{BCJ}, which states that we can effectively set one of the three numerators to zero. In this way the duality relation
implies that the other two numerators
must be equal  (up to a possible sign).  In order to achieve that, we absorb the propagator of the diagram
whose numerator is set to zero into the numerators of the
other two diagrams.  Specifically, consider the four-point all-gluon tree-level
amplitude,
\begin{equation}
i \mathcal{A}_4^\text{tree} = g^2 \Bigl( \frac{n_s c_s}{s} + \frac{n_t c_t}{t} + \frac{n_u c_u}{u} \Bigr) \,.
\end{equation}
Using the color Jacobi identity depicted in \fig{TreeFigure}, $c_s = c_t - c_u$, we can eliminate $c_s$ in favor of the other two color factors,
so that 
\begin{equation}
i \mathcal{A}_4^\text{tree} = g^2 \Bigl(\frac{n_t' c_t}{t} + \frac{n_u' c_u}{u} \Bigr)\,,
\end{equation}
where 
\begin{equation}
n_t' = n_t + n_s \frac{t}{s}\,, \hskip 2 cm 
n_u' = n_u - n_s \frac{u}{s} \,.
\end{equation}
Demanding that the numerators $n_t'$ and $n_u'$ (and $n_s' = 0$)
satisfy the duality relation of \fig{TreeFigure} then implies that 
\begin{equation}
n_t' = n_u' \,.
\label{IdenticalNumerator}
\end{equation}
The analysis is identical in the presence of matter.

\subsection{Setup of the calculation}
\label{CutsSection}

Our goal is to obtain the four-point one-loop amplitude with external
gravitons and with minimally-coupled massive spinning particles up to
spin 2 circulating in the loop.  Following the generalized-unitarity
method we first build the integrand-level generalized-unitarity cuts
in \eqn{GenericCut}.  For the one-loop four-point amplitude there are
three independent cuts, labeled by the Mandelstam invariant that can
be build out of the external graviton momenta on the tree-level
amplitudes, i.e. $s$-, $t$- and $u$-channel cuts. We consider all
spins up to spin 2 for the massive particles and we denote their mass
by $m$. While these masses can be different for each particle since
only a single particle at a time circulates in the loop there is no
need to put an index labeling the massive particle.  We take the
massive particles to be real.

We note that there is an ambiguity in the definition of the minimal coupling of a spin 2 particle to gravity. In this paper, we fix this ambiguity by demanding that we recover pure gravity in the appropriate massless limit.  This choice also preserves tree-level unitarity~\cite{treeLevelUnitarity} and causality~\cite{Bonifacio:2017nnt}. 
We discuss this ambiguity and the choice we make in this paper in Appendix~\ref{app:MinimalCoupling}.

For the amplitude in question there exist three independent helicity configurations. Specifically, we calculate 
\begin{equation}
\mathcal{M}_4^\text{1-loop}(1^{+},2^{-},3^{-},4^{+})\,, \hskip 1 cm 
\mathcal{M}_4^\text{1-loop}(1^{+},2^{+},3^{-},4^{+})\,, \hskip 1 cm
\mathcal{M}_4^\text{1-loop}(1^{+},2^{+},3^{+},4^{+})\,.
\label{HelicityDef1}
\end{equation}
We refer to these configurations respectively as
\textit{double-minus}, \textit{single-minus} and \textit{all-plus}.
All other amplitudes are related to these via relabelings and parity.


\begin{figure}[tb]
     \centering
     \includegraphics[width=\textwidth]{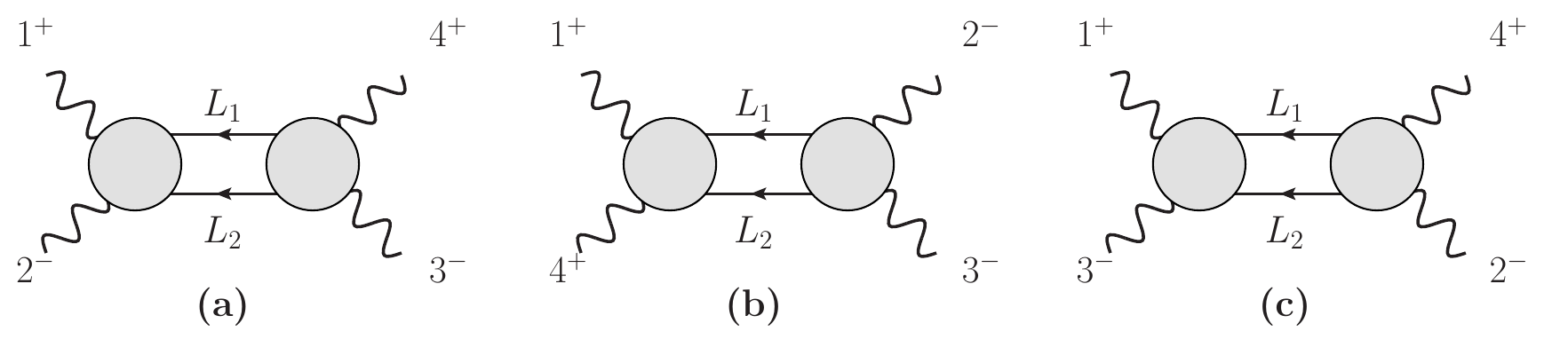}
\vskip - .3 cm 
     \caption{\captionlineskip The (a) $s$-, (b) $t$- and (c) $u$-channel two-particle
       cuts of a one-loop four-point amplitude with two negative-
       and two positive-helicity external gluons or gravitons (double-minus configuration). The
       internal lines represent massive spinning particles. The
       exposed lines are all on shell and the blobs represent
       tree-level amplitudes.}
      \label{TwoParticleCutHelicityFigure}
\end{figure}

We build the generalized-unitarity cuts from four-point
tree-level gravity amplitudes. The double copy implies that these tree-level
amplitudes can be directly obtained from the corresponding gauge-theory
ones, which can be described by the three diagrams shown in \fig{TreeFigure}. As we discussed in
the previous section we can use the
color Jacobi identity in the gauge-theory case to remove one diagram, at the expense of other
diagrams obtaining numerators that are nonlocal in the external kinematic
invariants.  The net effect is that after multiplying and dividing by
appropriate propagators every contribution to the cut can be assigned to cut box
diagrams.  Moreover, on the generalized cuts the four-point tree-level BCJ
numerator relations set the remaining numerators equal to each other, as noted in
\eqn{IdenticalNumerator}.  

\begin{figure}[tb]
     \centering
     \includegraphics[width=0.27 \textwidth]{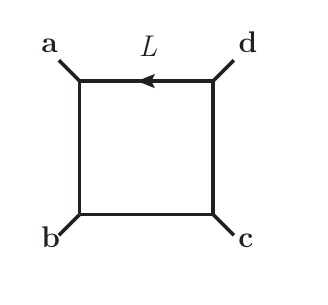}
\vskip - .8cm 
     \caption{\captionlineskip Generic box diagram whose color factor and denominator
       are given by Eqs.~(\ref{colorGen}) and (\ref{denGen})
       respectively. The external momenta are taken incoming while the
       direction of the loop momentum is indicated by the arrow. }
     \label{genericBox}
\end{figure}

For example, for the four-point gauge-theory amplitude the $s$-channel
cut in \fig{TwoParticleCutHelicityFigure}(a) is of the form,
\begin{equation}
i \mathcal{A}_4\Big\rvert_{s\rm\hbox{-}cut} = g^4
\int \frac{d^{4-2\epsilon} L}{(2\pi)^{4-2\epsilon}}  \, 
n_{\text{g},s}
\left( \frac{c_{1234}}{D_{1234}} + \frac{c_{1342}}{D_{1342}} \right)
\bigg\rvert_{s\rm\hbox{-}cut} \,.
\label{genAMHV}
\end{equation}
The box color factor (see \fig{genericBox}) is given by
\begin{equation}
c_{abcd} = \tilde{f}^{e_a g_1 g_4} \tilde{f}^{e_b g_2 g_1} \tilde{f}^{e_c g_3 g_2} \tilde{f}^{e_d g_4 g_3}\,,
\label{colorGen}
\end{equation}
where $abcd$ takes in the values indicated in \eqn{genAMHV}.  As usual,
repeated indices are summed.
The denominators are given by products of the usual Feynman propagators,
\begin{equation}
D_{abcd} = \Big( L^2 - m^2 \Big) \Big( (L+k_a)^2 - m^2 \Big) \Big( (L+k_a+k_b)^2 - m^2 \Big) 
\Big( (L+k_a+k_b+k_c)^2 - m^2 \Big)\,,
\label{denGen}
\end{equation}
where the $k_a$ are external momenta. Finally, $n_{\text{g},s}$ is a gauge-theory
kinematic factor common to both box diagrams.

The gravitational cuts are similar.  For example, the $s$-channel cut
is of the form,
\begin{equation}
i \mathcal{M}_4\Big\rvert_{s\rm\hbox{-}cut} = \left( \frac{\kappa}{2} \right)^4 
\int \frac{d^{4-2\epsilon} L}{(2\pi)^{4-2\epsilon}}  \,
n_{\text{GR},s}
\left( \frac{1}{D_{1234}} + \frac{1}{D_{1342}} \right)\bigg\rvert_{s\rm\hbox{-}cut} \,,
\label{genMMHV}
\end{equation}
where $n_{\text{GR},s}$ is the single $s$-channel gravitational kinematic
numerator. As we noted in \sect{ReviewSection}, the gravity amplitudes
follow from replacing the color factor with a gauge-theory kinematic
numerator.  As we describe in more detail below, the particle content
circulating in the loop is determined by the choice of the gauge-theory numerators.

As can be seen from Figs.~\ref{TwoParticleCutHelicityFigure}(a) and
\ref{TwoParticleCutHelicityFigure}(c), for the indicated helicity configuration,
the $u$-channel cut is obtained by relabeling the momenta and spinors
in the $s$-channel cut: $k_2 \leftrightarrow k_3$ and $\langle 2|
\leftrightarrow \langle 3|$. For the single-minus and all-plus configurations, all three cuts are given by appropriate relabelings of a single cut.

\subsection{Supersymmetric decompositions}
\label{susyDec}


We are interested in the problem of minimally-coupled massive matter
circulating in the loop.  A convenient way to organize this
calculation is by following the massless case where each particle can
be described as a linear combination of supersymmetric multiplets
circulating in the loop. Our organization is in direct correspondence
to this supersymmetric decomposition~\cite{SusyDecomposition,
  Dunbar:1994bn, Bern:1993tz}.  This allows us to recycle the results
of the calculation of the lower-spin particles circulating in the loop
into contributions for the higher-spin particles.  It also has the
advantage of grouping together terms that contain integrals of the
same tensor rank.  For the gauge-theory case such a decomposition has
already been used to organize the contributions of massive spin $0$,
$1/2$ and $1$ particles circulating in the loops~\cite{Bern:1993tz,
  BernMorgan}.  The double-copy construction will allow us to import
this into minimally-coupled gravitational theories with up to 
spin-2 massive particles.

\mysubsubsection{Gauge theory}

We start by examining the corresponding amplitudes in gauge
theory. For the case of one-loop four-point amplitudes with external
gluons and massive matter circulating in the loop,
Ref.~\cite{Bern:1993tz} showed that\footnote{\footnotelineskip We combine the `gauge
  boson' and `scalar' contributions of Ref.~\cite{Bern:1993tz} into
  the massive $S=1$ result.}
\begin{align}
\begin{split}
\mathcal{A}_4^{S=0} &= \mathcal{A}_4^{\newPiece{0}} \,,  \\
\mathcal{A}_4^{S=1/2} &= -2\mathcal{A}_4^{\newPiece{0}} + \mathcal{A}_4^{\newPiece{1/2}} \,,  \\
\mathcal{A}_4^{S=1} &= 3\mathcal{A}_4^{\newPiece{0}} - 4\mathcal{A}_4^{\newPiece{1/2}} + \mathcal{A}_4^{\newPiece{1}} \,.
\end{split}
\label{gaugeTheoryDecomposition}
\end{align}
where the notation $ \mathcal{A}_4^{\newPiece{\it S}}$ denotes the new piece we need to calculate at spin-$S$. 
In this way we express the amplitudes with a given spinning particle
circulating in the loop in terms of the simpler-to-calculate
new pieces. Inverting the above equations, we may think
of the new pieces as amplitudes with multiplets circulating in the
loop. These massive multiplets have the same degrees of freedom as the corresponding massless ones. Hence, in terms of on-shell supersymmetric representation theory they satisfy the BPS condition~\cite{susyRepReviews}. For recent calculations involving massive supersymmetric multiplets in gauge theory see Refs.~\cite{n4symCoulombBranch,Alday:2009zm}.

Before turning to the corresponding gravitational decomposition it is useful to first look at the massless limit.
For the gauge theory case we have that as $m\rightarrow 0$,
\begin{align}
\begin{split}
\mathcal{A}_4^{S=0} &\rightarrow \mathcal{A}_4^{S=0,\, \textit{m}=0} \,, \\
\mathcal{A}_4^{S=1/2} &\rightarrow \mathcal{A}_4^{S=1/2,\, \textit{m}=0} \,, \\
\mathcal{A}_4^{S=1} &\rightarrow \mathcal{A}_4^{S=1,\, \textit{m}=0} + \mathcal{A}_4^{S=0,\, \textit{m}=0} \,,
\label{masslessLimitGauge}
\end{split}
\end{align}
where we see that the $S=1$ case is nontrivial, which follows from the mismatch in the number of degrees of freedom in 
a massive and massless vector boson.

\mysubsubsection{Background field gauge}

A nice way to understand the above supersymmetric decomposition
is in terms of background field gauge~\cite{UnitarityReview1}.  While
we do not use background field gauge to compute the amplitudes, it does
offer useful insight into the structure of the amplitude.  For the
different particles circulating in the loop we can write the effective
action as
\begin{align}
\Gamma_{S=0}[A] & = \ln \det{}_{[0]}^{-1/2}\bigl(D^2 +m^2 \bigr)  \,, \nn \\
\Gamma_{S=1/2}[A] & = \ln \det{}_{[1/2]}^{1/4}\bigl(D^2 \, \mathbb{I} 
+ m^2 \, \mathbb{I} - g \frac{1}{2} \sigma^{\mu\nu} F_{\mu\nu} )  \,, \label{DetFormulas} \\
\Gamma_{S=1}[A] & = \ln \det{}_{[1]}^{-1/2}\bigl(D^2 \mathbb{I} 
           + m^2  \mathbb{I} - g \Sigma^{\mu\nu} F_{\mu\nu} \bigl)   
           + \ln \det{}_{[0]} \bigl( D^2 +m^2 \bigr) 
           + \ln \det{}_{[0]}^{-1/2} \bigl( D^2 +m^2 \bigr) \,, \nn
\end{align}
where $\mathbb{I}$ is the identity matrix, $\sigma_{\mu\nu}/2$ is the spin-$1/2$ Lorentz generator and
$\Sigma_{\mu\nu}$ is the spin-1 Lorentz generator.  Ignoring the
$F_{\mu\nu}$ terms and focusing on the $(D^2 + m^2)$ term, each state
corresponds to a power to which the determinant is raised: $-1/2$ for a bosonic state and $+1/2$ for a fermionic state.
For a massive real scalar there is precisely
one bosonic state corresponding the $-1/2$ power to which the first
determinant is raised.  For the $S=1/2$ fermion the determinant is
raised to the $1/4$ power to account for the $4 \times 4$ Dirac
determinant, effectively leaving a single power that corresponds to
the two states of a Majorana fermion.  Similarly, for the $S=1$
vector, the determinant is over a $4\times 4$ Lorentz generator space
so the exponent of $-1/2$ in the first term in $\Gamma_{S=1}[A]$
corresponds to 4 bosonic states.  This is then reduced to two states
by the ghost determinant corresponding to the second term and
increased by one state from the scalar longitudinal degree of freedom
required by a massive vector boson corresponding to the third
term.  This extra degree of freedom is incorporated into
\eqn{gaugeTheoryDecomposition} as well.  To make the supersymmetric
cancellations more manifest we rewrite the Dirac determinant as a
product of determinants so that the similarity to the 
bosonic case is clear,
\begin{align}
\det{}_{[1/2]}^{1/2}(\s D + i m) \det{}_{[1/2]}^{1/2}(\s D - i m) & = \det{}_{[1/2]}^{1/4}(\s D^2 + m^2) \nn\\
& = \det{}_{[1/2]}^{1/4}( D^2 +m^2 - g \frac{1}{2} \sigma^{\mu\nu} F_{\mu\nu}  ) \,,
\end{align}
where $\s D^2 = \frac{1}{2} \{ \s D, \s D\} + \frac{1}{2} [\s D, \s
  D]$.  This corresponds to using the second order formalism for
fermions described in Ref.~\cite{SecondOrderFermions}.  The fact that
the mass enters into \eqn{DetFormulas} in such a simple manner can
also be understood in terms of a Kaluza-Klein reduction of the
massless case from five dimensions, truncated to keeping only the
lowest massive state in the loop.

The effective-action determinants \eqref{DetFormulas} can be
straightforwardly applied to show that the supersymmetric
decomposition organizes contributions in terms of power counting of
the resulting diagrams.  The terms with leading power of loop momenta
come from the $D^2$ terms in \eqn{DetFormulas}, because $F_{\mu\nu}$
contains only the external gluon momenta.  If we set the $F_{\mu\nu}$
terms to zero then in all supersymmetric combinations the balance
between the bosons and fermions implies the leading powers of loop
momentum cancel.  Subleading terms in supersymmetric combinations come
from using one or more factors of $F_{\mu\nu}$ when generating a
graph; each $F_{\mu\nu}$ reduces the maximum power of momentum by one.
Terms with a lone $F_{\mu\nu}$ vanish, thanks to $ {\rm
  Tr}(\sigma^{\mu\nu}) = {\rm Tr}(\Sigma^{\mu\nu}) = 0$. This reduces
the leading power in an $m$-point one-particle-irreducible diagram
from $\ell^m$ down to $\ell^{m-2}$.  For
$\mathcal{A}_m^{\newPiece{1}}$, a comparison of the traces of products
of two and three $\sigma^{\mu\nu}$ and $\Sigma_{\mu\nu}$ shows that
further cancellations reduce the leading behavior all the way down to
$\ell^{m-4}$.  In gauges other than Feynman background field gauge,
these cancellations would be more more obscure. 

\mysubsubsection{Gravity}

Now consider the gravitational case.   In the $m\rightarrow 0$ 
limit we again have a mismatch in the number of degrees of freedom for
$S \geq 1$ between the massive and massless cases,
\begin{align}
\begin{split}
\mathcal{M}_4^{S=0} &\rightarrow \mathcal{M}_4^{S=0,\, \textit{m}=0} \,, \\
\mathcal{M}_4^{S=1/2} &\rightarrow \mathcal{M}_4^{S=1/2,\, \textit{m}=0} \,, \\
\mathcal{M}_4^{S=1} &\rightarrow \mathcal{M}_4^{S=1,\, \textit{m}=0} + \mathcal{M}_4^{S=0,\, \textit{m}=0} \,, \\
\mathcal{M}_4^{S=3/2} &\rightarrow \mathcal{M}_4^{S=3/2,\, \textit{m}=0} + \mathcal{M}_4^{S=1/2,\, \textit{m}=0} \,, \\
\mathcal{M}_4^{S=2} &\rightarrow \mathcal{M}_4^{S=2,\, \textit{m}=0} + \mathcal{M}_4^{S=1,\, \textit{m}=0} + \mathcal{M}_4^{S=0,\, \textit{m}=0} \,.
\label{masslessLimit}
\end{split}
\end{align}
In the massless case with spinning particles circulating in the loop
we can again decompose the amplitudes in terms of ones with
supermultiplets circulating in the loop~\cite{Dunbar:1994bn},
\begin{align}
\begin{split}
\mathcal{M}_4^{S=0,\, \textit{m}=0} &= \mathcal{M}_4^{\newPiece{0},\, \textit{m}=0} \, , \\
\mathcal{M}_4^{S=1/2,\, \textit{m}=0} &= - 2\mathcal{M}_4^{\newPiece{0},\, \textit{m}=0} + \mathcal{M}_4^{\newPiece{1/2},\, \textit{m}=0} \,, \\
\mathcal{M}_4^{S=1,\, \textit{m}=0} &=  2 \mathcal{M}_4^{\newPiece{0},\, \textit{m}=0}   - 4 \mathcal{M}_4^{\newPiece{1/2},\, \textit{m}=0} + \mathcal{M}_4^{\newPiece{1},\, \textit{m}=0} \,, \\
\mathcal{M}_4^{S=3/2,\, \textit{m}=0} &= - 2 \mathcal{M}_4^{\newPiece{0},\, \textit{m}=0} + 9 \mathcal{M}_4^{\newPiece{1/2},\, \textit{m}=0} 
                         - 6 \mathcal{M}_4^{\newPiece{1},\, \textit{m}=0} + \mathcal{M}_4^{\newPiece{3/2},\, \textit{m}=0} \,, \\
\mathcal{M}_4^{S=2,\, m=0} &=  2 \mathcal{M}_4^{\newPiece{0},\, \textit{m}=0} - 16 \mathcal{M}_4^{\newPiece{1/2},\, \textit{m}=0} 
           + 20 \mathcal{M}_4^{\newPiece{1},\, \textit{m}=0} -8 \mathcal{M}_4^{\newPiece{3/2},\, \textit{m}=0}  \\ &
\hskip .5 cm 
+\mathcal{M}_4^{\newPiece{2},\, \textit{m}=0}  \,.
\label{masslessSusyGR}
\end{split}
\end{align}
The $\{S\}$ pieces in each case are in direct correspondence to the
supermultiplets circulating in the loop, as defined in
Ref.~\cite{Dunbar:1994bn}\footnote{\footnotelineskip 
We use a real scalar while
  Ref.~\cite{Dunbar:1994bn} used a complex one. Hence there is
  relative factor of 1/2 for that contribution.}. For example, 
\begin{equation}
\mathcal{M}_4^{\newPiece{1/2},\, \textit{m}=0} = \mathcal{M}_4^{\mathcal{N}=1,\, \textit{m}=0} ,
\end{equation}
where $\mathcal{N}=1$ denotes the chiral multiplet consisting of a Weyl fermion
and two real scalars.  

Using the relation between the massive and massless amplitudes in
\eqn{masslessLimit} and the massless supersymmtric decomposition in \eqn{masslessSusyGR}, we
can organize our computation in a similar way as for gauge
theory in \eqn{gaugeTheoryDecomposition}. Specifically, we have
\begin{align}
\begin{split}
\mathcal{M}_4^{S=0} &= \mathcal{M}_4^{\newPiece{0}} \,,  \\
\mathcal{M}_4^{S=1/2} &= - 2\mathcal{M}_4^{\newPiece{0}} + \mathcal{M}_4^{\newPiece{1/2}} \,, \\
\mathcal{M}_4^{S=1} &= 3\mathcal{M}_4^{\newPiece{0}}  
  - 4 \mathcal{M}_4^{\newPiece{1/2}} + \mathcal{M}_4^{\newPiece{1}}\,, \\
\mathcal{M}_4^{S=3/2} &= - 4\mathcal{M}_4^{\newPiece{0}} 
  + 10 \mathcal{M}_4^{\newPiece{1/2}} -6\mathcal{M}_4^{\newPiece{1}} + \mathcal{M}_4^{\newPiece{3/2}} \,, \\
\mathcal{M}_4^{S=2} &=  5\mathcal{M}_4^{\newPiece{0}}  
 - 20 \mathcal{M}_4^{\newPiece{1/2}} +21\mathcal{M}_4^{\newPiece{1}} - 8 \mathcal{M}_4^{\newPiece{3/2}} + \mathcal{M}_4^{\newPiece{2}} \,.
\label{susyGR}
\end{split}
\end{align}
The massive multiplets circulating in the loop are `short', i.e. they have the same degrees of freedom as the corresponding massless ones. Hence, they obey the BPS condition~\cite{susyRepReviews}. While we have not tried directly embedding these amplitudes into
supergravity theories, it is an
interesting question to do so.  Here we use the relation 
to supermultiplets for a more modest aim of reorganizing the 
contributions, so that as the spin increases the new pieces become
simpler. Examples of supergravity calculations involving massive multiplets are given in Ref.~\cite{massiveSugra}.

We note that in general, care is required when using
dimensional regularization in conjunction with helicity methods and
supersymmetric decompositions.  To allow for different choices of regularization
scheme, we would need to correct the last line of
\eqn{gaugeTheoryDecomposition} to be~\cite{Bern:1993tz}
\begin{equation}
\mathcal{A}_4^{S=1} = (3-2\delta_R \epsilon) \mathcal{A}_4^{\newPiece{0}} 
    - 4\mathcal{A}_4^{\newPiece{1/2}} + \mathcal{A}_4^{\newPiece{1}},
\end{equation}
where $\delta_R = 0$ in the FDH scheme~\cite{Bern:1991aq} and $\delta_R =1$ in either
conventional dimensional regularization~\cite{Collins} or the 't
Hooft-Veltman scheme~\cite{tHooft:1972tcz}. One may then propagate
this correction to the gravitational amplitudes through the double
copy.
While the correction is of $\mathcal O (\epsilon)$, it can interfere with
infrared or ultraviolet singularities to give nontrivial
contributions.  However, for the massive amplitudes that we are
computing here, the distinction between different schemes is not
important because the four-graviton amplitudes with massive
matter in the loop are both ultraviolet and infrared finite (see
Sects.~\ref{uvSec} and \ref{Consistency}).

We also note that the coefficient of the scalar ($\mathcal{M}_4^{\newPiece{0}}$)
counts the degrees of freedom of the particle in question, modulo a
minus sign for the fermions. Recall that all particles are taken to be
real.  Also, given the general setup of our amplitudes
(Eqs. (\ref{genAMHV}) and (\ref{genMMHV})), we get a similar
decomposition for the corresponding numerators $n_{\text{g},\alpha}$
and $n_{\text{GR},\alpha}$.

Finally, the supersymmetric decomposition is simplified in the
case of the single-minus and all-plus configurations.  In these cases,
supersymmetric Ward identities~\cite{SWI} show that only the
$ \mathcal{M}_4^{\newPiece{0}}$ piece gives a nonzero contribution for each spin
particle in the loop.  Using this
observation, Eq.~(\ref{susyGR}) becomes
\begin{align}
\begin{split}
\mathcal{M}_4^{S=0}(1^\pm, 2^+, 3^+, 4^+) & = \mathcal{M}_4^{\newPiece{0}} (1^\pm, 2^+, 3^+, 4^+)\,, \\
\mathcal{M}_4^{S=1/2}(1^\pm, 2^+, 3^+, 4^+) & = - 2\mathcal{M}_4^{\newPiece{0}} (1^\pm, 2^+, 3^+, 4^+) \,, \\
\mathcal{M}_4^{S=1}(1^\pm, 2^+, 3^+, 4^+)  & = 3\mathcal{M}_4^{\newPiece{0}}(1^\pm, 2^+, 3^+, 4^+) \,, \\
\mathcal{M}_4^{S=3/2}(1^\pm, 2^+, 3^+, 4^+) & = - 4\mathcal{M}_4^{\newPiece{0}}(1^\pm, 2^+, 3^+, 4^+) \,, \\
\mathcal{M}_4^{S=2}(1^\pm, 2^+, 3^+, 4^+) & =  5\mathcal{M}_4^{\newPiece{0}}(1^\pm, 2^+, 3^+, 4^+)\, .
\label{allPlusSUSY}
\end{split}
\end{align}
Therefore for these two helicity configurations, it is sufficient to calculate
the $S=0$ amplitude where only a scalar circulates in the loop.

\subsection{Kinematic numerators through the double copy}

Following the double-copy construction we can directly obtain 
the gravitational unitarity-cut numerators from gauge-theory ones. 
We may express the double copy in terms of spinning particles or new pieces (\eqn{susyGR}) circulating in the loop. In the latter case, we find an especially compact representation for the numerators.

 Taking the
product of gauge-theory kinematic numerators we decompose them 
into cut numerators of the gravitational case. 
In terms of spin we have
\begin{align}
\begin{split}
n_{\text{g},\alpha}^{\tilde{S}} n_{\text{g},\alpha}^{S=0} & = n_{\text{GR},\alpha}^{\tilde{S}}\,,  \qquad \hbox{for $\tilde{S}= 0,1/2,1$} \,,\\
n_{\text{g},\alpha}^{S=1/2} n_{\text{g},\alpha}^{S=1/2} & = n_{\text{GR},\alpha}^{S=1} + n_{\text{GR},\alpha}^{S=0} \,, \\
n_{\text{g},\alpha}^{S=1} n_{\text{g},\alpha}^{S=1/2} & = n_{\text{GR},\alpha}^{S=3/2} + n_{\text{GR},\alpha}^{S=1/2} \,, \\
n_{\text{g},\alpha}^{S=1} n_{\text{g},\alpha}^{S=1} & = n_{\text{GR},\alpha}^{S=2} + n_{\text{GR},\alpha}^{S=1} + n_{\text{GR},\alpha}^{S=0} \,,
\end{split}
\end{align}
where $\alpha$ denotes the cut under consideration. These are in
direct correspondence to the Clebsch-Gordan decomposition.
In terms of the new pieces in \eqn{susyGR}, 
\begin{align}
\begin{split}
n_{\text{g},\alpha}^{\newPiece{\it S}} n_{\text{g},\alpha}^{\newPiece{0}} &= 
n_{\text{GR},\alpha}^{\newPiece{\it S}} \,,
 \qquad  \hbox{for $S= 0,1/2,1$ }, \\
n_{\text{g},\alpha}^{\newPiece{1/2}} n_{\text{g},\alpha}^{\newPiece{1/2}} &= 
n_{\text{GR},\alpha}^{\newPiece{1}} \,, \\
n_{\text{g},\alpha}^{\newPiece{1}} n_{\text{g},\alpha}^{\newPiece{1/2}} &= 
n_{\text{GR},\alpha}^{\newPiece{3/2}} \,, \\
n_{\text{g},\alpha}^{\newPiece{1}} n_{\text{g},\alpha}^{\newPiece{1}} &= 
n_{\text{GR},\alpha}^{\newPiece{2}} \,.
\end{split}
\end{align}

Observe that the gauge-theory numerator $n_{\text{g},\alpha}^{S=0}$ along with either 
$n_{\text{g},\alpha}^{S=1/2}$ or $n_{\text{g},\alpha}^{S=1}$ are
sufficient to construct all the gravitational numerators up to spin
2. We explicitly verified that both constructions yield the same result. Refs.~\cite{Bern:1993tz, BernMorgan} calculated the corresponding
amplitudes $\mathcal{A}_4^{S=0}$, $\mathcal{A}_4^{S=1/2}$ and
$\mathcal{A}_4^{S=1}$, from which we may extract the desired kinematic
numerators. As a consistency check, we match
$\mathcal{A}_4^{\newPiece{1}}$, which was calculated in
Ref.~\cite{Alday:2009zm}.

From the above construction we find a remarkably simple form
for the kinematic numerators. 
For the double-minus gauge-theory numerators we have,
\begin{equation}
n_{\text{g},\alpha}^{\newPiece{\it S}} = \big([14] \langle 23 \rangle \big)^2 (\psi_\alpha)^{2-2S} \,,
\end{equation}
while for the corresponding gravity numerators we have,
\begin{equation}
n_{\text{GR},\alpha}^{\newPiece{\it S}} = \big( [14] \langle 23 \rangle \big)^4 (\psi_\alpha)^{4-2S}\, ,
\end{equation}
where 
\begin{equation}
\psi_s \equiv \frac{\langle 2| \ell |1] \langle 3| \ell |4]}{s [14] \langle 23 \rangle}\,, \hskip 2 cm 
\psi_t \equiv \frac{(m^2+\mu^2)}{t}\,, \hskip 2 cm  
\psi_u \equiv \frac{\langle 3| \ell |1] \langle 2| \ell |4]}{u [14] \langle 32 \rangle} \,.
\end{equation}

For the single-minus and all-plus configurations it is sufficient to
calculate the numerators for a scalar circulating in the loop. For the
single-minus configuration for gauge theory and gravity we find
\begin{equation}
n_{\text{g}, s}^{S=0} = \frac{(m^2+\mu^2)}{s} \frac{[12]}{\langle 12 \rangle} \langle 3 | \ell | 4]^2,
\hskip 1.5 cm 
n_{\text{GR}, s}^{S=0} = \frac{(m^2+\mu^2)^2}{s^2} \frac{[12]^2}{\langle 12 \rangle^2} \langle 3 | \ell | 4]^4,
\end{equation}
while for the all-plus configuration we have
\begin{equation}
n_{\text{g}, s}^{S=0} = (m^2+\mu^2)^2 \frac{[12][34]}{\langle 12 \rangle \langle 34 \rangle} \,,
\hskip 1.5 cm
n_{\text{GR}, s}^{S=0} = (m^2+\mu^2)^4 \Bigg( \frac{[12][34]}{\langle 12 \rangle \langle 34 \rangle} \Bigg)^2\,.
\end{equation}
For these two configurations, we obtain the numerators for the $t$- and $u$-channel cuts by appropriate relabelings.

Following the
conventions of Ref.~\cite{BernMorgan}, we break the
$(4-2\epsilon)$-dimensional loop momentum $L$ into a
four-dimensional part $\ell$ and a $(-2\epsilon)$-dimensional part
$\mu$. We write $L = (\ell, \mu)$. Using this notation we have for example
\begin{equation}
\ell^2 = m^2 + \mu^2 \,,
\label{muDef}
\end{equation}
when the cut condition $L^2 = m^2$ is satisfied. We take $\epsilon <
0$ so that we can break the loop momentum in this fashion. Further,
whenever a four-dimensional vector $v \equiv (v,0)$ is contracted with
the $(4-2\epsilon)$-dimensional loop momentum, the
$(-2\epsilon)$-dimensional part is projected out,
\begin{equation}
v \cdot L = v \cdot \ell \,.
\end{equation}

Ref.~\cite{Arkani-Hamed:2017jhn} (Eqs.~(5.20) and (5.36)) calculated
Compton amplitudes for a massive particle in four dimensions of up to
spin 1 in gauge theory and up to spin 2 in gravity. We find similar
spin dependence in our double-minus numerators as for these Compton
amplitudes.

\subsection{Integral reduction and cut merging}
\label{MHVConstructionSection}

In this subsection we use standard integration-by-parts (IBP) methods  to reduce the generalized-unitarity cuts we
calculated in the previous section in terms of a basis of master
integrals. This allows us to fix all integral coefficients in
Eq.~(\ref{ReductionTarget}) other than those of the tadpole and the
bubble on external leg. We discuss these remaining pieces in Sect.~\ref{uvSec}.
We
show details for the double-minus configuration with a scalar in
the loop; the remaining helicity configurations and particle-in-the-loop
contributions are similar. In order to keep the expressions concise, we do not include the helicity labels.


The general strategy for constructing the full amplitudes is to
evaluate the cuts one by one in terms of the master integrals
appearing in Eq.~(\ref{ReductionTarget}).  If a given integral has a
generalized cut in the channel being evaluated then that channel fully
determines its coefficient.  By stepping through the three channels
in \fig{TwoParticleCutFigure} we obtain the coefficients
of all master integrals except $I_2(0)$ and $I_1$. The box 
integrals each have cuts in two channels so consistency requires
that they give the same coefficient.

We start with the $s$-channel cut of the $S=0$ double-minus helicity
configuration defined in \eqn{HelicityDef1}.  The discussion for the
$u$-channel cut follows in the same way, since it is given by a
relabeling of the $s$-channel one. We have
\begin{align}
i \mathcal{M}_4^{S=0} \bigg\rvert_{s\text{-cut}} 
= \left( \frac{\kappa}{2} \right)^4 (\langle 23 \rangle [14])^4
\int \frac{d^{4-2\epsilon} L}{(2\pi)^{4-2\epsilon}}  
\left( \frac{\langle 2| \ell |1] \langle 3| \ell |4]}{s [14] \langle 23 \rangle} \right)^4
\left( \frac{1}{D_{1234}} + \frac{1}{D_{1342}} \right)\bigg\rvert_{s\text{-cut}}.
\end{align}
We define 
\begin{equation}
v_1^\mu =  \langle 2| \gamma^\mu |1]\,,  \, \hskip 2 cm  v_2^\mu  =  \langle 3| \gamma^\mu |4] \,,
\end{equation}
which live in the four-dimensional subspace so that the $v_i$
effectively project out the $(4-2\epsilon)$-dimensional components.
This implies that $v_i \cdot \ell = v_i \cdot L$, as follows from the prescription~\cite{BernMorgan} that $\epsilon <0$ .  Next, we
lift the cut condition and regard our expression as part of the full
integrand that we would have obtained by Feynman rules,
\begin{equation}
i \mathcal{M}_4^{S=0, \,s\text{-channel}} = \left( \frac{\kappa}{2} \right)^4 (\langle 23 \rangle [14])^4
\int \frac{d^{4-2\epsilon} L}{(2\pi)^{4-2\epsilon}}  
\left( \frac{\langle 2| L |1] \langle 3| L |4]}{s [14] \langle 23 \rangle} \right)^4
\left( \frac{1}{D_{1234}} + \frac{1}{D_{1342}} \right),
\end{equation}
keeping in mind that it is only valid for contributions that
have an $s$-channel cut.  
We apply
IBP identities~\cite{IBP} in $4-2\epsilon$
dimensions to reduce the above integrals to the master integrals in
Eq.~(\ref{ReductionTarget}), using the
software FIRE6~\cite{Smirnov:2019qkx}.  Upon reducing to master
integrals we remove the ones that have no $s$-channel cut.

Next, we turn to the $t$-channel cut. We have
\begin{equation}
i \mathcal{M}_4^{S=0}\bigg\rvert_{t\text{-cut}} = \left( \frac{\kappa}{2} \right)^4 (\langle 23 \rangle [14])^4
\int \frac{d^4 \ell}{(2\pi)^4} \frac{d^{-2\epsilon} \mu}{(2\pi)^{-2\epsilon}}
\left( \frac{(m^2+\mu^2)}{t} \right)^4
\left( \frac{1}{D_{1234}} + \frac{1}{D_{1423}} \right)\bigg\rvert_{t\text{-cut}}, \quad
\end{equation}
where we subdivide the integration into four- and $(-2\eps)$-dimensional parts
\begin{equation}
\int \frac{d^{4-2\epsilon} L}{(2\pi)^{4-2\epsilon}}  = 
\int \frac{d^4 \ell}{(2\pi)^4} \frac{d^{-2\epsilon} \mu}{(2\pi)^{-2\epsilon}} \, .
\label{momBreak}
\end{equation}
As for the $s$ channel, we lift the cut conditions and regard our expression as part of the full integrand,
\begin{equation}
i \mathcal{M}_4^{S=0, \,t\text{-channel}} = \left( \frac{\kappa}{2} \right)^4 (\langle 23 \rangle [14])^4
\int \frac{d^4 \ell}{(2\pi)^4} \frac{d^{-2\epsilon} \mu}{(2\pi)^{-2\epsilon}} 
\left( \frac{(m^2+\mu^2)}{t} \right)^4
\left( \frac{1}{D_{1234}} + \frac{1}{D_{1423}} \right).
\end{equation}
As we discuss in Appendix~\ref{IBPsec}, the integrals with the
$(-2 \eps)$-dimensional components of loop-momentum $\mu$ in the
numerators can be expressed directly in terms of the master integrals
defined in Eq.~(\ref{4dMasters}).  After reducing our expression we
eliminate master integrals that vanish on the $t$-channel cut.

As noted above, an important consistency condition arises 
from the fact that the box integrals have cuts
in two channels, so we can determine their coefficients from
either channel.  We explicitly verified that the coefficients we
obtain for the box integrals from looking at the different channels are
the same. In this way we are able to extract all coefficients of
Eq.~(\ref{ReductionTarget}) other than $a_0$ and $b_0$, since the
corresponding integrals have no cuts in any channel; we obtain these
remaining two integral coefficients in Sect.~\ref{uvSec}.  


\subsection{Ultraviolet behavior and rational pieces}
\label{uvSec}

\begin{figure}
\centering
\includegraphics[scale=1.]{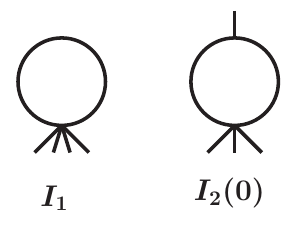}
\vskip -1. cm
\caption{\captionlineskip The tadpole and bubble-on-external-leg integrals. }
\label{unfixed}
\end{figure}

Analyzing the different generalized-unitarity cuts we obtain all the coefficients in Eq.~(\ref{ReductionTarget}) other than $a_0$ and $b_0$.
Tadpoles and bubbles on external legs (see Fig.~\ref{unfixed}) vanish on any cut, therefore their coefficients are not accessible through generalized unitarity. In this subsection we use the known UV properties of the amplitude to obtain these coefficients. Similarly to Sect.~\ref{MHVConstructionSection}, we discuss the double-minus configuration as a specific example. The other configurations follow in the same manner.

First, we observe that simply neglecting these two integrals leads to an inconsistent answer. Expanding to leading order in $\epsilon$ we get
\begin{equation}
\mathcal{M}_4^{S=0}\bigg\rvert_{a_0 \rightarrow 0, \, b_0 \rightarrow 0} =   (\langle 23 \rangle [14])^4 \frac{m^2 \tilde{Q}(s,t,m)}{\epsilon} + \mathcal{O}(\epsilon^0) \,,
\label{UVdiv}
\end{equation}
where $\tilde{Q}(s,t,m)$ is some rational function. We note that we expect no UV divergence to appear in the four-graviton amplitude since the only counterterm that we could write to absorb it is the Gauss-Bonnet term, which is evanescent in four dimensions~\cite{tHooft:1974toh}. Also, the UV divergence not coming out local hints that we neglected to include some integrals. 

We use the vanishing of the UV divergence to obtain the remaining coefficients $a_0$ and $b_0$. Note that since $a_0$ and $b_0$ are rational functions of $\epsilon$,  it seems that we do not have enough information to fully fix them. We surpass this difficulty by realizing that our problem admits a second integral basis where the integral coefficients do not depend on $\epsilon$ and $m$. This basis is overcomplete in that it contains ($4-2\epsilon$)- and
higher-dimensional integrals, while the one introduced in Eq.~(\ref{ReductionTarget})
only includes ($4-2\epsilon$)-dimensional integrals.
We discuss this basis in detail and provide an algorithmic way of switching between the two bases in Appendix~\ref{IBPsec}.

We start by bringing the quantity we get from cut merging to the basis containing higher-dimensional integrals. In doing so, we remove any integrals that vanish on all cuts. We refer to this piece expressed in this basis as the `on-shell-constructable' piece. 
Then, we need to figure out which integrals without unitarity cuts we need to include in our expression, since in principle one could add infinitely many higher-dimensional integrals.

We consider the pieces that would arise in a Feynman-rules calculation. For the amplitude at hand we would find integrals up to quartically divergent. After we IBP reduce them and express them in the basis that contains the higher-dimensional integrals, we would in principle find all scalar integrals up to quartically divergent appearing. Hence we need to consider:
\begin{equation}
I_1, \qquad I_2(k^2)\,, \qquad I_1^{6-2\epsilon}\,, \qquad I_2^{6-2\epsilon}(k^2)\,, \qquad I_2^{8-2\epsilon}(k^2)\,,
\end{equation}
where we imagine the limiting case $k^2\rightarrow 0$. For our definition of higher-dimensional integrals see \eqn{eq:genericIntegralDef}. These are all the integrals that have no unitarity cuts and are up to quartically divergent. The reason why we consider the above limiting case is because the coefficients of these integrals might have a $1/k^2$ pole.

Then, the unknown piece in our amplitude up to normalization takes the form
\begin{equation}
\lim_{k^2\rightarrow 0} \Big( \beta_1 I_1 + \beta_2 I_2(k^2) + \beta_3 I_1^{6-2\epsilon} + \beta_4 I_2^{6-2\epsilon}(k^2) + \beta_5 I_2^{8-2\epsilon}(k^2) \Big) 
= \gamma_1 I_1 + \gamma_2 I_2(0) + \gamma_3 I_1^{6-2\epsilon} \,,
\label{UVpiece}
\end{equation}
where we use
\begin{equation}
I_2^{6-2\epsilon}(k^2) =  I_1 + \frac{k^2}{6} I_2(0) + \mathcal{O}(k^4) \,, \quad \quad
I_2^{8-2\epsilon}(k^2) =  I_1^{6-2\epsilon} + \frac{k^2}{6} I_1 + \mathcal{O}(k^4) \,,
\end{equation}
following Appendix~B of Ref.~\cite{BernMorgan}.  The unknown coefficients
$\gamma_i$ are rational functions of the kinematics that do not
contain $\epsilon$ or $m$ dependence. In writing this expression we
assume that the coefficients of the integrals are at worse divergent as
$1/k^2$ and that the expression is finite in the $k^2\rightarrow 0$
limit when a massive particle is circulating in the loop.\footnote{\footnotelineskip We drop ill-defined `cosmological constant' tadpole
diagrams with a $1/ \left( \sum_{i=1}^4 k_i \right)^2$ propagator.}

Upon these considerations, our amplitude takes the form
\begin{equation}
\mathcal{M}_4^{S=0} = \mathcal{M}_4^{S=0}\bigg\rvert_\text{on-shell-constr.} + \frac{1}{(4\pi)^{2-\epsilon}} \left( \frac{\kappa}{2} \right)^2 \mathcal{M}^\text{tree}_4 \big( \gamma_1 I_1 + \gamma_2 I_2(0) + \gamma_3 I_1^{6-2\epsilon} \big) \,,
\label{EqFullAmpAnsatz}
\end{equation}
where the $\gamma_i$ are coefficients are determined from the requirement that ultraviolet divergences cancel, the integrals 
$I_1$ and $I_2$ are given in \eqns{TadpoleIntegral}{Bubble0Integral}, and
\begin{equation}
\mathcal{M}^\text{tree}_4 \equiv \mathcal{M}^\text{tree}_4(1^{+},2^{-},3^{-},4^{+}) =  \left( \frac{\kappa}{2} \right)^2 \frac{(\langle 23 \rangle [14])^4}{s t u} \,.
\label{mtree}
\end{equation}
The Mandelstam variables are defined below \eqn{4dMasters}. Note that the little-group scaling for the unknown terms is fixed to be $(\langle 23 \rangle [14])^4$. Since the $\gamma_i$ are rational functions of the kinematics, we are free to multiply and divide by $(s t u)$ to introduce $\mathcal{M}^\text{tree}_4$.

Next, demanding that the amplitude has no UV divergence and that all
three $\gamma_i$ are independent of the mass $m$ uniquely fixes
$\gamma_1$ and $\gamma_3$ to nonzero values while setting $\gamma_2 =
0$. The results for the amplitudes collected in Appendix~\ref{AmplitudesResultsAppendix}
the values of the $\gamma_i$ are all chosen to make the amplitudes UV finite.

As a simple consistency check, we repeat this analysis adding integrals
that have no unitarity cuts and are divergent worse than quartically
(namely ($8-2\epsilon$)- and higher-dimensional tadpoles). We verify
that the answer is the same, i.e. the coefficients of these new
integrals are set to zero.

We now briefly comment on the $m\rightarrow 0$ limit. In this limit,
both $I_1$ and $I_2(0)$ vanish, and the amplitude has no UV
divergence, as can be seen from Eq.~(\ref{UVdiv}). Hence our amplitude
has the correct UV behavior in this limit as well. Moreover, all
integrals that have no unitarity cuts become scaleless and hence zero
in dimensional regularization. Therefore, there are no unfixed
coefficients and our construction based on the unitarity cuts captures
the full amplitude.

Finally, we want to clarify why we chose the overcomplete basis with
higher-dimensional integrals. The coefficients $a_0$ and $b_0$ are in
principle arbitrary rational functions of $\epsilon$ and $m$, and of
the kinematic variables $s$ and $t$. The existence of a basis where
the integral coefficients do not contain $\epsilon$ or $m$
significantly restricts the functional dependence of $a_0$ and $b_0$
on $\epsilon$ and $m$. The existence of such a basis is a nontrivial
fact that may be understood via analyzing the calculation in a
covariant gauge, as we explain in Appendix~\ref{IBPsec}. In this way
the two integral bases are not completely equivalent, since the latter one is
exploiting more of the specific properties of the problem at hand.

\subsection{Further ultraviolet properties}

\mysubsubsection{Quadratic and quartic divergence}

Next, we want to analyze the quadratic and quartic divergence of our amplitude. In dimensional regularization, the signature of these divergences are poles around $\epsilon = 1$ and $\epsilon = 2$ in the final answer. Since we compute our amplitude to all order in $\epsilon$ we may probe these poles. We tackle this question in the basis that contains only ($4-2\epsilon$)-dimensional integrals. In the basis that includes higher-dimensional integrals, many of the integrals are quadratically or quartically divergent, which obscures the analysis. We discuss the double-minus configuration for a spin-0 particle in the loop; the other cases are similar. As we demonstrate, our amplitude has no quadratic or quartic divergences.

We start with the quadratic divergence. In the chosen basis the only
quadratically divergent integral is the tadpole. However, the
coefficient of the tadpole is linear in $(\epsilon - 1)$, hence there
is no contribution to the quadratic divergence from it. It then
suffices to expand all coefficients around $\epsilon = 1$ and only
keep the divergent piece. We get
\begin{align}
\mathcal{M}_4^{S=0} &= 
-\frac{1}{(4\pi)^{2-\epsilon}} \left( \frac{\kappa \null}{2} \right)^2 \mathcal{M}^\text{tree}_4
\frac{s u}{(\epsilon-1)t^6}
(3 s^2 u^2 + 10m^2 s t u + 6m^4 t^2)
\Big( \iota_2(s) + \iota_2(u) \Big) \nn \\[4pt]
 & \hskip 3 cm \null 
+ \mathcal{O}\Big( (\epsilon-1)^0 \Big) \,,
\end{align}
where we introduced 
\begin{equation}
\iota_2(s) = 
s {} \big( 2 m^2 I_3(s) - I_2(s) \big)
\,.
\end{equation}
Note that $I_n \equiv I_n^{4-2\epsilon} = I_n^{2-2(\epsilon-1)}$. It appears that our amplitude has a quadratic divergence which is non local due to the $1/t^6$ factor. However, using Eq.~(\ref{dimShift}) with $D=2-2(\epsilon-1)$ we get
\begin{equation}
I_3^{4-2(\epsilon-1)}(s) = 
\frac{1}{2(\epsilon-1)}
\big( -2m^2 I_3^{2-2(\epsilon-1)}(s) + 
I_2^{2-2(\epsilon-1)}(s) \big)
\,,
\end{equation}
which gives
\begin{equation}
\iota_2(s) = 
-2 s {} (\epsilon-1) I_3^{4-2(\epsilon-1)}{(s)}
\, .
\end{equation}
Since $I_3^{4-2(\epsilon-1)}(s)$ is finite as $\epsilon \rightarrow 1$, we see that there is no quadratic divergence.

The analysis of the quartic divergence follows in a similar manner. In this basis there are no quartically-divergent integrals. We expand our result around $\epsilon = 2$ to get
\begin{align}
\mathcal{M}_4^{S=0} = 
\frac{1}{(4\pi)^{2-\epsilon}} \left( \frac{\kappa \null}{2} \right)^2 \mathcal{M}^\text{tree}_4 
\frac{s u}{(\epsilon-2)t^5}
(5 s u +6m^2 t)
\Big( \iota_4(s) + \iota_4(u) \Big) 
+ \mathcal{O}\Big( (\epsilon-2)^0 \Big)
\,,
\end{align}
where 
\begin{equation}
\iota_4(s) =
 2 s I_1 + s {} (s-6m^2)I_2(s) + 
4 m^4 s I_3(s)
\,.
\end{equation}
Using  Eq.~(\ref{dimShift}) two consecutive times we get 
\begin{equation}
\iota_4(s) = 
- 4 s {} (\epsilon -2) I_3^{4-2(\epsilon-2)}{(s)}
\,,
\end{equation}
which shows that there is no quartic divergence.

In our problem we demonstrated that the coefficients of the integrals
which have no unitarity cuts in \eqn{EqFullAmpAnsatz} contain no $m$
dependence. This property was crucial in order to fix them using the
vanishing of the (logarithmic) UV divergence. In a more general
situation we expect this property to no longer be true. In such a
scenario, analyzing the higher UV divergences offers an alternative
method for obtaining these coefficients. For example, if we demand the
vanishing of the quadratic and quartic divergences along with the
logarithmic one for our problem, we may fix the coefficients
$\gamma_i$ to the values we found above, without needing to impose that they do not contain
$m$ dependence.

\mysubsubsection{Ultraviolet divergences in higher dimensions}

We can also inspect the ultraviolet properties in higher dimensions.
This is straightforward because we obtain expressions for the
amplitudes valid to all orders in $\epsilon$.  In the calculations we
keep the external kinematics and helicities fixed in four dimensions.
In addition, we use the FDH scheme~\cite{FDHScheme} which keeps the
number of physical states circulating in the loop fixed at their
four-dimensional values. However, we can still analytically continue
the loop momentum to any dimension and study the divergence
properties.  We can use this as a rather nontrivial check on our
expressions and also to point out a simple relation between
appropriately defined divergences in higher dimensions and
coefficients of terms in the large-mass expansion. We discuss the double-minus configuration with a scalar in the loop for concreteness, however our results hold for all cases considered in this paper.

Since our expressions are valid to all orders in the dimensional-regularization parameter $\epsilon$, we
may shift the spacetime dimension of the loop momentum by $2 \sigma$
via $\epsilon \rightarrow \epsilon - \sigma$. For example for
$\sigma = 1$ the spacetime dimension is shifted to
$D=6-2\epsilon$. In the shifted dimension we define the coefficients of
the ultraviolet divergences as
\begin{equation}
\mathcal{M}_4^{S=0} \big \vert_{\epsilon \rightarrow \epsilon - \sigma} =
\frac{1}{(4 \pi)^\sigma} \frac{1}{(\sigma-1)!} \frac{1}{\epsilon} 
\sum_{n=0}^{\sigma-1} \left(m^2\right)^n \delta_{\sigma,\,n}^{UV}(s,t) 
 + \mathcal{O}\left(\epsilon^0\right) ,
\label{UVShiftDim}
\end{equation}
where we choose the normalization in a particular way to
account for the angular loop-integration in different
dimensions.  Note that there is no UV divergence for $\sigma =0$,
corresponding to $D=4-2\eps$, due to the lack of a corresponding
counterterm matrix element.  For $\sigma = 1$ the counterterm is an
$R^3$-type operator.  In this case, up to a numerical factor the
coefficient of the divergence matches the tree amplitude generated
from an $R^3$ insertion, which we describe in
\sect{EFTMatchingSection}.  For convenience we restrict ourselves to
even dimensions (for $\eps \rightarrow 0$) because in these cases the
dimension-shifting formulas (\eqn{dimShift}) bring the higher-dimensional
integrals back to four dimensions.  For $ 2 \le \sigma \le 6$ we have
explicitly confirmed that the divergences are local and therefore
correspond to appropriate derivatives of $R^4$-type operators, as
expected.
 We note that because individual integrals have nonlocal
coefficients the fact that the divergences are local provides a rather
nontrivial check on our expressions.  Furthermore, starting at $D=8-2
\eps$ all the integrals are divergent, so their coefficients feed into
this check.

Finally, we point out a relation between
the ultraviolet divergences in Eq.~(\ref{UVShiftDim}) and terms
in the large-mass expansion in four dimensions.  
Specifically, defining 
\begin{equation}
\mathcal{M}_4^{S=0} = \sum_{n=1}^{\infty} \frac{\delta^{IR}_n(s,t)}{m^{2n}} \,,
\end{equation}
we have explicitly checked through $\sigma = 4$ or equivalently $D =12 - 2\eps$ that 
\begin{equation}
\delta^{UV}_{\sigma,0} = \delta^{IR}_\sigma \,.
\label{eq:UVIRequiv}
\end{equation}
Similarly, we find that $\delta^{UV}_{\sigma,n}$ is proportional to
$\delta^{UV}_{\sigma-n,0} = \delta^{IR}_{\sigma-n}$ with a common
proportionality constant for all multiplets.  
For this
correspondence to hold it is important to keep both the external and internal states at their four dimensional value; only the loop momentum is analytically
continued to higher spacetime dimensions.  
It is quite remarkable
that such simple relations exist between the coefficients of
the divergences in higher dimensions and the terms in the
large-mass expansion of the amplitudes in four dimensions.

\subsection{Consistency checks}
\label{Consistency}

We have carried out a variety of checks on our amplitudes.  Basic
self-consistency checks are that ultraviolet and infrared or mass
singularities be of the right form.

Ultraviolet singularities must be local.  In general this is
nontrivial and happens only after the ultraviolet singularities are
combined.  In our case, the coefficient of the divergences vanishes.
We also verified that the $1/(\epsilon -1)$ pole cancels, consistent
with the fact that the Ricci-scalar counterterm vanishes by the
equation of motion.  Similarly for the $1/(\epsilon-2)$ pole, the
expression is not only local but it vanishes. Further, we verified
that the divergences obtained by analytically continuing the loop
momentum to higher dimensions while keeping the state counts and
external kinematics to their four-dimensional values are also local.

Another nontrivial check comes from looking at the $m \rightarrow 0$
limit.  Since the internal lines are massive, there is no IR
divergence for our amplitude. However, we may regard the mass of the
internal lines as an infrared regulator for the corresponding massless
amplitude and study the IR divergence as $m \rightarrow 0$.  In
gravity the infrared singularities are quite simple since there are
only soft singularities and no collinear or mass
singularities~\cite{GravIR}.  The soft singularities arise only from
gravitons circulating in the loop. This implies that as $m \rightarrow
0$ all contributions must be infrared finite except for the
$\mathcal{M}_4^{\newPiece{2}}$ piece (corresponding to $\mathcal{N} = 8$ supergravity in the
massless limit), since this is the only piece that has a graviton
circulating in the loop in this limit.

To carry out this check we start with the on-shell-constructable
piece. We start from the ($4-2\epsilon$)-dimensional integral basis,
where only the boxes and the triangles have IR singularities. The
triangles only contain simple logarithms (see Eq.~(\ref{triangles}))
while the boxes contain both logarithms and dilogarithms (see
Eq.~(\ref{boxes})). To simplify the check we use Eq.~(\ref{dimShift})
to trade the boxes for triangles and ($6-2\epsilon$)-dimensional
boxes. Importantly, the latter have no IR singularities. In this form
the infrared singularities are all pushed into triangle integrals,
hence it suffices to verify that their coefficients vanish as
$m \rightarrow 0$, which we confirm for all but the
$\mathcal{M}_4^{\newPiece{2}}$ piece.  Next we look at the contributions with no
unitarity cuts. These pieces are zero if we take $m \rightarrow 0$
before we expand in $\epsilon$. On the other hand, divergent terms
appear if we first expand in $\epsilon$ and then take $m \rightarrow
0$. However, given that the ultraviolet divergence vanishes in
four-dimensions these infrared divergent $\log(m)$ pieces also cancel
among themselves.  For the $\mathcal{M}_4^{\newPiece{2}}$ piece 
there is indeed a infrared
singularity that develops as $m\rightarrow 0$.  In this case, we
recover the known infrared divergence of pure Einstein
gravity~\cite{GravIR}.

As another check, we have explicitly verified that as $m\rightarrow 0$
our massive results match the massless ones given in
Ref.~\cite{Dunbar:1994bn}, up to an overall sign in
$\mathcal{M}_4^{\newPiece{1/2}}$, as noted in
Ref.~\cite{Abreu:2020lyk}.  One simple way to implement this check is
to start with the expressions for the amplitudes in the
$(4-2\epsilon)$-dimensional integral basis, set $m \rightarrow 0$ in
the integral coefficients, and replace the massive integrals with
massless ones.

Finally, we note that contributions from individual integrals do not
decay at large mass as required by decoupling, while the amplitudes
have the required property. This involves nontrivial cancellation
between the pieces, providing yet another check.

\section{Amplitudes in the low-energy effective field theory}
\label{EFTMatchingSection}

In this section we study four-graviton scattering in a general parity-even low-energy EFT. Such EFTs start from the Einstein-Hilbert action and
extend to systematically include higher-dimension operators.  We include a massless scalar field to our analysis, corresponding to the dilaton found in string theory. 

We match this EFT to the one-loop amplitudes determined in \sect{AmplitudeConstructionSection} and collected
in Appendix~\ref{AmplitudesResultsAppendix}. In this context, the EFT is valid for
energies significantly smaller than the mass of the spinning particle
in the loop.  In this way we determine the modification to the
low-energy theory of gravity due to the presence of a heavy particle.
We take this as a nontrivial model of ultraviolet physics feeding
into low-energy physics.

For the lowest-dimension operators, we calculate the four-point
tree-level amplitudes in this EFT and compare them to the expansion of our
one-loop amplitudes in the large-mass limit in order to obtain their Wilson
coefficients. More generally,  since we later put bounds on the coefficients
appearing in the amplitudes themselves, there is no need to relate
these back to a Lagrangian.  For comparison to the bounds derived in
subsequent sections we also present the one-loop scattering amplitudes
expanded in the large-mass limit. In Appendix~\ref{HighOrdersAppendix}
we present the expansions to much higher orders, which should be
useful for further studies of the bounds.

Finally,  we obtain the Regge limits of our one-loop amplitudes. These are useful later in analyzing low-energy coefficients via dispersion relations.

\subsection{Setup of the effective field theory}

The first few terms of the EFT describing low-energy gravitational scattering are
\begin{equation}
S_{\rm EFT} = \int d^4x \sqrt{-g} \,\biggl[
	-\frac{2}{\kappa^2}R + \frac{1}{2} \partial_\mu \phi \partial^\mu \phi + 
	\frac{2\beta_{\phi}}{\kappa^2}  \phi \, C +
 \frac{8}{\kappa^3} \frac{\beta_{R^3}}{3!} \, R^3 +  
 \frac{2\beta_{R^4}}{\kappa^4} C^2 + \frac{2\tilde{\beta}_{R^4}}{\kappa^4} \tilde{C}^2 
 + \ldots \biggr] \,,
\label{eft}
\end{equation}
 where $R$ is the Ricci scalar,
$\kappa$  is given in terms of Newton's constant $G$ via
$\kappa^2 = 32\pi G$, and the metric is $g_{\mu\nu} = \eta_{\mu\nu} +
\kappa h_{\mu\nu}$ in terms of the graviton field $h_{\mu\nu}$.  To describe the most general parity-even theory that captures low-energy four-graviton scattering, we include the massless scalar field $\phi$. The
factors of ${1 \over \kappa}$ are chosen to normalize the kinetic term
canonically and to remove the factors of $\kappa$ that would appear in
the three-point tree-level amplitudes built out of a single insertion of $\phi \, C$ or $R^3$ and the four-graviton tree-level amplitudes built out of a single insertion of $C^2$ or $\tilde{C}^2 $. The $\beta_\phi$, $\beta_{R^3}$, $\beta_{R^4}$
and $\tilde{\beta}_{R^4}$ are Wilson coefficients that depend on the details of the UV physics. The composite operators are defined by
\begin{align}
R^3 \equiv R^{\mu \nu \kappa \lambda} R_{\kappa \lambda \alpha \gamma} R^{\alpha \gamma}_{\quad \mu \nu} \,,&
\qquad
C \equiv R^{\mu \nu \kappa \lambda}R_{\mu \nu \kappa \lambda} \,,
\qquad
\tilde{C} \equiv  \frac{1}{2}  R^{\mu \nu \alpha \beta}  \epsilon_{\alpha \beta}^{\quad \gamma \delta} R_{\gamma \delta \mu \nu} \,, 
\label{eq:definitionsRiemannInvariants}
\end{align}
where $R_{\lambda \mu \nu \kappa}$ is the Riemann tensor. 
One can systematically add higher-dimension operators; we choose not to do so here since our later analysis is at the amplitude level, so 
the mapping back to Lagrangian coefficients is not necessary.

In writing the effective action \eqref{eft} we apply the equations of
motion and integrate by parts to reduce the number of terms to a
minimum independent set. In particular, in constructing the
higher-dimensional operators we replace instances of the Ricci scalar
and tensor, $R$ and $R_{\mu \nu}$, with appropriate contractions of
the matter stress-energy tensor.
We drop such terms since they give rise to higher-point matter
interactions, which do not affect our analysis.
For example, we do not include $R^2$-type terms because the squares of
$R$ and $R_{\mu \nu}$ do not contribute due to the equations of
motion, while the contraction of two Riemann tensors $C$ can be traded
for the Gauss-Bonnet contribution, which is equal to a total
derivative in four dimensions~\cite{tHooft:1974toh}. Furthermore, we
do not include operators that our calculation is not sensitive to.
Specifically, we do not consider the other possible contraction of
three Riemann tensors, since it does not contribute to four-graviton
scattering~\cite{vanNieuwenhuizen:1976vb,Broedel:2012rc}.  We restrict
ourselves to parity-even interactions and neglect the parity-odd
operators $\phi \, \tilde{C}$ and $C
\tilde{C}$~\cite{Sennett:2019bpc,Endlich:2017tqa}. The possible
parity-even contractions of four Riemann tensors were obtained in
Ref.~\cite{Fulling:1992vm}.  Recent studies that use similar
Lagrangians are found in Refs.~\cite{Endlich:2017tqa, Dunbar:2017szm,
  Brandhuber:2019qpg, AccettulliHuber:2020oou, Emond:2019crr}.

\subsection{Scattering amplitudes in the effective field theory}

To describe the amplitudes it is useful to extract overall dependence on the spinors, leaving only 
functions of $s,u$ with simple crossing properties. Specifically, for the
independent helicity configurations we define,
\begin{align}
\mathcal{M}_4(1^{+},2^{-},3^{-},4^{+})\, &=  (\langle 23 \rangle [14])^4 \, f (s,u) \,,  \nn \\
\mathcal{M}_4(1^{+},2^{+},3^{-},4^{+})\, &= 
\left( [12] [14] \langle 13 \rangle \right)^4 \, g(s,u) \,,  \nn \\
\mathcal{M}_4(1^{+},2^{+},3^{+},4^{+}) &=  \left( \frac{[12][34]}{\langle 12 \rangle \langle 34 \rangle} \right)^2 \, h(s,u) \,,
\label{HelicityDef}
\end{align}
corresponding to the \textit{double-minus}, \textit{single-minus} and \textit{all-plus} helicity configurations.
As usual we do not include the overall $i$ that normally would appear in 
Feynman diagrams.
All other amplitudes are given by permutations and complex conjugation,
\begin{equation}
\label{eq:hermitiananalyticity}
\mathcal{M}_4(1^{h_1},2^{h_2},3^{h_3},4^{h_4}) = \mathcal{M}_4^*(1^{-h_1},2^{-h_2},3^{-h_3},4^{-h_4}) \, .
\end{equation}
where we define the complex conjugation to not act on the $i \epsilon$  prescription.\footnote{\footnotelineskip More precisely, given complex conjugation $\gamma: z \mapsto z^*$, we define $f^*(z)$ as $f^* \equiv \gamma \circ f \circ \gamma: z \mapsto (f(z^*))^*$. For identical scalar particles \eqref{eq:hermitiananalyticity} becomes the familiar hermitian analyticity.}
In parity-preserving theories complex conjugation acts only on the spinors swapping the positive and negative helicity 
spinors,
\begin{equation}
\mathcal{M}_4(1^{h_1},2^{h_2},3^{h_3},4^{h_4}) = \mathcal{M}_4 (1^{-h_1},2^{-h_2},3^{-h_3},4^{-h_4}) \Bigr|_ {\lambda_i \leftrightarrow \tilde \lambda_i} \,,
\end{equation}
so that the $f$, $g$ and $h$ functions are unaltered.

In general relativity the leading-order results for the amplitudes above take the form
\begin{align}
\label{eq:oneloopGR}
f_{\text{GR}}(s,u) &=  \left( \frac{\kappa \null}{2} \right)^2 {1 \over s t u} + \ldots ,  \,,\\
g_{\text{GR}}(s,u) &= \frac{1}{(4\pi)^{2}} \left( \frac{\kappa \null}{2} \right)^4 \frac{1}{s^2 t^2 u^2} \frac{s^2 + t^2 + u^2}{360} + \ldots \,,  \\
h_{\text{GR}}(s,u) &= -\frac{1}{(4\pi)^{2}} \left( \frac{\kappa \null}{2} \right)^4  \frac{s^2 + t^2 + u^2}{120} + \ldots \, .
\end{align}
Below we do not consider loop effects in general relativity itself
(due to gravitons circulating in the loops) but focus on the
properties of the higher-derivative operators in the gravitational EFT
generated by integrating out massive degrees of freedom. For the same
reason IR divergences are not an issue for our analysis since the
corrections of interest are manifestly IR finite.


Permutation symmetry of the amplitudes plays a crucial role in our
analysis and manifests itself in the following crossing relations
\begin{align}
\label{eq:crossing}
f(s,u) &= f(u,s) \, ,  \\
g(s,u) &= g(u,s) = g(s,t) \,,  \\
h(s,u) &= h(u,s) = h(s,t) \,. 
\end{align} 
where $s+t+u=0$.

With our normalizations the three-point amplitude arising from the Einstein term
is\footnote{\footnotelineskip We implicitly use complex momenta
so that the three-point amplitude does not vanish from kinematic constraints.}
\begin{equation}
\mathcal{M}_3^{\rm GR}(1^{+},2^{+},3^{-}) = \frac{\kappa}{2}
\biggl( \frac{ [ 12 ]^3} {[23] [31]} \biggr)^2 \,.
\end{equation}
The three-points amplitudes with an insertion of the $\phi \, C$ or the  $R^3$ operator are~\cite{Broedel:2012rc}
\begin{equation}
\label{eq:threepointR3}
\mathcal{M}_3^{\phi \, C}(1^{+},2^{+},3^\phi) = \beta_\phi [12]^4
 \,, ~~~
\mathcal{M}_3^{R^3}(1^{+},2^{+},3^{+}) = 
\beta_{R^3}
 \Big( [12] [23] [31] \Big)^2 \,,
\end{equation}
where $\beta_\phi$ and $\beta_{R^3}$ are the Wilson coefficient for these operators appearing in \eqn{eft}.

At four points $\phi \, C$ contributes to the double-minus and all-plus configurations,
\begin{align}
\mathcal{M}_4^{\phi \, C}(1^+,2^+,3^+,4^+) &= -3 (\beta_\phi)^2 s t u \left( \frac{[12][34]}{\langle 12 \rangle  \langle 34 \rangle} \right)^2 
\,, \nn \\
\mathcal{M}_4^{\phi \, C}(1^+,2^-,3^-,4^+) &= - (\beta_\phi)^2 \frac{\bigl(\langle 23 \rangle [14] \bigr)^4}{t}
 \,.
\end{align}
On the other hand,
there are two independent helicity configurations, the
all-plus and the single-minus configurations, that contain a single
insertion of $R^3$. We obtain these amplitudes following
Ref.~\cite{Broedel:2012rc}.  We find
\begin{align}
\mathcal{M}^{R^3}_4(1^+,2^+,3^-,4^+)
& = \beta_{R^3} \Bigl(\frac{\kappa}{2}\Bigr)  \frac{1}{s t u}  \left( [12] [14] \langle 13 \rangle \right)^4\,,
\end{align}
and 
\begin{align}
\mathcal{M}^{R^3}_4(1^+,2^+,3^+,4^+)
&= 10 \beta_{R^3} \Bigl(\frac{\kappa}{2}\Bigr) s t u \left( \frac{[12][34]}{\langle 12 \rangle  \langle 34 \rangle} \right)^2 \,,
\end{align}
which is slightly rearranged compared to Ref.~\cite{Broedel:2012rc}.  We build a double-minus contribution out of two insertions of $R^3$~\cite{BiaHigherOrder},
\begin{equation}
\mathcal{M}_4^{R^3}(1^{+},2^{-},3^{-},4^{+}) = (\beta_{R^3})^2 \frac{su}{t} \bigl(\langle 23 \rangle [14] \bigr)^4 \,.
\end{equation}
For the $R^4$-type operators the amplitudes are~\cite{AccettulliHuber:2020oou} 
\begin{align}
\begin{split}
\mathcal{M}_4^{R^4}(1^{+},2^{+},3^{+},4^{+}) &= \beta_{R^4}^-  \frac{(s^2 +t^2+u^2)^2}{2}
\left( \frac{[12]\, [34]}{\langle 12 \rangle \, \langle 34 \rangle} \right)^2 \,, \\
\mathcal{M}_4^{R^4}(1^{+},2^{-},3^{-},4^{+}) &= \beta_{R^4}^+ \bigl(\langle 23 \rangle [14] \bigr)^4 \,,
\end{split}
\end{align}
where by $\mathcal{M}_4^{R^4}$ we refer to the amplitudes build out of both $C^2$ and $\tilde{C}^2$, with
\begin{equation}
\label{eq:R4definition}
\beta_{R^4}^{\pm} \equiv \beta_{R^4} \pm \tilde{\beta}_{R^4} \,,
\end{equation} 
Using these four-point amplitudes we may extract the coefficients
$\beta_{R^3}$, $\beta_{R^4}$ and $\tilde{\beta}_{R^4}$ by matching
to our one-loop calculation in the large-mass limit. Since we did not include the massless scalar field in the construction of our one-loop amplitudes, we have $\beta_\phi=0$ in this case.

Next, we bring our one-loop amplitudes in a form suitable to compare
to the EFT amplitudes listed above. 
We start with the double-minus amplitude.
As usual, we organize the
contributions according to the supersymmetric decomposition \eqref{susyGR},
\begin{equation}
\mathcal{M}^{\newPiece{\it S}}_4(1^{+},2^{-},3^{-},4^{+}) = (\langle 23 \rangle [14])^4 \ f^{\newPiece{\it S}} (s,u) \,.
\end{equation}
In the large-mass limit, for the double-minus amplitudes
we have,
\begin{align}
f&^{\newPiece{0}}
=  \mathcal K  \biggl(
 \frac{1}{6300 m^4} + \frac{t}{41580 m^6} 
+ \frac{81(s^2+u^2) + 155 s u}{15135120 m^8} 
+ \frac{ \big(161 (s^2+u^2) + 324 s u \big) t}{151351200 m^{10}}
+ \cdots \biggr) 
\,, \nn\\
f&^{\newPiece{1/2}}
=  \mathcal K  \biggl(
\frac{1}{1120 m^4} + \frac{t}{8400 m^6}
+ \frac{15(s^2+u^2) + 28 s u}{554400 m^8}  
+ \frac{ \big(153 (s^2+u^2) + 313 s u \big) t}{30270240 m^{10}}	
+ \cdots \biggr) 
\,, \nn\\
f&^{\newPiece{1}}
=  \mathcal K  \biggl(
\frac{1}{180 m^4} + \frac{t}{1680 m^6} 
+ \frac{22(s^2+u^2) + 39 s u}{151200 m^8}
+ \frac{ \big(20 (s^2+u^2) + 43 s u \big) t}{831600 m^{10}}	
+ \cdots \biggr) 
\,, \nn\\
f&^{\newPiece{3/2}} 
=  \mathcal K  \biggl(
\frac{1}{24 m^4} + \frac{t}{360 m^6} 
+ \frac{9(s^2+u^2) + 14 s u}{10080 m^8}
+ \frac{ \big(8 (s^2+u^2) + 21 s u \big) t}{75600 m^{10}}
+ \cdots \biggr) 
\,, \nn\\
f&^{\newPiece{2}}=
\mathcal K   \biggl( 
\frac{1}{2 m^4} + \frac{s^2+s u+u^2}{120 m^8} 
+ \frac{ s t u }{504 m^{10}} 
+ \cdots \biggr) 
\,, \hskip .5 cm 
\label{fExpansionShort}
\end{align}
where 
\begin{equation}
\mathcal K = \frac{1}{(4\pi)^{2}}
\left( \frac{\kappa}{2} \right)^4 \,.
\label{eq:KDef}
\end{equation}

For the all-plus and single-minus configurations it suffices to give
the result for the spin-0 contribution since we obtain the remaining
amplitudes via Eq.~(\ref{allPlusSUSY}). For the single-minus configuration
we have
\begin{equation}
\mathcal{M}_4^{S=0}(1^{+},2^{+},3^{-},4^{+}) = 
\left([12] \langle 13 \rangle [14] \right)^4  \, g(s, u) \,,
\end{equation}
where                                                             
\begin{equation}
g(s, u) =
\mathcal K \bigg(
 \frac{1}{5040 m^2 s t u} + \frac{1}{6306300 m^8} 
    + \frac{(s^2 + s u + u^2)}{441080640 m^{12}} 
+ \ldots \bigg)
\,.
\label{gExpansionShort}
\end{equation}
Finally, for the all-plus configuration we have,
\begin{equation}
\mathcal{M}_4^{S=0}(1^{+},2^{+},3^{+},4^{+}) =
\left( \frac{[12][34]}{\langle 12 \rangle \langle 34 \rangle} \right)^2 \,
h(s,u) \,,
\end{equation}
where 
\begin{align}
h(s,u) & =  
\mathcal K \biggl(
    \frac{s t u}{504 m^2} + \frac{(s^2 + s u + u^2)^2}{3780 m^4} 
    + \frac{(s^2 + s u + u^2) s t u}{7920 m^6}\nn \\[3pt]
    & \hskip 1 cm 
    + \frac{75(s^6+u^6) + 225 (s^5 u + s u^5) + 559 ( s^4 u^2 + s^2 u^4) + 743 s^3 u^3}{7207200 m^8}  \nn \\[3pt]
    & \hskip 1 cm 
    + \frac{3 (s^2 + s u + u^2)^2 s t u}{400400 m^{10}}  
+ \ldots \biggr)
\,.
\label{hExpansionShort}
\end{align}

The fact that the highest power of $m$ appearing in the large-mass
expansion is $m^{-2}$ may be contrasted to the high powers of the $m$
in the coefficients of the integrals in the
($4-2\epsilon$)-dimensional integral basis (see
Eqs.~(\ref{allPlusLargeMass}), (\ref{singleMinusLargeMass1}) and
(\ref{singleMinusLargeMass2})). It is a nontrivial consistency check
that our amplitudes vanish in the large-mass limit, as expected from
decoupling. Indeed, the above large-mass behavior hinges on
nontrivial cancellations between all pieces of the amplitude.

Now consider the matching and extraction of the Wilson coefficients
$\beta_{R^3}$, $\beta_{R^4}$ and $\tilde{\beta}_{R^4}$ (or, equivalently,
$\beta_{R^3}$ and $\beta_{R^4}^{\pm}$). Since the relation between the Wilson
coefficients and the amplitudes is linear, the Wilson coefficients
satisfy the supersymmetric decomposition
(Eq.~(\ref{susyGR})). Hence, we may organize our results in terms of
the multiplets circulating in the loop. One may then assemble the
corresponding coefficients for any spinning particle circulating in
the loop using Eq.~(\ref{susyGR}).

Since the all-plus and single-minus amplitudes are nonzero only for the $\newPiece{0}$ piece, we have
\begin{align}
(\beta_{R^3})^{\newPiece{0}} =  \frac{1}{(4\pi)^2} \left(\frac{\kappa}{2}\right)^3 \frac{1}{m^2} \frac{1}{5040}\,,  \qquad
(\beta_{R^4}^-)^{\newPiece{0}} &=  \frac{1}{(4\pi)^2} \left(\frac{\kappa}{2}\right)^4 \frac{1}{m^4}  \frac{1}{7560}\,,  \nn \\
(\beta_{R^3})^{{\newPiece{\it S}} \neq 0} =  (\beta_{R^4}^-)^{{\newPiece{\it S}} \neq 0} &= 0 \,.
\end{align}
where our notation $(\beta_{X})^{\newPiece{S}}$ means the value of $\beta_{X}$ as determined by the data for the new piece for a given spin $S$.
The double-minus configuration is nonzero for any multiplet circulating in the loop. We find
\begin{alignat}{2}
(\beta_{R^4}^+)^{\newPiece{0}} &= \frac{1}{(4\pi)^2}  \left(\frac{\kappa}{2}\right)^4 \frac{1}{m^4} \frac{1}{6300}\,, \hspace{1.5 cm} 
(\beta_{R^4}^+)^{\newPiece{1/2}} &&= \frac{1}{(4\pi)^2}  \left(\frac{\kappa}{2}\right)^4 \frac{1}{m^4}  \frac{1}{1120} \,, \nonumber \\[3pt]
(\beta_{R^4}^+)^{\newPiece{1}} & = \frac{1}{(4\pi)^2}  \left(\frac{\kappa}{2}\right)^4 \frac{1}{m^4}  \frac{1}{180} \,, \hspace{1.7 cm}
(\beta_{R^4}^+)^{\newPiece{3/2}} &&= \frac{1}{(4\pi)^2} \left(\frac{\kappa}{2}\right)^4 \frac{1}{m^4}  \frac{1}{24} \,,  \nonumber \\[3pt]
(\beta_{R^4}^+)^{\newPiece{2}} &= \frac{1}{(4\pi)^2} \left(\frac{\kappa}{2}\right)^4 \frac{1}{m^4}  \frac{1}{2} \,.
\end{alignat}

\subsection{Regge limits of the amplitudes}

For our analysis of the amplitudes with dispersion relations in the
next sections, we need the behavior of the amplitudes for $t
\rightarrow \infty$ and $s \in \mathbb{R} $ fixed, with $|s| < 4 m^2$
and for $s \rightarrow \infty$ and $t \in \mathbb{R} $ fixed, with
$|t| <4m^2$. We extract these directly from the explicit values of the
amplitudes in Appendix~\ref{HighOrdersAppendix}.

We start with the $t \rightarrow \infty$ limit. For the new-pieces in the supersymmetric
decomposition we have
\begin{alignat}{2}
f^{\newPiece{\it S}}(s,u) &\sim \frac{\log(t)}{t^2}\,, \qquad &&S=0,\,1/2,\,1, \nn \\
f^{\newPiece{3/2}}(s,u) &\sim \frac{\log^2(t)}{t^2}\,, \qquad
&&f^{\newPiece{2}}(s,u) \sim \frac{1}{t}\,, \nn \\
g^{\newPiece{0}}(s,u) &\sim \frac{1}{t^2}\,, \qquad
&&h^{\newPiece{0}}(s,u) \sim t^2\,,
\label{ReggeLimitNewPiece}
\end{alignat}
where the $f, g$, and $h$ function are related to the amplitude via
\eqn{HelicityDef}.  Using \eqn{susyGR} we assemble the contributions 
for each particle of a given spin,
\begin{alignat}{2}
f^{{S}}(s,u) &\sim \frac{\log(t)}{t^2}\,, \qquad &&S=0,\,1/2,\,1, \nn \\
f^{{3/2}}(s,u) &\sim \frac{\log^2(t)}{t^2}\,, \qquad
&&f^{{2}}(s,u) \sim \frac{1}{t}\,, \nn \\
g^{S}(s,u) &\sim \frac{1}{t^2}\,, \qquad
&&h^{S}(s,u) \sim t^2 \,,  \qquad  S=0,\,1/2,\,1, \, 3/2, \, 2.
\label{ReggeLimit}
\end{alignat}
For the $f^{{S}}$ functions corresponding to the case of no helicity
flips between incoming and incoming states the spin 2 contribution
dominates as expected.

Next, we consider the  $s \rightarrow \infty$ limit. We find 
\begin{alignat}{2}
f^{\newPiece{\it S}}(s,u) &\sim \frac{1}{s}\,, \qquad 
	&&S=0,\,1/2,\,1, \,3/2, \,2,\nn \\
g^{\newPiece{0}}(s,u) &\sim \frac{1}{s^2}\,, \qquad
&&h^{\newPiece{0}}(s,u) \sim s^2\,,
\label{ReggeSLimitNewPiece}
\end{alignat}
which gives 
\begin{equation}
f^{S}(s,u) \sim \frac{1}{s}\,, \qquad 
g^{S}(s,u) \sim \frac{1}{s^2}\,, \qquad
h^{S}(s,u) \sim s^2\,,  \hskip 1.5 cm  S=0,\,1/2,\,1, \, 3/2, \, 2.
\label{ReggeSLimit}
\end{equation}
Note that the limits of the functions $g^S$ and $h^S$ in \eqns{ReggeSLimitNewPiece}{ReggeSLimit} follow from those in \eqns{ReggeLimitNewPiece}{ReggeLimit}, since these functions are crossing symmetric.

\section{Properties of gravitational amplitudes}
\label{Sec:GravAmplitudes}

We now turn to the properties of the low-energy effective field
theory, arising from taking the large-mass expansion of the one-loop
four-graviton amplitudes calculated in
\sect{AmplitudeConstructionSection}.  Here we do not consider loop
effects in general relativity itself (due to gravitons circulating in
the loops) but focus on the properties of the leading-order
higher-derivative operators in a weakly-coupled gravitational EFT
generated by integrating out massive degrees of freedom. For the same
reason IR divergences are not an issue for our analysis since the
corrections of interest are manifestly IR finite.  
We also note that we do not need to deal with UV divergences, since
the one-loop four graviton amplitude with polarization tensors
restricted to four dimensions considered here is UV finite~\cite{tHooft:1974toh}.

To make the analysis more complete we also include the example of tree-level
graviton scattering in string theory (see Appendix~\ref{StringAppendix}).  As recently discussed in \cite{Chowdhury:2019kaq}, in this case the scattering amplitudes have a great degree of universality; e.g. to leading order considered here they do not depend on the details of the string compactification. 

A general question we ask in this paper is the following: where do physical theories land in the space of couplings that satisfy the bounds from causality, unitarity and crossing?
In this section we review general properties of the gravitational amplitudes relevant for the derivation of the bounds. In subsequent sections we then proceed with the derivation of various bounds on the Wilson coefficients, following the recent developments of Refs.~\cite{Tolley:2020gtv,Caron-Huot:2020cmc,Arkani-Hamed:2020blm,Alberte:2020bdz,Sinha:2020win}. Finally, we check that the results presented in the paper satisfy the expected bounds and analyze the region in the space of couplings covered by known theories.  

\subsection{Low-energy expansion}
\label{sec:lowenergy}

The functions $f$, $g$ and $h$ defined in Eq.~\eqref{HelicityDef}
correspond to the independent helicity configurations. We consider their low-energy expansion\footnote{\footnotelineskip Here we allowed for parity-odd and parity-violating effects which render certain coefficients in the low-energy expansion complex.}
\begin{align}
f(s,u) &=\left( \frac{\kappa \null}{2} \right)^2 {1 \over s t u} +  |\beta_{R^3}|^2 {s u \over t}  - |\beta_\phi|^2  {1 \over t}  + \sum_{i=0}^\infty f_{2 i,i} s^i u^i 
   + \sum_{i=1}^\infty \sum_{j=0}^{[{i \over 2}]} f_{i,j} (s^{i-j} u^j + s^j u^{i-j}) \,, \nn \\
g(s,u) &=  \Bigl(\frac{\kappa}{2}\Bigr)   {\beta_{R^3} \over s t u} + \sum_{p,q =0}^\infty g_{2p+3q,q} \sigma_2^p \sigma_3^q \,, \nn \\
h(s,u) &= [ 10  \Bigl(\frac{\kappa}{2}\Bigr)  \beta_{R^3} - 3 \beta_{\phi}^2 ] s t u +   \sum_{p,q=0, 2p+3q \geq 4}^\infty h_{2p+3q,q} \sigma_2^p \sigma_3^q \,,
\label{eq:lowenergyexp}
\end{align}
where $[x]$ means the integer part of $x$, and we introduced
\begin{equation}
\sigma_k \equiv (-1)^k {s^k + t^k + u^k \over k } .
\end{equation}

In \eqn{eq:lowenergyexp} we explicitly write the massless exchange
poles and assume that the rest of the amplitude admits a simple
low-energy expansion. This is a structure expected from integrating
out massive degrees of freedom and indeed, the amplitudes analyzed in
the present paper are of this type. It does not however capture
correctly the structure of the amplitude once the loops involving
massless particles are included. For example, consider
the case of the one-loop correction due to gravitons circulating in
the loop; see Eq.~\eqref{eq:oneloopGR} and
Appendix~\ref{sec:puregravity}. In this case, we see that for the
all-plus amplitude one-loop Einstein gravity generates a non-zero
$h_{2,0}$. Similarly, for the single-minus amplitude $g(s,u)$ the
one-loop result in Einstein gravity has poles 
in each of the Mandelstam variables. Finally, the double-minus amplitude $f(s,u)$,
see Eq. \eqref{eq:doubleminusGR1loop}, contains IR divergences,
logarithms, as well as higher-order singularities in $1/t$ compared to
the formulas above. In the present paper we focus on the effect
produced by integrating out massive degrees of freedom and do not
analyze the effects from one-loop massless exchanges. Mathematically,
this is simply due to the fact that at leading order that we are
interested in, the two effects lead to additive contributions to the
scattering amplitude and can be analyzed separately. 
Moreover, the
corrections to graviton scattering due to integrating out massive
degrees of freedom satisfy all the basic properties that we discuss
later in the section and as such the low-energy expansion generated in
this way satisfies consistency bounds. Physically, integrating out
massive degrees of freedom leads to effects that are localized in the
impact parameter space $b \lesssim {1 / m_{\rm gap}}$, which are
encoded via higher-derivative operators in a gravitational EFT,
whereas the one-loop effect due to graviton exchange contributes at
any impact parameter, which also manifests itself through the fact
that such corrections do not admit the representation
\eqref{eq:lowenergyexp}. It would be very interesting to develop a
systematic and unified approach to treat both effects in the context
of gravitational scattering, but this is beyond the scope of the
present paper and we leave it for future work. 

In writing \eqn{eq:lowenergyexp} we take into account the crossing
relations \eqref{eq:crossing}. Note that $\sigma_3 = - s t u$ for
$u=-s-t$. In the expansion \eqref{eq:lowenergyexp} $f_{i,j}$ are
real. For parity-preserving theories $h_{k,j}$ and $g_{k,j}$ are real
as well. In the formulas above $\beta_{R^3}$ encodes the unique
non-minimal correction to the three-point amplitude of gravitons as
defined in \eqn{eq:threepointR3}. Similarly $\beta_\phi$ encodes the
non-minimal coupling of a scalar to two gravitons.  In
parity-preserving theories these are real.

For completeness we take into account the possibility of non-minimal coupling to a massless scalar in the amplitude above with the three-point amplitude given in \eqn{eq:threepointR3}.
  Curiously the non-minimal correction to the three-point graviton amplitude and the non-minimal
coupling to a massless scalar mix in the first term in the low-energy
expansion of $h(s,u)$ \eqref{eq:lowenergyexp}. This point was
discussed in detail in Ref.~\cite{Broedel:2012rc}.

We list the explicit results for these functions obtained for the
amplitudes considered in the present paper as an expansion 
in large mass in
Eqs.~\eqref{fExpansionShort}, \eqref{gExpansionShort} and \eqref{hExpansionShort},
as well as in Appendix~\ref{HighOrdersAppendix}.  The exact form 
of the amplitudes is found in Appendix~\ref{AmplitudesResultsAppendix}.

\subsection{Unitarity constraints}
\label{sec:unitarity}

Here we consider the constraints that arise from unitarity. Since the
gravitational EFTs of interest are weakly coupled, we limit ourselves
 to perturbative unitarity as discussed for example in
Refs.~\cite{Arkani-Hamed:2017jhn,Arkani-Hamed:2020blm}. It would be  of
course be very interesting to implement unitarity nonperturbatively
along the lines of Refs.~\cite{Hebbar:2020ukp,KelianMaster}. In four
dimensions this would also require understanding the constraints of
unitarity at the level of IR-finite observables; see
e.g. Refs.~\cite{Strominger:2017zoo,Arkani-Hamed:2020gyp} for a recent discussion. Luckily, for
our purpose of investigating the leading-order corrections to the
gravitational EFT these subtle but important issues are irrelevant.
Here we primarily follow the discussion of
Ref.~\cite{Arkani-Hamed:2020blm} but modify it to account for
differing helicity configurations.

To discuss the constraints coming from unitarity we note that the general
incoming two-graviton state is a superposition of amplitudes with
different helicities.  In total there are four choices for the incoming
and four choices for the outgoing states. We therefore consider the
following matrix of possible amplitudes
\begin{align}
\label{eq:defM}
\begin{pmatrix}
\mathcal{M}_{(+,-,-,+)} & \mathcal{M}_{(+,-,-,-)}  & \mathcal{M}_{(-,-,-,+)} & \mathcal{M}_{(-,-,-,-)}  \\
\mathcal{M}_{(+,-,+,+)} & \mathcal{M}_{(+,-,+,-)}  & \mathcal{M}_{(-,-,+,+)} & \mathcal{M}_{(-,-,+,-)}  \\
\mathcal{M}_{(+,+,-,+)} & \mathcal{M}_{(+,+,-,-)}  & \mathcal{M}_{(-,+,-,+)} & \mathcal{M}_{(-,+,-,-)}  \\
\mathcal{M}_{(+,+,+,+)} & \mathcal{M}_{(+,+,+,-)}  & \mathcal{M}_{(-,+,+,+)} & \mathcal{M}_{(-,+,+,-)}  
\end{pmatrix} ,
\end{align}
where we labeled the helicities of the gravitons using an all-incoming convention.


Consider now scattering in the physical $t$-channel $14 \to 23$. To
describe this situation we consider the center-of-mass frame and
choose helicity spinors as
follows~\cite{Arkani-Hamed:2017jhn,Arkani-Hamed:2020blm}
\be
\lambda_1 &= t^{1/4} \begin{pmatrix}
1 \\
0
\end{pmatrix} , ~~~ \lambda_4 = t^{1/4} \begin{pmatrix}
0 \\
1
\end{pmatrix} , \nn \\
\lambda_2 &=i t^{1/4}  \begin{pmatrix}
\cos {\theta \over 2} \\
\sin{ \theta \over 2}
\end{pmatrix}, ~~~ \lambda_3 = i t^{1/4}  \begin{pmatrix}
\sin {\theta \over 2} \\
- \cos{ \theta \over 2}
\end{pmatrix}  .
\ee
Since we consider particles $1$ and $4$ to be incoming we take $\tilde \lambda_{1,4} = \lambda_{1,4}^*$. Particles $2$ and $3$ are outgoing, therefore $\tilde \lambda_{2,3} = -\lambda_{2,3}^*$. With this choice we get $t = \< 14 \> [14] $, $s = \<12 \> [12] =- t \sin^2 {\theta \over 2}$, so that $\cos \theta = 1 + {2 s \over t}$.  
Evaluating the matrix (\ref{eq:defM}) for this kinematics we obtain
\be
\mathcal{M}(s,t) \equiv \begin{pmatrix}
t^4 f(s,u) & s^2 t^2 u^2 g^*(s,u)  & s^2 t^2 u^2  g^*(s,u)   & h^*(s,u) \\
s^2 t^2 u^2 g(s,u) & u^4 f(s,t) & s^4 f(t,u)  &  s^2 t^2 u^2  g^*(s,u) \\
s^2 t^2 u^2 g(s,u)& s^4  f(t,u)  & u^4  f(s,t) &   s^2 t^2 u^2  g^*(s,u) \\
h(s,u) & s^2 t^2 u^2 g(s,u) & s^2 t^2 u^2 g(s,u)  & t^4  f(s,u)
\end{pmatrix} .
\label{Mmatrix}
\ee

Unitarity restricts the form of the discontinuity of various amplitudes. We introduce the $t$-channel discontinuity as
\be
\label{eq:unitarityT}
{\rm Disc}_t \mathcal{M}(s,t) \equiv {\mathcal{M}(s, t+i \eps) - \mathcal{M}(s, t-i \eps) \over 2 i } \, .
\ee
Through the optical theorem or unitarity, the discontinuity \eqref{eq:unitarityT} is related to the square of the $2 \to n$ amplitude  where we insert a complete set of intermediate states. It is convenient to decompose intermediate states into the irreducible representation of the Poincare group, which are therefore
labeled by the total energy $\sqrt{t}$, spin $J$ and potentially other quantum numbers $i$. The simplest example is when we have an exchange by a single particle of mass $\sqrt{t}$ and spin $J$. In this case \eqref{eq:unitarityT} produces the square of the corresponding three-point couplings multiplied by a kinematical polynomial.\footnote{\footnotelineskip In the case of external scalars these are familiar Legendre polynomials. For spinning particles the analogous polynomials in $D=4$ are Wigner d-functions, which are written explicitly below.} A slightly more general situation is when we have an exchange by multiple particles of the same mass and spin. In this case the discontinuity produces \eqref{eq:unitarityT} a sum over the products of the corresponding three-point couplings. Finally, we can also have a multi-particle state as an intermediate state of total energy $\sqrt{t}$ and spin $J$. It is convenient to think about it again as a single-particle state with the continuous label for the species (which corresponds to the distribution of the total energy among the constituent particles). The result is always the same: we can write Eq.~\eqref{eq:unitarityT} as a sum of kinematical polynomials multiplied by various {\it spectral densities} $\rho_J(t)$ which encode the sum over the products of couplings to the intermediate states of energy $\sqrt{t}$ and $J$ in a given theory. What we just said is simply a restatement of the standard partial-wave expansion in a language which is perhaps slightly more intuitive. For a more complete and detailed derivation see Refs.~\cite{Hebbar:2020ukp,KelianMaster}. 
In this way we can write the discontinuity in terms of the spectral density.  For example, 
\begin{equation}
{\rm Disc}_t \mathcal{M}_{(+,-,-,+)}(s,t) = \sum_{J=0}^\infty {1 + (-1)^J \over 2} \rho_J^{++}(t) d_{0,0}^J(x) \,.
\end{equation}

Applying this to all helicity configurations in the matrix of amplitudes \eqref{Mmatrix} we get 
\be
{\rm Disc}_t \mathcal{M}(s,t) &= \sum_{J=0}^\infty {1 + (-1)^J \over 2} \begin{pmatrix}
\rho_J^{++}(t) d_{0,0}^J(x) & 0  & 0  & \left[ \tilde \rho_J^{++}(t) \right]^*  d_{0,0}^J(x) \\
0 & 0  & 0  &  0 \\
0 & 0 & 0 &  0 \\
\tilde \rho_J^{++}(t)   d_{0,0}^J(x) & 0 & 0  & \rho_J^{++}(t) d_{0,0}^J(x)
\end{pmatrix}  \nn \\
&+ \sum_{J=4}^\infty \begin{pmatrix}
0& 0  & 0  & 0 \\
0 & \rho_J^{+-}(t)  d_{4,4}^J(x)   & \rho_J^{+-}(t) (-1)^J d_{4,-4}^J(x)  &  0 \\
0& \rho_J^{+-}(t) (-1)^J d_{4,-4}^J(x) & \rho_J^{+-}(t) d_{4,4}^J(x)   &   0 \\
0 & 0 & 0  &0
\end{pmatrix} \nn \\
&+ \sum_{J=4}^\infty {1 + (-1)^J \over 2} \begin{pmatrix}
0 & [\tilde \rho_J^{+-}(t)]^* d_{4,0}^J(x)  & [\tilde \rho_J^{+-}(t)]^* d_{4,0}^J(x)  & 0 \\
\tilde \rho_J^{+-}(t) d_{4,0}^J(x) & 0  & 0  &  [\tilde \rho_J^{+-}(t)]^* d_{4,0}^J(x) \\
\tilde \rho_J^{+-}(t)  d_{4,0}^J(x) & 0 & 0 &   [\tilde \rho_J^{+-}(t)]^* d_{4,0}^J(x) \\
0 & \tilde \rho_J^{+-}(t) d_{4,0}^J(x) & \tilde \rho_J^{+-}(t)  d_{4,0}^J(x)  & 0
\end{pmatrix} ,
\label{eq:unitaritydisc}
\ee
where we introduced $x \equiv \cos \theta = 1+{2 s \over t}$ and we
used Wigner d-matrices $d_{\lambda,\lambda'}^J(x)$ which we list explicitly in Appendix~\ref{app:wignerd}. The formula above can be derived by explicitly analyzing
the effect of an exchange by a particle, or equivalently an
irreducible representation, of given mass $\sqrt{t}$ and spin $J$. The
relevant three-point amplitudes are fixed up to a number and their
products are encoded in the various spectral densities
$\rho_J(m^2)$. The indices $++$ and $+-$ denote the helicities of the
corresponding incoming gravitons.  

To discuss dispersion relations we also need to understand the properties of the $u$-channel discontinuity which is defined as
\be
{\rm Disc}_u \mathcal{M}(s,t) \equiv {\mathcal{M}(s,-s-t-i \eps) - \mathcal{M}(s,-s-t+i \eps) \over 2 i } .
\ee
By the same argument it takes the following form\footnote{We thank Alessandro Vichi for pointing out a typo in Eq.~(\ref{eq:unitaritydiscUchannel}) of an earlier version of the paper.}
\be
\label{eq:unitaritydiscUchannel}
{\rm Disc}_u \mathcal{M}(s,t) &= \sum_{J=0}^\infty {1 + (-1)^J \over 2} \begin{pmatrix}
0 & 0  & 0  & \left[ \tilde \rho_J^{++}(u) \right]^*  d_{0,0}^J(\tilde{x}) \\
0 & \rho_J^{++}(u) d_{0,0}^J(\tilde{x})  & 0  &  0 \\
0 & 0 &  \rho_J^{++}(u) d_{0,0}^J(\tilde{x}) &  0 \\
\tilde \rho_J^{++}(u)   d_{0,0}^J(\tilde{x}) & 0 & 0  &0
\end{pmatrix}  \nn \\
&+ \sum_{J=4}^\infty \begin{pmatrix}
\rho_J^{+-}(u)  d_{4,4}^J(\tilde{x}) & 0  & 0  & 0 \\
0 & 0  & \rho_J^{+-}(u) (-1)^J d_{4,-4}^J(\tilde{x})  &  0 \\
0 & \rho_J^{+-}(u) (-1)^J d_{4,-4}^J(\tilde{x})  & 0 &   0 \\
0 & 0 & 0   & \rho_J^{+-}(u) d_{4,4}^J(\tilde{x})
\end{pmatrix} \nn \\
&+ \sum_{J=4}^\infty {1 + (-1)^J \over 2} \begin{pmatrix}
0 & [\tilde \rho_J^{+-}(u)]^* d_{4,0}^J(\tilde{x})  & [\tilde \rho_J^{+-}(u)]^* d_{4,0}^J(\tilde{x})  & 0 \\
\tilde \rho_J^{+-}(u) d_{4,0}^J(\tilde{x}) & 0  & 0  &  [\tilde \rho_J^{+-}(u)]^* d_{4,0}^J(\tilde{x}) \\
\tilde \rho_J^{+-}(u)  d_{4,0}^J(\tilde{x}) & 0 & 0 &   [\tilde \rho_J^{+-}(u)]^* d_{4,0}^J(\tilde{x}) \\
0 & \tilde \rho_J^{+-}(u) d_{4,0}^J(\tilde{x}) & \tilde \rho_J^{+-}(u)  d_{4,0}^J(\tilde{x})  & 0
\end{pmatrix}  ,
\ee
where $\tilde{x}=1+{2s\over u}$.

The spectral densities that enter the unitarity relation are given in terms of the product of the couplings to the corresponding intermediate states. The diagonal terms being given by the absolute value square of the couplings are nonnegative
\be
\label{eq:unitarityCauchyS1}
\rho_J^{++}(m^2) &\geq 0,  ~~~ \rho_J^{+-}(m^2) \geq 0 \,.
\ee
The off-diagonal terms satisfy simple Cauchy-Schwartz inequalities\footnote{\footnotelineskip In terms of couplings to the intermediate state of energy $\sqrt{t}$ and spin $J$ these inequalities simply state that $\left| \sum_i \lambda_{++ i} \lambda_{+- i}^* \right|^2 \leq ( \sum_i |\lambda_{++ i} |^2 ) ( \sum_i | \lambda_{+- i} |^2 ) $ and $| \sum_{i} \lambda_{++i}^2 | \leq  \sum_{i} |\lambda_{++i}|^2$ respectively.}
\be
\label{eq:unitarityCauchyS}
| \tilde \rho_J^{+-}(m^2) |^2 &\leq \rho_J^{++}(m^2) \rho_J^{+-}(m^2) ,~~~ | \tilde \rho_J^{++}(m^2) | \leq \rho_J^{++}(m^2) \,. 
\ee

Below we discuss bounds on the possible form of the low-energy coefficients stemming from unitarity and growth of the amplitude at infinity.
In particular, it is convenient for us to consider combinations of amplitudes that have nonnegative semi-definite discontinuities both in the $t$- and $u$-channel.
We consider the following convenient choices that are sufficient for our purposes
\be
\label{eq:matrixh}
 \mathcal{M}_h (s,t) &\equiv \begin{pmatrix}
t^4 f(s,u) + u^4 f(s,t) - s^4 f(t,u) & h^*(s,u)   \\
h(s,u) & t^4 f(s,u) + u^4 f(s,t) - s^4 f(t,u) 
\end{pmatrix} ,  \\[7pt]
 \mathcal{M}_g (s,t) & \equiv \begin{pmatrix}
t^4 f(s,u) + u^4 f(s,t) & 2 s^2 t^2 u^2 g^*(s,u)   \\
2 s^2 t^2 u^2 g(s,u) & t^4 f(s,u) + u^4 f(s,t)  
\end{pmatrix}  .
\label{eq:matrixg}
\ee
These matrices are $t$--$u$ crossing symmetric
\be
 \mathcal{M}_{h,g} (s,t) =  \mathcal{M}_{h,g} (s,u) \,.
\ee
Using the form of the discontinuity dictated by unitarity above together with the Cauchy-Schwartz inequalities \eqref{eq:unitarityCauchyS} one can check that
\be
\label{eq:unitarityNew}
\partial_s^n {\rm Disc}_t \mathcal{M}_{h,g} |_{s=0}  =  \partial_s^n {\rm Disc}_u \mathcal{M}_{h,g} |_{s=0}\succeq 0 , ~~~~~ n \geq 0 \,,
\ee
where the notation `$\succeq 0$' means that the matrix is positive semi-definite.\footnote{\footnotelineskip A hermitian $n \times n$ matrix $M$ is called positive semi-definite if $z^* M z \geq 0$ for all $z \in \mathbb{C}^n$.} 
We obtain these inequalities from the following properties of the Wigner d-matrices together with \eqref{eq:unitarityCauchyS}
\be
&\partial_s^n d_{0,0}^J(x)|_{s=0} \geq 0 \,, \label{eq:legendrepos} \\
&\partial_s^n \left( d_{4,4}^J(x) - (-1)^J d_{4,-4}^J(x) \right) |_{s=0}   \geq 0 \,,  \label{eq:legendrepos2}   \\
&\partial_s^n \left( \rho_J^{++}(t) d_{0,0}^J(x) +  \rho_J^{+-}(t)  d_{4,4}^J(x) - 2 | \tilde \rho_J^{+-}(t) | d_{4,0}^J(x)  \right)|_{s=0}   \geq 0 \,, \label{eq:legendrepos3}
\ee
where $s = 0$ corresponds to $x=1$ (see definition of $x$ below \eqn{eq:unitaritydisc}). The first property \eqref{eq:legendrepos} is a well-known property of Legendre polynomials.
We have not attempted to prove \eqref{eq:legendrepos2} and \eqref{eq:legendrepos3} that rely on the properties of the relevant $d_{\lambda,\lambda'}^J(x)$, however we checked them explicitly up to $J=30$. We also checked using formulas from Appendix~\ref{app:wignerd} the conditions above for any $J$ and $n=0,1,2$; these are the cases that we consider below in more detail. For forward scattering a closely related discussion can be found in Ref.~\cite{Trott:2020ebl}.

The matrices above admit simple eigenvectors in the parity-preserving case, where we have $g(s,u)=g^*(s,u)$ and $h(s,u)=h^*(s,u)$. In this case the matrices above have eigenvalues
\be
\label{eq:functionsparity}
\mathcal{M}_{h,\pm} (s,t) &= u^4 f(s,t)+t^4 f(s,u)-s^4 f(t,u) \pm h(s,u) \,, \nn \\
\mathcal{M}_{g,\pm} (s,t) &= u^4 f(s,t)+t^4 f(s,u) \pm 2 s^2 t^2 u^2 g(s,u) \,.
\ee
Unitarity \eqref{eq:unitarityNew} then becomes the statement about nonnegativity of the discontinuity of \eqn{eq:functionsparity}.
In this case we can apply dispersion relations directly to the functions \eqref{eq:functionsparity}.

\subsection{Causality}

Causality and unitarity put constraints on the high-energy behavior of the amplitude. The corresponding bounds are well known in the case of gapped QFTs~\cite{Martin:1965jj}, but are on a less-rigorous footing in gravitational theories.

At tree level the gravitational amplitudes are expected to satisfy 
\be
\label{eq:CRG}
\lim_{|t| \to \infty} | \mathcal{M}_{\text{tree}} (s,t) | \leq t^2 \,, \hskip 1.5 cm s<0 \,.
\ee
This result is very intuitive but hard to establish rigorously. It naturally emerges from various considerations \cite{Camanho:2014apa,Maldacena:2015waa,Chandorkar:2021viw}.

The situation is less clear nonperturbatively, however the simple qualitative picture of high-energy scattering in gravity together with
unitarity and causality again imply that a similar bound exists.  The bound is usually assumed to be
\be
\label{eq:QRG}
\lim_{|t| \to \infty} | \mathcal{M}_{\text{full}} (s,t) | < t^2\,,  \hskip 1.5 cm s<0 \,.
\ee
This can be used to write dispersion relations in $D \geq 5$ where the amplitudes are IR finite and make sense nonperturbatively. In $D=4$ the situation is less clear due to the IR divergences, but presumably a similar bound exists for the IR safe observables. It is also possible to satisfy \eqref{eq:QRG} at tree level but this requires an infinite number of particles of arbitrary high spin to be exchanged. A famous example of this type is given by the tree-level string amplitudes. 

In this paper, we are interested in particular in the properties of the one-loop scattering amplitudes in $D=4$. These do not have to satisfy \eqref{eq:CRG} or \eqref{eq:QRG}, therefore we do not use them. Indeed, let us imagine that the bound \eqref{eq:CRG} is saturated at tree-level. This corresponds to exchange of a particle of spin $2$. When we go to 
one-loop we can exchange a pair of spin-$2$ particles and the resulting amplitude will grow like $t^3$.
Therefore it is natural that the following bound holds at one loop
\be
\label{eq:ReggeLoop}
\lim_{|t| \to \infty} | \mathcal{M}_{1-\text{loop}} (s,t) | \leq t^{3}, \hskip 1.5 cm s<0 \,.
\ee
 The explicit amplitudes computed in the paper indeed satisfy
\eqn{eq:ReggeLoop}. Below, in writing dispersion relations we
assume the Regge bound to be as given in \eqn{eq:ReggeLoop} and choose the
number of subtractions accordingly. Note that since \eqn{eq:ReggeLoop} is weaker than \eqn{eq:CRG}, bounds derived in
this way apply both to the tree and one-loop amplitudes. They
also apply to full nonperturbative amplitudes in $D \geq 5$.

It is a special feature of gravitational amplitudes that an infinite number of particles has to be exchanged at tree-level in order for the amplitudes to satisfy \eqref{eq:CRG}. This follows from the fact that such exchanges at tree-level correspond to non-minimal couplings, which are known to lead to violations of \eqn{eq:CRG}~\cite{Camanho:2014apa,Belin:2019mnx,Simons-Duffin-Zhiboedov}. The only exception to this rule is general relativity, which corresponds to the minimal self-coupling of the graviton. 

The tree-level Regge growth bound \eqref{eq:CRG} is the reason why it is so hard to construct tree-level gravitational amplitudes different from the ones of general relativity. This famous achievement of string theory leads to causal and unitarity amplitudes with infinitely many poles that corresponds to exchanges of infinitely many particles of arbitrarily large spin. In fact, every such modification must contain strings \cite{Caron-Huot:2016icg}.\footnote{\footnotelineskip Curiously, as explained in some detail in Appendix~\ref{app:Mstumodel}, if we allow for accumulation points in the spectrum we can consider much simpler amplitude functions. These functions, however, contain an infinite number of particles of a given mass.}

\subsection{Dispersive sum rules}

Given the gravitational amplitudes that satisfy unitarity and the Regge bounds we can consider various dispersion relations. 
One class of dispersion relations that we find useful recasts the vanishing of the (subtracted) amplitude in the Regge limit  
in terms of its discontinuity.

Such relations are known as superconvergence relations~\cite{Simons-Duffin-Zhiboedov}, or dispersive sum rules \cite{Caron-Huot:2020adz}.
We follow the latter terminology and consider the following integrals
\be
\label{eq:superconvergence}
B_k^+(s) &=\oint_{\infty} {d t \over 2 \pi i} \mathcal{M} (s,t) {1 \over t} {1  \over (t (s+t))^{k} } = 0 \, , \hskip 1.5 cm k \geq 2 \,,
\ee
where the condition $k \geq 2$ originates from the Regge bound (\ref{eq:ReggeLoop}) and guarantees that the arc at infinity produces a vanishing contribution. 

\begin{figure}
\centering
\includegraphics[scale=.45]{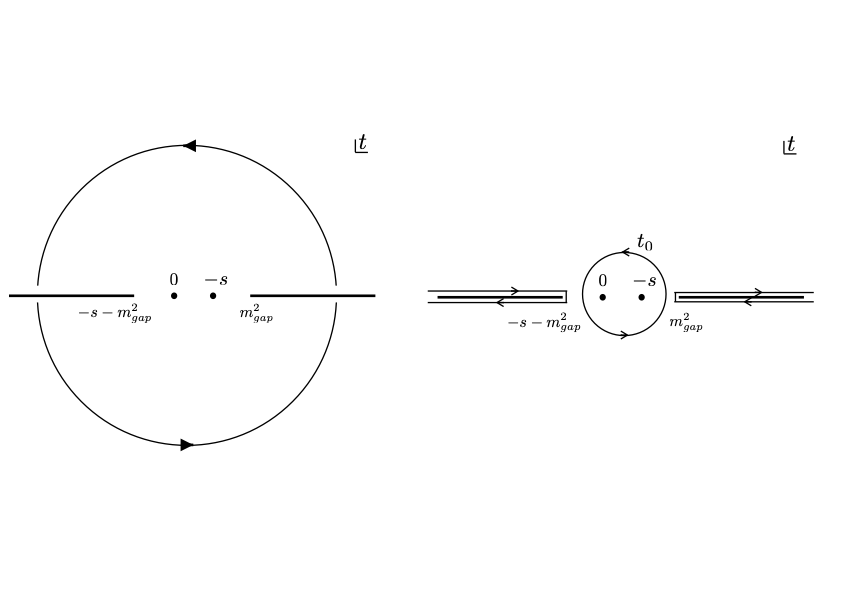}
\caption{\captionlineskip The vanishing of the arc integral at infinity on the left panel, cf. \eqref{eq:superconvergence}, can be restated as an equality of the integrals on the right panel (see \eqref{eq:superconvergenceE}.) }
\label{fig:dispsr}
\end{figure}

By deforming the contour we get the following formula
\begin{align}
\label{eq:superconvergenceE}
B_{k}^+(s) &:     \oint_{t_0} {d t \over 2 \pi i} \mathcal{M}(s,t) {1 \over t} {1 \over (t (s+t))^{k} } \\
 & \hskip 2 cm \null
 = \int_{m_{\rm gap}^2}^\infty {d t \over \pi} {1  \over (t (s+t))^{k} } \left( {1 \over t} {\rm Disc}_t \mathcal{M}(s,t) + {1 \over s+t} {\rm Disc}_u \mathcal{M}(s,t) \right)  , \nn
\end{align}
where the integrals are depicted in \fig{fig:dispsr}. 

We evaluate the LHS of \eqn{eq:superconvergenceE} using the low-energy expansion, and we use the RHS to establish some nontrivial properties that the coefficients of the low-energy expansion should satisfy. More precisely, we consider  the expansion of (\ref{eq:superconvergenceE}) around $s=0$.

A few comments are in order. Firstly, note that the minimal tree-level graviton-exchange term in $f(s,u) \sim {1 \over s t u}$ does not contribute to the sum rules of interest here, $B_{k\geq 2}^+(s)$. In particular, we are able to expand the sum rules in powers of $s$. Assuming a more stringent Regge bound, $B^+_{1}(s)$ can be studied \cite{Caron-Huot:2021rmr} and used to bound various coefficients in terms of $G$. We do not consider this sum rule in the present paper. Secondly, the non-minimal coupling $\beta_{R^3}$ contributes to $B_2^+$ and we utilize this fact in Sect. \ref{sec:R3bound}  in order to bound $\beta_{R^3}$ in terms of other EFT data.

\subsection{The theory islands}

What is the space of gravitational EFTs that admit a consistent UV completion? Answering this question nonperturbatively is a formidable task, but we can consider the simpler question of understanding perturbatively in $G$ possible consistent UV completions of gravity. It is in this spirit that in this paper we focus on the leading-order-in-$G$ correction to general relativity.

Imagine that we label all possible perturbatively consistent theories of gravity by an index $i$ so that the four-graviton scattering amplitudes in a given theory are given by the matrix $\mathcal{M}_i$ (see \eqn{eq:defM}).\footnote{\footnotelineskip For simplicity we keep the index discrete, but of course it can be continuous, e.g. denoting the mass of the particle exchanged in the loop. It does not affect the discussion.} It is clear then that by taking a superposition of such amplitudes with non-negative coefficients we again get an amplitude that satisfies all the constraints discussed in the sections above
\be
\label{eq:superposition}
\mathcal{M}_{\text{theory}}  = \sum_{i} c_i \mathcal{M}_i , ~~~~~ c_i \geq 0 \,.
\ee
By expanding such amplitudes at low energies, as in \sect{sec:lowenergy}, we get a set of higher-derivative corrections. The low-energy Wilson coefficients then form a convex cone which we can think of as the {\it theory space},\footnote{\footnotelineskip A convex cone is a subset of vector space closed under linear combinations with positive coefficients.} which is generated by vectors $\mathcal{M}_i$.\footnote{\footnotelineskip Nonperturbatively in $G$, Eq.\eqref{eq:superposition} is not necessarily consistent since is can violate nonperturbative unitarity $|S_J| \leq 1$. Since we are interested in the leading-order correction in the regime when gravity is weakly coupled, this issue is irrelevant.}  

It has been recently established that causality, crossing and unitarity constrain Wilson coefficients both from above and from below  \cite{Tolley:2020gtv,Caron-Huot:2020cmc,Arkani-Hamed:2020blm,Sinha:2020win}. In the space of couplings formed by weakly-coupled theories it is thus natural to talk about {\it the theory island}, namely the region spanned by known perturbative UV completions of gravity. By definition we have 
\be
\text{Theory island} \subseteq \text{EFThedron} ,
\ee
where by the EFThedron we call the set of amplitudes satisfying the constraints reviewed in this section (in particular we only impose perturbative unitarity). It is interesting to ask to what extent we can populate the space of allowed couplings by known perturbative UV completions. Recall that by a perturbative UV completion we call an $S$-matrix that satisfies unitarity, causality and crossing perturbatively in Newton's constant $G$ and for any process $m \to n$.

There are two classes of perturbative UV completions we consider here: tree-level corrected theories and one-loop corrected theories. In tree-level corrected theories the leading-order correction to general relativity enters at tree-level, in other words the higher-derivative corrections in the Lagrangian are suppressed by a new scale, e.g. string scale, which can be much lower than the Planck scale. Perturbative string theories are famous examples of this type. 
It is a well-known fact that gravitational amplitudes in string theory have a great degree of universality, as discussed recently in Ref.~\cite{Chowdhury:2019kaq}.  In Appendix~\ref{StringAppendix} we review the cases of superstring, heterotic and bosonic strings at tree level.\footnote{\footnotelineskip Strictly speaking the bosonic string is not part of the theory space due to the presence of the tachyon in the spectrum, but it is useful to keep it to check various formulas.}  Alternatively, we can consider theories where the higher-derivative operators come with an extra power of $G$. The one-loop amplitudes computed in the present paper are precisely of this type. By choosing the mass spectrum and spins of the particles propagating in the loop we can get various amplitudes. 

Combining these two examples we consider the following set of amplitudes in the present paper
\be
\mathcal{M}_{\text{theory}}^{\text{here}} = c_{(\text{ss})} \mathcal{M}_{\text{tree}}^{(\text{ss})} +  c_{(\text{hs})} \mathcal{M}_{\text{tree}}^{(\text{hs})} +  c_{(\text{bs})} \mathcal{M}^{(\text{bs})}_{\text{tree}} + \sum_{S=0,{1 \over 2},1,{3 \over 2},2} c_{S} \mathcal{M}_S^{1-\text{loop}} \, .
\ee
Obviously, we do not claim that this space is complete. Let us emphasize that in the analysis below we allow taking arbitrary superpositions of string amplitudes with various tensions as well as one-loop amplitudes with arbitrary choices of the masses and spins, $S \leq 2$, of particles circulating in the loop.

On the string side, for example we can consider the $\eps$ deformation of the superstring amplitude discussed in the conclusions of Ref.~\cite{Arkani-Hamed:2020blm}. It is not clear however that these deformations can be promoted to fully consistent perturbative S-matrices. Moreover, we checked that adding this correction does not affect the theory island considered in the present paper in a noticeable way. 

On the field-theory side we can consider gapped strongly-coupled theories of matter coupled to gravity. One interesting example is large-$N$ QCD coupled to gravity \cite{Kaplan:2020tdz}. In this case we expect the tree-level corrections due to non-minimally coupled glueballs and one-loop corrections due to minimally-coupled glueballs to both enter at the same order $G^2$.  Finally, we do not discuss here amplitudes in  models with extra dimensions~\cite{ArkaniHamed:1998rs,Randall:1999ee,Randall:1999vf}. Of course, it would be very interesting to extend our analysis to include these examples as well. 

In writing the contribution of matter we can imagine integrating out a multiple number of particles of given spin with various masses. For the constraints discussed below one can check that the effect of this freedom can be absorbed into rescaling of the coefficients $c_{S}$. In particular, the bounds that we describe below hold for any choice of masses and number of species for particles that we integrate out.

In the following sections we use these amplitudes to populate the theory island. Surprisingly, we find that 
\be
\text{Theory island} \ll \text{EFThedron} \,,
\ee
in a sense that should become clear below.
In other words, the set of known weakly-coupled theories occupy a much smaller space in the space of couplings than is allowed by the general constraints coming from the analysis of $2 \to 2$ scattering. This is, of course, in accord with the ongoing landscape vs swampland debate but is also different from that. Indeed, the theory island as defined here does not guarantee the existence of the nonperturbative completion of gravity. We only study the consistency of the leading-order corrections to the Einstein-Hilbert theory perturbatively in $G$ and already in this setting we seem to find a much smaller space of possibilities than follows from general constraints.

\section{Deriving bounds: elastic amplitude}
\label{Sec:Bounds}

In this section we analyze bounds on the low-energy expansion of the elastic amplitude $f(s,u)$ using the techniques of Ref.~\cite{Arkani-Hamed:2020blm} and~\cite{Tolley:2020gtv,Caron-Huot:2020cmc}.\footnote{\footnotelineskip See also Ref.~\cite{Zhang:2020jyn} for a closely related discussion.} We focus on couplings of the same dimensionality and derive two-sided bounds on them by explicitly identifying the facets of the relevant polytopes and then taking their crossing-symmetric section. 

As observed in Ref.~\cite{Arkani-Hamed:2020blm}, identifying the relevant boundaries are particularly simple in case of the one-channel dispersion relations when the polytopes in question are cyclic.\footnote{\footnotelineskip We refer the reader to Ref.~\cite{Arkani-Hamed:2020blm} for the definition of this term.} In case of the two-channel dispersion relations the cyclicity property is lost. However, as Ref.~\cite{Arkani-Hamed:2020blm} observed, a set of new boundaries that appear in this case involve low-spin partial waves and can be explicitly identified upon inspection. This is indeed what we observed in the examples below. To the best of our knowledge there is no proof that the list of boundaries obtained in this way is complete (and therefore that the bounds are optimal), and we do not attempt such a proof in the present work. 

We will not attempt to relate couplings of different dimensionality to each other either. This problem was recently analyzed in Ref.~\cite{Tolley:2020gtv,Caron-Huot:2020cmc,Sinha:2020win} and the expected dimensional-analysis scaling of various couplings with order $\mathcal O(1)$ coefficients was rigorously established. It would be very interesting to apply these techniques to the gravitational amplitudes discussed here but we defer this to the future.

\subsection{Strategy}

We first briefly describe the basic strategy to derive bounds~\cite{Arkani-Hamed:2020blm}. We consider the vector of low-energy couplings of the same dimensionality $\vec F$ which via dispersion relations is given by the sum of $s$- and $u$-channel partial waves
\be
\label{eq:couplingcone}
\vec F = \sum_J  \left( c_{J,s} \vec V_{J,s} + c_{J,u} \vec V_{J,u} \right) , ~~~~~ c_{J,s} , c_{J,u} \geq 0 \, .
\ee
In this formula $c_{J,s}$ and $c_{J,u}$ encode spectral densities discussed in \sect{sec:unitarity} and $\vec V_{J,s}$, $\vec V_{J,u}$ are known functions related to partial waves---see e.g. \eqref{eq:dispcouplingrepresentation} for the precise formula. However for the present discussion we can simply think of $\vec V_{J,s}$ and $\vec V_{J,u}$ as some abstract, given vectors and $ c_{J,s}$ , $c_{J,u}$ being non-negative numbers. We would like to characterize the space \eqref{eq:couplingcone}, which is a polytope. A convenient way to do it is by identifying its facets
\be
\label{eq:boundarypolytope}
\vec F \cdot \vec W_i \geq 0 \,, 
\ee
where $\vec W_i$ is a normal to a given facet. From \eqn{eq:couplingcone} we see that all the facets are of the form\footnote{\footnotelineskip Strictly speaking, this statement is only true for finite-dimensional sums in \eqn{eq:couplingcone}. For infinite-dimensional sums we can also have limiting points. For us the limiting point is $J=\infty$. We find, however, that its existence does not play a role in the analysis in the following sections. The reason for it is that in practice we search for the boundaries \eqref{eq:boundarypolytope} by first truncating the sum over spins up to some $J_{\rm max}$ and then extrapolating to $J_{\rm max} = \infty$.}
\be
(W_i)_{I_0} = \eps_{I_0  I_1 ... I_d} V_{a_1}^{I_1} ... V_{a_d}^{I_d} \,,
\ee
where we assumed that the vectors $\vec V_{a}$ are $(d+1)$-dimensional and we contracted them using the $(d+1)$-dimensional $\epsilon$-tensor.\footnote{\footnotelineskip Here $d$ has nothing to do with the dimensionality of spacetime. Instead, it is the dimensionality of the relevant subset of the low-energy couplings that we would like to bound.} The index $a$ labels both spin and channel. Therefore, we can characterize the space \eqref{eq:couplingcone}
by a set of determinants $\langle F , a_1 , .... , a_d \rangle \geq 0$. The task is then to find all $(a_1 , .... , a_d)$, such that \eqn{eq:boundarypolytope} holds.
In general, this is a formidable task since the space of vectors in \eqn{eq:couplingcone} is infinite dimensional. 

As explained in Ref.~\cite{Arkani-Hamed:2020blm}, remarkably, the set of vectors $ \vec V_{J,s}$ defines a cyclic polytope,
which for us simply means that the set of its boundaries can be written down explicitly very easily. They essentially take the form of determinants built out of pairs of consecutive vectors in the sum over spins. For example, for $d=4$ the relevant determinants take the form $(i , i+1 , j , j+1)$ with $j>i$, where $i$ and $j$ label various spins in the sum \eqref{eq:couplingcone}.

The cyclicity property is lost when considering the sum of the $s$- and $u$-channel as in \eqn{eq:couplingcone}. By inspecting the resulting polytope 
Ref.~\cite{Arkani-Hamed:2020blm} observed that it is ``almost cyclic'' in the sense that the only boundaries which are not of the type $(i , i+1 , j , j+1)$ in the example considered above involve low-spin partial waves in \eqn{eq:couplingcone} and can be found by direct inspection. This is what we also find and this is the strategy of finding bounds that we follow in this paper. In particular, we do not prove that the set of bounds found in this way is complete (even though we suspect it to be the case). It would be very interesting to rigorously demonstrate this.

After the allowed region \eqref{eq:boundarypolytope} is identified we impose crossing symmetry, which is the subspace defined by linear relations between various components of the coupling vector $\vec F$. By taking the crossing-symmetric slice of the allowed space of couplings we get two-sided bounds on the low-energy Wilson coefficients.

\subsection{Non crossing-symmetric dispersive representation of low-energy couplings}
\label{sec:nonsusymmetric}

Consider again the double-minus amplitude
\be
\label{eq:amplitudeMHV}
\mathcal{M}_4(1^-,2^-,3^+,4^+) = (\langle 12 \rangle [34])^4 \ f (t,u) \, .
\ee
Following Ref.~\cite{Arkani-Hamed:2020blm}, we introduce the expansion\footnote{\footnotelineskip Note that we performed an $s \leftrightarrow t$ transformation compared to \eqn{eq:lowenergyexp}.}
\be
\label{eq:fexpa}
f(t,u)= f(t,-s-t) = \left( \frac{\kappa \null}{2} \right)^2 {1 \over s t u} +  |\beta_{R^3}|^2 {t u \over s}   - |\beta_\phi|^2  {1 \over s} + \sum_{k \geq j \geq 0 } a_{k,j} s^{k-j} t^j \, .
\ee
From \eqn{eq:fexpa} we see that for fixed $k$ all $a_{k,j}$ have the same dimensionality. 
Crossing symmetry $f(t,u)=f(u,t)$ leads to linear relations among $a_{k,j}$ which will play an important role below.

Of course, in \eqn{eq:lowenergyexp} we are expanding the same function, so the $f_{i,j}$ introduced there and the $a_{k,j}$ in \eqn{eq:fexpa} are all related to each other in a trivial fashion. We get for the first few coefficients
\be
\label{eq:firstfewcoefficient}
a_{0,0} = f_{0,0}\,, \hskip 1.2 cm  a_{1,0}=-f_{1,0}\,, \hskip 1.2 cm a_{1,1} = 0 \,. 
\ee
The vanishing of $a_{1,1}$ is a consequence of $t$--$u$ crossing symmetry of $f(t,u)$ which is not manifest in \eqn{eq:fexpa}.

The function of interest $f(t,u)$ admits a simple dispersive representation
\be
f(t,-s-t) &= \oint {d s' \over 2 \pi i} {f(t,-s'-t) \over s-s'} =\left( \frac{\kappa \null}{2} \right)^2 {1 \over s t u} +  |\beta_{R^3}|^2 {t u \over s}   - |\beta_\phi|^2  {1 \over s}   \nn \\[4pt]
& \null \hskip .3 cm 
 - \int_{m_{\rm gap}^2}^\infty {d m^2 \over \pi}  
\biggl( \sum_{J=0}^\infty {1+(-1)^J \over 2}  {\rho_{J}^{++}(m^2) d_{0,0}^J(1+{2t \over m^2} ) \over m^8} {1 \over s-m^2}  \nn\\
& \null\hskip 4 cm
+ \sum_{J=4}^\infty {\rho_{J}^{+-}(m^2) d_{4,4}^J \left(1+{2t \over m^2} \right) \over (t+m^2)^4} {1 \over -s-t-m^2} \biggr) \,. 
\label{eq:dispersionrelationsf}
\ee
In deriving (\ref{eq:dispersionrelationsf}) we used the Regge bound (\ref{eq:ReggeLoop}) to drop the arcs at infinity. When applied to \eqn{eq:amplitudeMHV} it implies that $|f(t,-s-t)| \leq {1/|s|}$ at large $s$ and fixed $t$. Indeed, this is the case for all the one-loop amplitudes that we consider. 


An important feature of \eqn{eq:dispersionrelationsf} is that crossing symmetry $f(t,u)=f(u,t)$ is not manifest (hence the title of this section). Therefore imposing it leads to interesting constraints. In fact, as we will see below it leads to two-sided bounds on the couplings.

By expanding \eqn{eq:dispersionrelationsf} at small $s$ and $t$ we get the dispersive representations for the couplings $a_{k,j}$. They take the following form
\be
\label{eq:dispcouplingrepresentation}
\boxed{ a_{k,j} =\int_{m_{\rm gap}^2}^\infty {d m^2 \over \pi} {1 \over (m^2)^{k+1}}  \left( \sum_{J=0}^\infty {1+(-1)^J \over 2} {\rho_{J}^{++}(m^2) \over m^8}  P_{++}^j({\cal J}^2) + \sum_{J=4}^\infty {\rho_{J}^{+-}(m^2) \over m^8} P_{+-}^{k,j}({\cal J}^2) \right), }
\ee
where we introduced the spin Casimir ${\cal J}^2 = J (J+1)$. We call \eqn{eq:dispcouplingrepresentation} the dispersive representation of low-energy couplings. We again emphasize that crossing symmetry leads to linear relations among $a_{k,j}$ which are not manifest in \eqn{eq:dispcouplingrepresentation}.

In the formula above we introduced polynomials $P_{++}^j({\cal J}^2)$ and $P_{+-}^{k,j}({\cal J}^2)$ which have the following properties\footnote{\footnotelineskip In the $++$ channel the closed-form expression takes the form $P_{++}^j({\cal J}^2) ={1 \over \Gamma(j+1)^2} \prod_{n=1}^{j} ( {\cal J}^2 - n (n-1) )$.}
\be
P_{++}^0({\cal J}^2) &= 1\,, \qquad P_{+-}^{k,0}({\cal J}^2)= (-1)^{k} \,, \nn \\
P_{++}^1({\cal J}^2) &= {\cal J}^2\,, \qquad P_{+-}^{k,1}({\cal J}^2)= (-1)^{k+1} ( {\cal J}^2 - 20 - k) \,, \nn \\
\lim_{J \to \infty} P_{++}^j ({\cal J}^2)&= {1 \over \Gamma(j+1)^2} J^{2 j} + \ldots \,, \qquad \lim_{J \to \infty}  P_{+-}^{k,j} ({\cal J}^2) =  {(-1)^{k+j} \over \Gamma(j+1)^2} J^{2 j} + \ldots \,.
\ee
Based on the formulas above we see that couplings $a_{k,j}$ written in \eqn{eq:dispcouplingrepresentation} roughly probe the $k$'th moment of the spectral density with respect to $m^2$ and $(2j)$'th moment with respect to spin. In particular, we expect that higher-$j$ coefficients to be more sensitive to higher-spin spectral densities and higher-$k$ couplings to have the large ${m^2 \over m_{\rm gap}^2}$ region more suppressed. This is indeed what we will find in the explicit examples below.

Crossing symmetry leads to the set of sum rules which were dubbed null constraints in Ref.~\cite{Caron-Huot:2020cmc}. Effectively, they express the low-spin partial-wave data in terms of the higher-spin data. Consider for example the relation $a_{1,1}=0$, listed in \eqn{eq:firstfewcoefficient}. Using the formulas above we can write it as follows
\be
\label{eq:nullconstraint}
\int_{m_{\rm gap}^2}^\infty {d m^2 \over (m^2)^{2}}  {\rho_{4}^{+-}(m^2) \over m^8} = \int_{m_{\rm gap}^2}^\infty {d m^2 \over (m^2)^{2}} \left( \sum_{J=2}^\infty {1+(-1)^J \over 2} {\rho_{J}^{++}(m^2) \over m^8}  {\cal J}^2 + \sum_{J=5}^\infty {\rho_{J}^{+-}(m^2) \over m^8} ({\cal J}^2 - 21) \right) ,
\ee
where all the terms in the RHS of \eqn{eq:nullconstraint} are non-negative. In other words, crossing symmetry allows us to express the moment of $\rho_{4}^{+-}(m^2)$ in terms of all other spectral densities.

Let us next describe the first few bounds for $k \leq 6$. We do not attempt to derive the bounds across the couplings of different dimensionalities as it was done in Ref.~\cite{Tolley:2020gtv,Caron-Huot:2020cmc,Sinha:2020win,Caron-Huot:2021rmr}. Instead we focus on the geometry of couplings of the same dimensionality as was studied in Ref.~\cite{Arkani-Hamed:2020blm}. For $k=1,3,5$ we do not get any nontrivial bounds. For even $k$ the results are presented below.

 ${\bf k=0\!:}$

We first start with $k=0$. From (\ref{eq:dispersionrelationsf}) it immediately follows that 
\be
a_{0,0} \geq 0 \,. 
\ee
Moreover, we have
\be
\label{eq:momentsak0}
a_{k,0} ={1 \over \pi} \int_{m_{\rm gap}^2}^\infty {d m^2 \over m^{2k + 10}} \left( \sum_{J=0}^\infty {1+(-1)^J \over 2}  \rho_{J}^{++}(m^2)   + (-1)^k \sum_{J=4}^\infty \rho_{J}^{+-}(m^2) \right) . 
\ee
This immediately implies that $a_{k,0}$ for even $k$ can be interpreted as even moments $\mu_{k}$
\be
\mu_k \equiv {1 \over \pi} \int_{m_{\rm gap}^2}^\infty {d m^2 \over m^{2k + 10}} \left( \sum_{J=0}^\infty {1+(-1)^J \over 2}  \rho_{J}^{++}(m^2)   + \sum_{J=4}^\infty \rho_{J}^{+-}(m^2) \right).
\ee
As a result the bounds of  Ref.~\cite{Bellazzini:2020cot} apply. For odd $k$, we obviously have $|a_{k}| \leq \mu_{k}$.

 ${\bf k=2\!:}$

Next consider $k=2$. Here the couplings of the same dimensionality are ${\bf a}_2 = (a_{2,0}, a_{2,1}, a_{2,2})$. One obvious constraint is
\be
\label{eq:a20positivity}
a_{2,0} \geq 0 \,. 
\ee
On top of that we get the following list of constraints which specify the boundary of the region of allowed couplings,\footnote{\footnotelineskip The constraint \eqref{eq:a20positivity} can be understood as arising from the limiting point when $i_{s} \text{\ or } i_{u} \to \infty$ .}

\be
\label{eq:nonsymlev2}
\{  \< {\bf a}_2 , i_s , i_s+2 \>_{i_s \geq 2} \,, \hskip 1 cm \< {\bf a}_2 , (i_u+1)^{(2)} , i_u^{(2)} \>_{i_u \geq 4} \,, \hskip 1 cm \< {\bf a}_2 , 4_u^{(2)} , 2_s \> \} \geq 0 \, .
\ee
Here we follow the notation of Ref.~\cite{Arkani-Hamed:2020blm}, which is defined as
\be
\<  {\bf a}_2 , i , j \> \equiv \det  \begin{pmatrix}
 a_{2,0} & v_{i,0} & v_{j,0}  \\
 a_{2,1} & v_{i,1} & v_{j,1} \\
 a_{2,2} & v_{i,2} & v_{j,2}  
\end{pmatrix} ,
\label{eq:determinant}
\ee
where we also define the following vectors
\be
v_{J,m}^{(s)} =   (J_{s} )_m &\equiv   {1 \over m! }\partial_t^m d^{0,0}_J (1+2 t) |_{t=0}   \,,  \\
v_{J,k,m}^{(u)} = ( J_u^{(k)} )_m &\equiv   {1 \over m! }\partial_t^m {1 \over (k-m)!} \partial_s^{k-m} {1 \over (1+s+t)} {d^{4,4}_J \left(1+ 2 t  \right) \over (1+t)^4} |_{s,t=0}  \, , 
\label{eq:vsnonsym}
\ee
where in the formula above we set the arbitrary mass scale $m^2 = 1$ since it leads to trivial overall rescaling of the vectors that enter into the determinant \eqref{eq:determinant} and as such does not affect the positivity bounds \eqref{eq:nonsymlev2}.
Note that the vector $v_{i,k}^{(u)}$ depends on the order $k$ of the corrections that we are studying, which we denote by $i_u^{(k)}$. For example, to evaluate (\ref{eq:nonsymlev2}) we should set $k=2$ in \eqn{eq:vsnonsym}. The constraints \eqref{eq:nonsymlev2} have an intuitive explanation. By plugging the dispersive representation of the couplings into the determinant $\<  {\bf a}_2 , i , j \>$ we get a sum over partial waves where the coefficients of partial wave $i$ and $j$ are zero. The plane $\<  {\bf a}_2 , i , j \>$ then separates the region in the coupling space where these coefficients are positive and negative. 

We see that \eqn{eq:nonsymlev2} involves cyclic constraints from the $s$-channel $\< {\bf a}_2 , i_s , i_s+2 \>_{i_s \geq 2}$, similarly from the $u$-channel $ \< {\bf a}_2 , (i_u+1)^{(2)} , i_u^{(2)} \>_{i_u \geq 4}$, as well as a mixed constraint $ \< {\bf a}_2 , 4_u^{(2)} , 2_s \>$ that involves low-spin partial waves. This phenomenon observed in Ref.~\cite{Arkani-Hamed:2020blm} continues to hold for higher $k$ as well. It allows us to easily identify the set of conditions that carve out the region of allowed couplings analytically rather straightforwardly by inspecting the low-spin boundaries. Of course, in practice we only look for possible boundaries up to some finite spin $J_{{\rm max}}$ and assume that the observed pattern continues all the way to $J_{{\rm max}} = \infty$.

In deriving the bounds above we did not impose crossing symmetry. At the level $k=2$ we are working it implies that 
\be
\text{Crossing}: ~~~ a_{2,1} = a_{2,2} \,.
\ee 
Taking the crossing-symmetric slice of the constraints \eqref{eq:nonsymlev2} gives the two-sided bound on the Wilson coefficients
\be
\label{eq:nonsymlev2bound}
  - {90 \over 11} \leq {a_{2,1} \over a_{2,0}} = {a_{2,2} \over a_{2,0}}  \leq 6 \,.
\ee

\begin{figure}
\centering
\includegraphics[scale=.85]{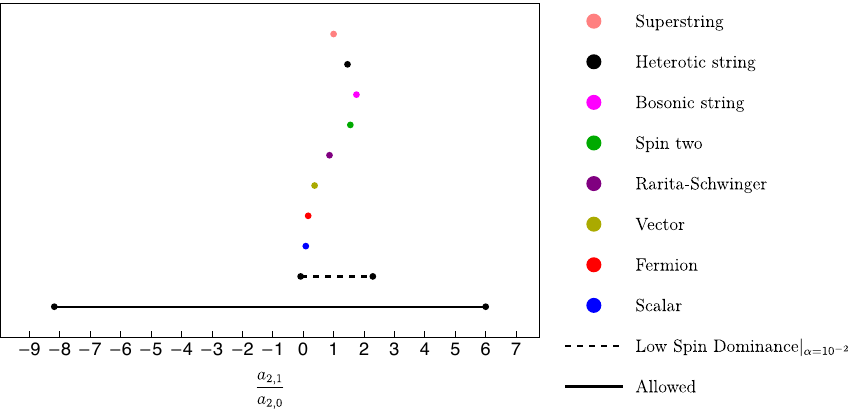}
\caption{\captionlineskip The allowed region for ${a_{2,1} \over a_{2,0}}$ given by \eqn{eq:nonsymlev2bound} is depicted in black. The explicit amplitudes that emerge from integrating out the one-loop matter or tree-level string theories are depicted in various colors. Assuming LSD in the form \eqref{eq:lowspindomAlpha} with $\alpha = 10^{2}$ one can derive stronger bounds which we depict by the dashed line.}
\label{fig:nonsymk2}
\end{figure}

It is very striking that for the explicit UV completions studied here we find a much narrow window of
possibilities that we depict in \fig{fig:nonsymk2}. We now try to understand the origin of this fact.
To do this it will be useful to use the null constraint $a_{2,1}-a_{2,2}=0$ to derive a bound similar to \eqn{eq:nonsymlev2bound}. 
By adding the null constraint to the dispersive representations for $a_{2,1}$ and $a_{2,2}$ it is straightforward to show that
\be
\label{eq:rigorouslow}
  - {108 \over 13} { \langle \rho_{5}^{+-} \rangle_2 \over \langle  \rho_{0}^{++} \rangle_2  + \langle \rho_{4}^{+-} \rangle_2  + \langle \rho_{5}^{+-} \rangle_2  } \leq {a_{2,1} \over a_{2,0}} = {a_{2,2} \over a_{2,0}}  \leq {6 \langle \rho_{2}^{++} \rangle_2 + {16 \over 7} \langle \rho_{4}^{+-} \rangle_2 \over \langle \rho_{0}^{++} \rangle_2  + \langle \rho_{4}^{+-} \rangle_2+ \langle \rho_{2}^{++} \rangle_2} \,,
\ee  
where we introduced $\langle \ldots \rangle$ for the relevant integrals over intermediate energies $m^2$
\be
\label{eq:momentsrho}
 \langle \rho_{J} \rangle_{k} \equiv {1 \over \pi} \int_{m_{\rm gap}^2}^\infty {d m^2 \over m^{2k + 10}} \rho_{J}(m^2) \,.
\ee
To derive the upper bound in \eqn{eq:rigorouslow} we used the dispersive representation for $a_{2,1} - {2 \over 7} (a_{2,2} - a_{2,1})$. To derive the lower bound we considered $a_{2,1} + {4 \over 13} (a_{2,2} - a_{2,1})$.

The bound \eqref{eq:rigorouslow} is rigorous.  Its advantage however is that it only contains low-spin spectral densities. By adding the null constraint $a_{2,1}-a_{2,2}=0$ with an appropriate coefficient we made sure that all the higher-spin contributions are sign-definite and the bound \eqref{eq:rigorouslow} then follows. It is clear then how can we come close to the saturation of the bounds \eqref{eq:nonsymlev2bound}. The upper bound saturation requires that $\langle \rho_{2}^{++} \rangle_2$ is dominant, whereas the lower-bound saturation requires that $\langle \rho_{5}^{+-} \rangle_2$ dominates. Note that each of these is not the minimal spin that appears in the corresponding channel, which are $\langle \rho_{0}^{++} \rangle_2$ and $\langle \rho_{4}^{+-} \rangle_2$ correspondingly. 

This is not what happens in the known physical theories as we will show in more detail. In the examples that we analyze in this paper the higher-spin contributions to the spectral densities are suppressed compared to the minimal spin ones. The fact that at large spin the spectral densities decay exponentially in spin is something familiar from the fact that the scattering is local in the impact parameter space. Analyzing the concrete examples we see that this hierarchy continues all the way to minimal spin. We call this phenomenon {\it low-spin dominance} (LSD) of partial wave  and mathematically we can express it as follows:
\be
\label{eq:lowspindom}
\text{Low-spin dominance (weak)} : ~~~ \langle \rho_{4}^{+-} \rangle_k &\geq \langle \rho_{J>4}^{+-} \rangle_k ,  ~~~ \langle \rho_{0}^{++} \rangle_k \geq \langle \rho_{J>0}^{++} \rangle_k \, .
\ee

In fact, \eqn{eq:lowspindom} is very conservative and in the explicit examples the suppression of higher-spin partial waves is much stronger than \eqn{eq:lowspindom}. Nevertheless, already using \eqns{eq:lowspindom}{eq:rigorouslow} we can strengthen the bound \eqref{eq:nonsymlev2bound} to obtain
\be
\label{eq:k2lowspinboundweak}
\eqref{eq:rigorouslow}+\eqref{eq:lowspindom} : ~~~ - {54 \over 13} \leq {a_{2,1} \over a_{2,0}} = {a_{2,2} \over a_{2,0}}  \leq 3 \,. 
\ee
We clarify that here we derived \eqn{eq:k2lowspinboundweak} to illustrate the point that LSD implies stronger bounds. We have not tried to find the optimal bound that follows from \eqn{eq:lowspindom} and dispersive representations of the couplings. It would be interesting to study this more systematically.

It is also instructive to see what happens if we assume LSD in the strong form
\be
\label{eq:lowspindomSTR}
\text{Low-spin dominance (strong)} : ~~~ \langle \rho_{4}^{+-} \rangle_k &\gg \langle \rho_{J>4}^{+-} \rangle_k ,  ~~~ \langle \rho_{0}^{++} \rangle_k \gg \langle \rho_{J>0}^{++} \rangle_k \,.
\ee
We will discuss in more detail the expected form of the hierarchy below in the section dedicated to spectral densities, but for now it is sufficient to say that we will find that \eqn{eq:lowspindomSTR} correctly captures the region occupied by the known theories. To make \eqref{eq:lowspindomSTR} more precise let us introduce its quantitative version
\be
\label{eq:lowspindomAlpha}
\text{Low-spin dominance ($\alpha$-factor)} : ~~~  \langle \rho_{4}^{+-} \rangle_k &\geq \alpha \langle \rho_{J>4}^{+-} \rangle_k ,  ~~~  \langle \rho_{0}^{++} \rangle_k \geq  \alpha \langle \rho_{J>0}^{++} \rangle_k \,.
\ee
For illustrating bounds, below we choose for concreteness $\alpha = 10^{2}$.  This choice is not accidental as the perturbative examples considered in this paper, tree-level string amplitudes and one-loop matter amplitudes, happen to be of this type.\footnote{\footnotelineskip One exception is the heterotic string amplitude where we have $\langle \rho_{2}^{++} \rangle_{k} \sim {1 \over 10} \langle \rho_{0}^{++} \rangle_k$.}

Let us now consider the bounds coming from \eqn{eq:rigorouslow} with the additional assumption of $\alpha = 10^{2}$ LSD \eqref{eq:lowspindomAlpha}.
In this way we obtain
\be
\label{eq:stronglowspink2}
\eqref{eq:rigorouslow} + \eqref{eq:lowspindomAlpha} |_{\alpha = 10^{2}} : ~~~ -0.083 \leq {a_{2,1} \over a_{2,0}} = {a_{2,2} \over a_{2,0}}  \leq 2.286 \,. 
\ee
It is also interesting to consider what happens if we consider $\alpha \to \infty$. In this case one can again use the trick of adding a null constraint $(a_{2,2}-a_{2,1})$ with some non-zero coefficient to dispersive representation of the couplings to show that
\be
\label{eq:LSDinftyk2}
\text{LSD}_{\alpha \to \infty} : ~~~ 0 \leq {a_{2,1} \over a_{2,0}} = {a_{2,2} \over a_{2,0}}  \leq 2  \,.
\ee

Remarkably, we find that the region occupied by known theories in Fig.~\ref{fig:nonsymk2} lies inside \eqn{eq:stronglowspink2} and even \eqn{eq:LSDinftyk2}. We will see below that the phenomenon of the data being well explained by the strong version of LSD continues for higher $k$'s as well.

 ${\bf k=4\!:}$

\begin{figure}
\centering
\includegraphics[scale=.7]{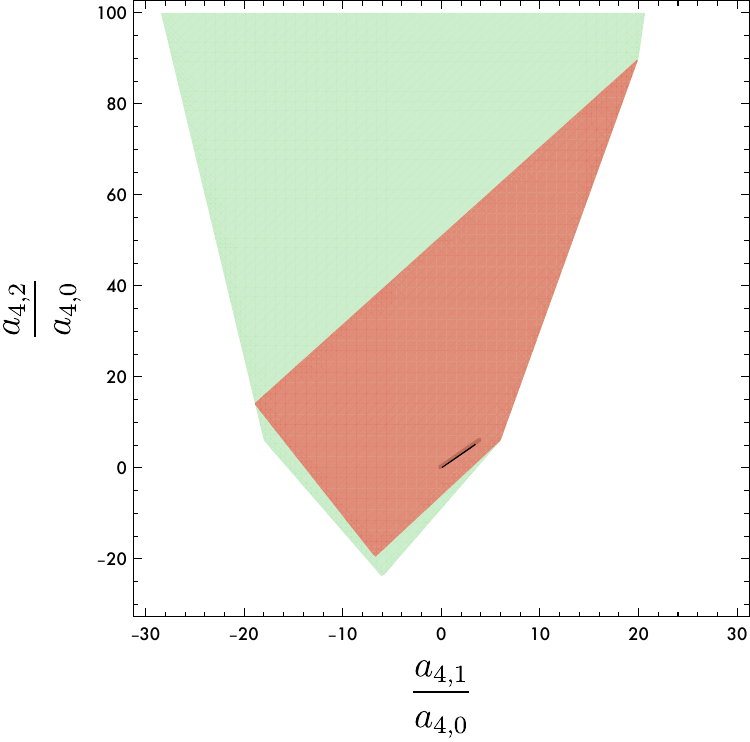}
\caption{\captionlineskip The allowed region for $({a_{4,1} \over a_{4,0}} , {a_{4,2} \over a_{4,0}})$.  The lightly shaded (green) region corresponds to the bounds \eqref{eq:boundsgreen}.
The darkly shaded (red) region corresponds to the bounds \eqref{eq:redregion}. 
The black theory island (which is so narrow that on this scale it looks like a line segment) is the region covered by known amplitudes.
The small gray-shaded region that surrounds the black theory island corresponds to the LSD $\alpha = 10^{2}$ bounds. 
}
\label{fig:a412}
\end{figure}

For $k=4$ we first consider a subset of couplings $ {\bf \hat a_4} = (a_{4,0}, a_{4,1} , a_{4,2})$. From the dispersive representation of the couplings it immediately follows that
\be
\label{eq:simplea40positivity}
a_{4,0} \geq 0 \,.
\ee
Analyzing the sums over the partial waves we get the following list of constraints\footnote{\footnotelineskip By taking $i_{s} \text{\ or } i_{u} \to \infty$ limit in \eqn{eq:boundsgreen} we recover the simple bound \eqref{eq:simplea40positivity}.}
\be
\label{eq:boundsgreen}
\bigl\{  \< {\bf \hat a}_2 , i_s , i_s+2 \>_{i_s \geq 2}\,, ~~ \< {\bf \hat a}_4 , i_u^{(4)}+1 , i_u^{(4)} \>_{i_u \geq 5}\,, ~~   \< {\bf  \hat a}_4 , 5_u^{(4)} , 2_s \> \bigr\} \geq 0 \,.
\ee
These include the usual cyclic constraints, as well as an extra fixed
constraint. We plot the allowed region in the as shaded (green) in
\fig{fig:a412}. A characteristic feature of the allowed region is
that it is unbounded.

In deriving the bounds above we have not used crossing symmetry which
leads to improved two-sided bounds. To do that we can consider a
complete set of couplings at $k=4$, namely ${\bf a_4} = (a_{4,0},
a_{4,1} , a_{4,2}, a_{4,3} , a_{4,4})$.  It is then straightforward to
check by inspection that the following set of constraints define the
boundary region in the space of couplings: 
\be
\label{eq:redregion}
\bigl\{ &\< {\bf a}_4 , i_s, i_s+2 , j_s , j_s + 2 \>_{j > i \geq 2} \,,  ~~ \< {\bf a}_4 , i_u^{(4)}, i_u^{(4)}+1 , j_u^{(4)}, j_u^{(4)}+1 \>_{j > i \geq 5}\, , \nn \\
 &\< {\bf a}_4 , i_s, i_s+2 , j_u^{(4)}+1 , j_u^{(4)} \>_{i \geq 4, j \geq 5}\,, ~~ \< {\bf a}_4 , 2_s, 4_s , j_u^{(4)}+1 , j_u^{(4)} \>_{j \geq 7} \,, \nn \\
& \< 2_s, {\bf a}_4 , i_s , i_s+2  , 5^{(4)}_u \>_{i \geq 4} \,, ~~ \< 5_u^{(4)},  {\bf a}_4 , j_u^{(4)} , j_u^{(4)} + 1, 2_s \>_{j \geq 7} \,, \nn \\
 & \< {\bf a}_4 , 4_u^{(4)}, 5_u^{(4)}, 6_u^{(4)}, 7_u^{(4)} \> \,, ~~ \< {\bf a}_4 , 2_s, 4_s,  4_u^{(4)}, 5_u^{(4)} \>,  \< {\bf a}_4 , 2_s, 4_s,  7_u^{(4)}, 4_u^{(4)} \> \,, \nn \\
 & \<   {\bf a}_4, 4_u^{(4)} , 5_u^{(4)} ,  7_u^{(4)} , 2_s\>  \,, \<  4_u^{(4)},  {\bf a}_4 , 5_u^{(4)} ,  6_u^{(4)} , 4_s\>  , \<  4_u^{(4)},  {\bf a}_4 , 6_u^{(4)} ,  7_u^{(4)} , 4_s\> \bigr\} \geq 0 \,.
\ee
It is easy to recover the previous constraints \eqref{eq:boundsgreen} from \eqn{eq:redregion} by taking some of the spins to infinity. For example $\lim_{i_s \to \infty}  \< 2_s, {\bf a}_4 , i_s , i_s+2  , 5^{(4)}_u \>_{i \geq 4} \geq 0$ reduces to $\< {\bf  \hat a}_4 , 5_u^{(4)} , 2_s \> \geq 0$. 

A somewhat new feature of this case compared to the ones considered above that mixed $s$--$u$ channel boundaries come in infinite families. However, these families are again either cyclic, as in $\< {\bf a}_4 , i_s, i_s+2 , j_u^{(4)}+1 , j_u^{(4)} \>_{i \geq 4, j \geq 5}$ or involve low-spin partial wave only and therefore easily identifiable. We find it quite remarkable that the allowed region can be found analytically!


We then consider the section of this region by the crossing symmetry relations that take the form
\be
\label{eq:crossingk4}
\text{Crossing}: ~~~ a_{4,3} = 2 (a_{4,2} - a_{4,1})\,, ~~~~~ a_{4,4} = a_{4,2} - a_{4,1} \, .
\ee
As a result we get the region of allowed couplings depicted in red in \fig{fig:a412}. To generate the plot we considered bounds \eqref{eq:redregion} where we truncated the maximal spin to $i_{\rm max},j_{\rm max}, k_{\rm max} = 20$. We also checked explicitly that the all determinants obtained in this way are non-negative given dispersive representation for the couplings truncated to $J_{\rm max}=200$. 

The bounds coming from imposing \eqn{eq:redregion} together with crossing symmetry \eqref{eq:crossingk4} are two-sided, both from above and from below. The theory island occupied by the known theories forms a tiny black slit inside the allowed region.

As we did for $k=2$ we can again understand the structure of the island using the idea of LSD. To this extent we can use the null constraints coming from crossing \eqref{eq:crossingk4} to derive the following rigorous bounds
\be
\label{eq:boundsk4lowspin}
-{7.01  \langle \rho_{5}^{+-} \rangle_4 + 20.18 \langle \rho_{6}^{+-} \rangle_4 \over \langle \rho_{0}^{++} \rangle_4  + \langle \rho_{4}^{+-} \rangle_4+  \langle \rho_{5}^{+-} \rangle_4 +  \langle \rho_{6}^{+-} \rangle_4}\leq &{a_{4,1} \over a_{4,0}}
\leq  {6 \langle \rho_{2}^{++} \rangle_4 + 20 \langle \rho_{4}^{++} \rangle_4 + 4.01 \langle \rho_{4}^{+-} \rangle_4 \over \langle \rho_{0}^{++} \rangle_4  + \langle \rho_{4}^{+-} \rangle_4+ \langle \rho_{2}^{++} \rangle_4 + \langle \rho_{4}^{++} \rangle_4} \,,  \\
-{20.16  \langle \rho_{5}^{+-} \rangle_4 \over \langle \rho_{0}^{++} \rangle_4  + \langle \rho_{4}^{+-} \rangle_4+  \langle \rho_{5}^{+-} \rangle_4 +  \langle \rho_{6}^{+-} \rangle_4} \leq &{a_{4,2} \over a_{4,0}}   \nn\\
& \hskip - 2 cm 
\leq  {6 \langle \rho_{2}^{++} \rangle_4 + 90 \langle \rho_{4}^{++} \rangle_4 + 6.43 \langle \rho_{4}^{+-} \rangle_4 + 15.32 \langle \rho_{6}^{+-} \rangle_4 \over \langle \rho_{0}^{++} \rangle_4  + \langle \rho_{4}^{+-} \rangle_4+ \langle \rho_{2}^{++} \rangle_4 + \langle \rho_{4}^{++} \rangle_4 +  \langle \rho_{6}^{+-} \rangle_4} \,. 
\label{eq:boundsk4lowspin2}
\ee
Again we see that the corners of the red regions in Fig.~\ref{fig:a412} can be obtained from \eqn{eq:boundsk4lowspin} by assuming that the sub-leading spins in the corresponding channels are dominant. This is not what happens in the physical theories and again assuming weak LSD \eqref{eq:lowspindom} we can get a tighter bounds in which physical theories reside. In practice, we see that physical theories occupy even smaller region which can be understood using the strong version of LSD \eqref{eq:lowspindomSTR}.


\begin{figure}
\centering
\includegraphics[scale=.75]{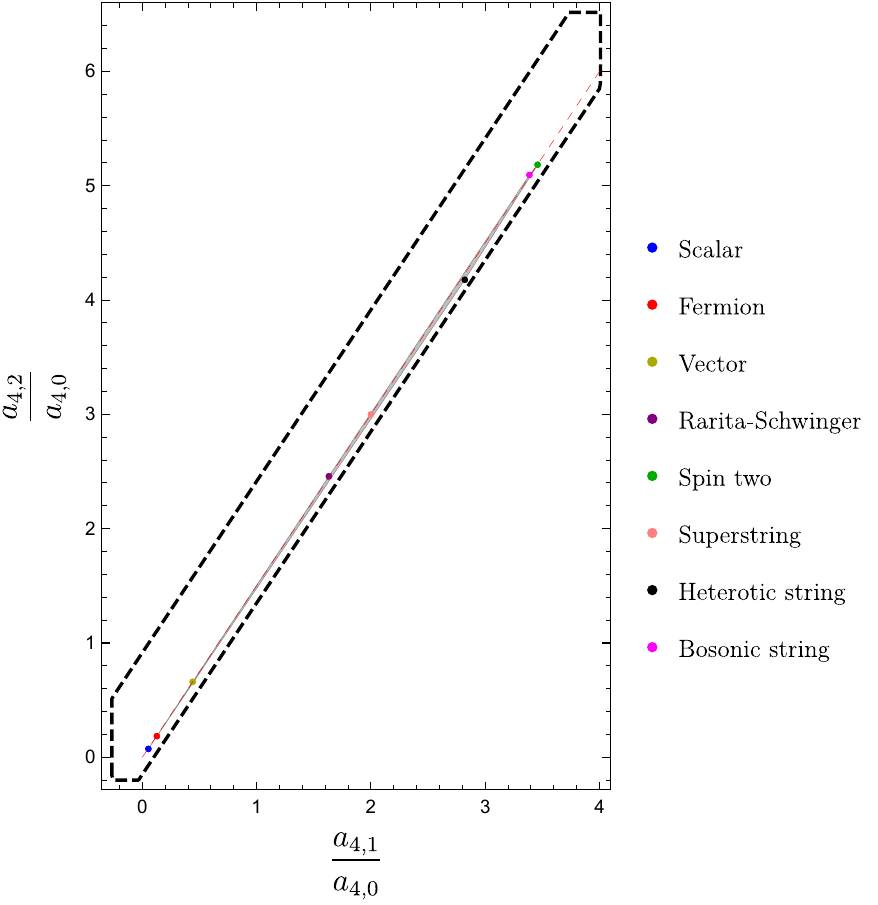}
\caption{\captionlineskip A scaled version of the theory island from \fig{fig:a412}. The dashed black line bounds the region which is found from \eqns{eq:boundsk4lowspin}{eq:differenceboundk4} using the LSD assumption \eqref{eq:lowspindomAlpha} with $\alpha = 10^{2}$. The dashed red line $a_{4,2}={3 \over 2} a_{4,1}$  corresponds to $\alpha=\infty$ bound \eqref{eq:LSDinfty}. Remarkably, all known theories lie in this small region where they populate a small region around the $\alpha = \infty$ curve.
}
\label{fig:a412scaled}
\end{figure}

A striking feature of  \fig{fig:a412scaled} is that the known theories align closely with the straight line with the slope ${3 / 2}$. To understand this we can derive the bound analogous to \eqn{eq:boundsk4lowspin} for the difference $a_{4,2} - {3 \over 2} a_{4,1}$. It takes the following form
\be
\label{eq:differenceboundk4}
-{3 \langle \rho_{2}^{++} \rangle_4 +  15  \langle \rho_{5}^{+-} \rangle_4 \over \langle \rho_{0}^{++} \rangle_4 + \langle \rho_{4}^{+-} \rangle_4  + \langle \rho_{2}^{++} \rangle_4+  \langle \rho_{5}^{+-} \rangle_4} \leq &{a_{4,2} - {3 \over 2} a_{4,1} \over a_{4,0}} \leq  {60 \langle \rho_{4}^{++} \rangle_4 + 0.47 \langle \rho_{4}^{+-} \rangle_4 + 45.58 \langle \rho_{6}^{+-} \rangle_4 \over \langle \rho_{0}^{++} \rangle_4  + \langle \rho_{4}^{+-} \rangle_4 + \langle \rho_{4}^{++} \rangle_4 +  \langle \rho_{6}^{+-} \rangle_4} \,.
\ee
We emphasize that at this point the bound \eqref{eq:differenceboundk4} is rigorous and no additional assumptions have been made. 
To derive an upper bound we have considered the dispersive representation for $a_{4,2} - {3 \over 2} a_{4,1} + {1687 \over 7205}(a_{4,3} - 2 a_{4,4}) + {411 \over 524} (a_{4,3} - 2(a_{4,2} - a_{4,1})) $, where we made use of crossing \eqref{eq:crossingk4}. The 
lower bound follows directly from the dispersive representation of $a_{4,2} - {3 \over 2} a_{4,1}$.

To make use of it, we can now assume $\alpha = 10^{2}$ LSD and apply it to the bounds \eqref{eq:boundsk4lowspin} and \eqref{eq:differenceboundk4}. In this way we get
\be
\eqref{eq:boundsk4lowspin} + \eqref{eq:boundsk4lowspin2}+\eqref{eq:differenceboundk4}+\text{LSD}_{\alpha =10^2} : ~~~ -0.27 \leq &{a_{4,1} \over a_{4,0}} \leq 4.01, \\
-0.20 \leq &{a_{4,2} \over a_{4,0}} \leq 6.52 , \nn \\
-0.15 \leq &{a_{4,2} - {3 \over 2} a_{4,1} \over a_{4,0}} \leq 0.92 .
\ee
We plot the result in Fig.~\ref{fig:a412scaled} and again observe that the known examples neatly land in the predicted region. Moreover, it is straightforward to see that
if we increase $\alpha$ the bound \eqref{eq:differenceboundk4} becomes not optimal and instead we get
\be
\label{eq:LSDinfty}
\text{LSD}_{\alpha \to \infty} : ~~~~ {a_{4,2} - {3 \over 2} a_{4,1}  \over a_{4,0}} &=0 , ~~~~0 \leq {a_{4,1} \over a_{4,0}} \leq 4 , ~~~~ 0 \leq {a_{4,2} \over a_{4,0}} \leq 6 \,. 
\ee
which is the line along which our explicit examples cluster. Intuitively, this result can be understood as follows:
We write down explicitly the dispersive representation for the couplings $a_{4,i}$, 
\be
a_{4,0} &=\int_{m_{\rm gap}^2}^\infty {d m^2 \over \pi }  {\rho_{++0}(m^2) \over (m^2)^9}+ \int_{m_{\rm gap}^2}^\infty {d m^2 \over \pi }  {\rho_{+-4}(m^2) \over (m^2)^9}  + {\rm higher \ spin} \,, \hskip 2 cm \null \nn\\
a_{4,1} &= 4 \int_{m_{\rm gap}^2}^\infty {d m^2 \over \pi }  {\rho_{+-4}(m^2) \over (m^2)^9}  + {\rm higher \ spin} \,,  \nn \\
a_{4,2} & = 6 \int_{m_{\rm gap}^2}^\infty {d m^2 \over \pi }
{\rho_{+-4}(m^2) \over (m^2)^9} + {\rm higher \ spin} \,, 
\label{eq:lowspink4} 
\ee 
where `higher spin' denotes infinitely
many partial-wave contributions. Suppose that the lowest-spin partial waves in each
channel dominates. Note that the leading spin $J=0$ contribution in
the $\rho_{++J}$ channel drops out from $a_{4,1}$ and
$a_{4,2}$. Notice that if we set the `higher-spin' terms to $0$ in the formulas
above we get the region \eqref{eq:LSDinfty}. A priori it is not clear
that this follows from sending $\alpha \to \infty$ in
\eqn{eq:lowspindomAlpha} since we still have infinitely many spins
that can potentially compensate for the smallness of
$1/\alpha$. Crossing symmetry, however, guarantees that it is indeed the
case and higher-spin contributions cannot compensate for smallness of
$1/\alpha$ as can be seen from the explicit analysis using the finite
number of partial-wave bounds similar to \eqn{eq:differenceboundk4}.

Note that there are two distinct ways in which points can be close to
the straight line in \fig{fig:a412scaled}. First, as explained above
by assuming the strong version of LSD in $a_{4,1}$ and
$a_{4,2}$. Another mechanism is to have $a_{4,0} \gg a_{4,1} ,
a_{4,2}$ which happens when the contribution of $\rho_{++0}$ to
$a_{4,0}$ dominates over the partial waves that enter to $a_{4,1}$ and
$a_{4,2}$. In this case the points appear very close to the origin on
the $({a_{4,1} \over a_{4,0}} , {a_{4,2} \over a_{4,0}})$ plot and the
slope defined by the ratio ${a_{4,2} \over a_{4,1}}$ is not directly
visible, nor does it need to be ${3 \over 2}$.  In fact, $a_{4,1}$ and
$a_{4,2}$ have nontrivial higher-spin corrections in
\eqn{eq:lowspink4}, i.e. $\rho^{++}_2$ is sizable compared to
$\rho^{+-}_4$ and contributes ruining the $3/2$ ratio for ${a_{4,2}
  \over a_{4,1}}$.  This is precisely what happens for the scalar and
fermion contributions in \fig{fig:a412scaled}. In this case, by
looking at Fig.~\ref{fig:a412scaled} it is not apparent that
\eqn{eq:lowspink4} is not an accurate description of these points
since they are close to the origin, which is on the line.  When this
happens the simple model of dropping higher-spin contributions in
\eqn{eq:lowspink4} is too crude. For example, consider the one-loop
massive scalar amplitude which lies very close to the origin in
Fig.~\ref{fig:a412scaled}. In this case to accurately capture
$a_{4,1}$ we also need to include $\rho^{++}_2(m^2)$ (which is
comparable to $\rho^{+-}_4(m^2)$ in this case) in the formulas
\eqref{eq:lowspink4}. For $a_{4,2}$ to get $1\%$ precision
$\rho^{++}_4(m^2)$ required as well.  We re-iterate that in deriving
the bounds using LSD we did not make any assumptions about the
contributions of infinitely many higher-spin terms in
\eqn{eq:lowspink4} and instead used crossing symmetry to derive
rigorous and tight bounds based on \eqn{eq:lowspindomAlpha}. In
contrast, when we use \eqn{eq:lowspink4} and try to estimate the
contribution of higher-spin partial waves we observe that depending on
the model and the coupling at hand, the number of terms required to be
kept in the expansion to reach good precision varies.

We observe that the relation \eqref{eq:LSDinfty} generates a hierarchy $\sim
10^{-2}$ between certain coefficients in the low-energy EFT. A naive
low-energy observer could have been puzzled by the fact that $|
a_{4,2} - {3 \over 2} a_{4,1} | \ll a_{4,0}$. We see that this
hierarchy is generated by unitarity, namely it appears due to the
dominance of the dispersive integrals by the low-spin partial waves.
We discuss this point further in \sect{sec:hierarchyfromunitarity}.

${\bf k=6 \!: }$

\begin{figure}[tbh]
\centering
\includegraphics[scale=.65]{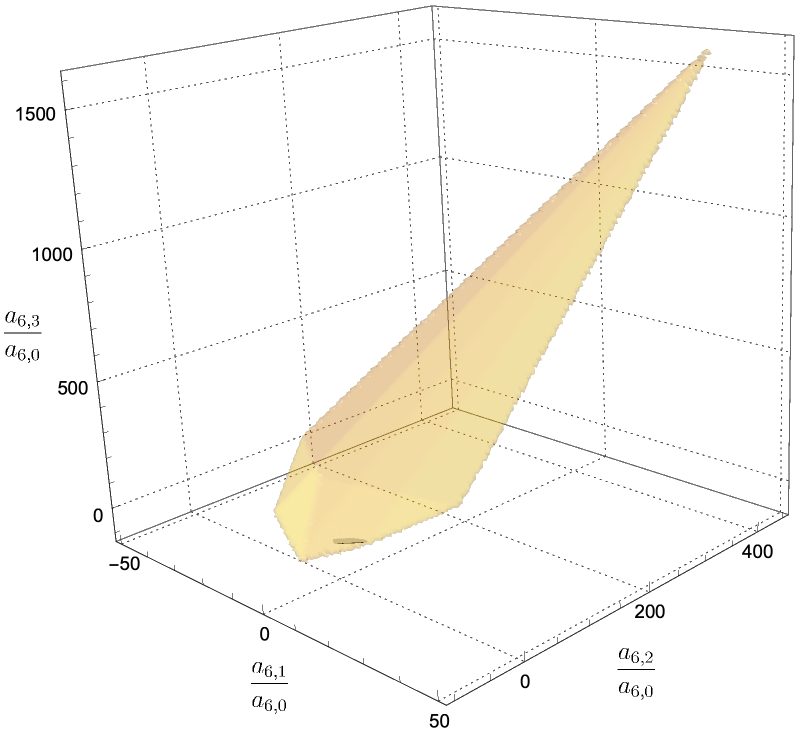}
\caption{\captionlineskip The allowed region for couplings $({a_{6,1} \over a_{6,0}}, {a_{6,2} \over a_{6,0}}, {a_{6,3} \over a_{6,0}})$ derived from constraints \eqref{eq:k6region}. 
The black little island (barely visible) is the space occupied by known perturbative amplitudes.
The small gray shaded region that surrounds it corresponds to the LSD $\alpha = 10^{2}$ bound. The details on how the plot was generated can be found in Appendix~\ref{app:fullk6analysis}.
}
\label{fig:k6full}
\end{figure}

The analysis gets more and more complicated as we go to higher $k$. For $k=6$ we get the $7$-dimensional coupling vector ${\bf a_6} = (a_{6,0}, a_{6,1} , a_{6,2}, a_{6,3}, a_{6,4}, a_{6,5}, a_{6,6})$. Out of seven couplings only four are independent due to the crossing-symmetry relations
\be
\label{eq:crossingk6}
\text{Crossing}: ~~~ a_{6,4} &= 5 (a_{6,1} - a_{6,2}) + 3 a_{6,3} \,, \nn \\
a_{6,5} &=6 (a_{6,1} - a_{6,2}) + 3 a_{6,3}\,, \nn \\
a_{6,6} &= 2 (a_{6,1} - a_{6,2}) + a_{6,3} \,.
\ee
Working out the boundary of the coupling region is more laborious, but follows the same pattern that we observed before
\be
\label{eq:k6region}
\bigl\{ &\< {\bf a}_6 , i_s, i_s+2 , j_s , j_s + 2, k_s, k_s+2 \>_{k>j > i \geq 2}\,,  \nn \\
&\< {\bf a}_6 , i_u^{(6)}+1, i_u^{(6)} , j_u^{(6)}, j_u^{(6)}+1 ,  k_u^{(6)}, k_u^{(6)}+1  \>_{k> j > i \geq 5}\,, \nn \\
 &\< {\bf a}_6 , 5_u^{(6)} , 4_u^{(6)} , 6_u^{(6)} , 7_u^{(6)} , 8_u^{(6)} , 9_u^{(6)} \>\,, ~~~ \text{mixed $s$--$u$ constraints} \bigr\} \geq 0 \,,
\ee
where we list the explicit mixed constraints that we found in Appendix~\ref{app:fullk6analysis}.  After the boundaries are identified we take the crossing-symmetric slice \eqref{eq:crossingk6}.
The resulting region of allowed couplings is plotted in \fig{fig:k6full}. The region covered by the known theories is again a small island in the space of couplings. 

As in the analysis above the island occupied by the explicit examples can be understood using the idea of strong LSD. As before the first step is to derive a set of rigorous bounds in terms of the low-spin partial waves. We do not present the complete analysis here but only present some of the relevant bounds
\be
\label{eq:extraboundsk6low}
-{9 \langle \rho_2^{++} \rangle_6 + 25  \langle \rho_{5}^{+-} \rangle_6 \over \langle \rho_{0}^{++} \rangle_6  + \langle \rho_{4}^{+-} \rangle_6 +  \langle \rho_{2}^{++} \rangle_6 + \langle \rho_{5}^{+-} \rangle_6 }\leq &{a_{6,2} - {5 \over 2} a_{6,1} \over a_{6,0}} \nn \\
& \null \hskip - 2 cm 
\leq  {40 \langle \rho_{4}^{++} \rangle_6 + 315 \langle \rho_{6}^{++} \rangle_6 + 32.56 \langle \rho_{6}^{+-} \rangle_6 + 220.41 \langle \rho_{7}^{+-} \rangle_6  \over \langle \rho_{0}^{++} \rangle_6  + \langle \rho_{4}^{+-} \rangle_6+ \langle \rho_{4}^{++} \rangle_6+ \langle \rho_{6}^{++} \rangle_6 + \langle \rho_{6}^{+-} \rangle_6+ \langle \rho_{7}^{+-} \rangle_6} \,,  \\
-{20 \langle \rho_{2}^{++} \rangle_6 + 66.67  \langle \rho_{5}^{+-} \rangle_6 \over  \langle \rho_{0}^{++} \rangle_6  + \langle \rho_{4}^{+-} \rangle_6 +  \langle \rho_{2}^{++} \rangle_6 + \langle \rho_{5}^{+-} \rangle_6} \leq &{a_{6,3} - {10 \over 3} a_{6,1} \over a_{6,0}} \nn\\
& \null \hskip - 2 cm
\leq  {73.34 \langle \rho_{4}^{++} \rangle_6 + 1540 \langle \rho_{6}^{++} \rangle_6 + 163.49 \langle \rho_{6}^{+-} \rangle_6 + 495.64 \langle \rho_{7}^{+-} \rangle_6 \over \langle \rho_{0}^{++} \rangle_6  + \langle \rho_{4}^{+-} \rangle_6+ \langle \rho_{4}^{++} \rangle_6+ \langle \rho_{6}^{++} \rangle_6 + \langle \rho_{6}^{+-} \rangle_6+ \langle \rho_{7}^{+-} \rangle_6} .
\ee

From the formulas above and a similar analysis for $a_{6,i}$ we also see that
\be
\label{eq:LSDinftyk6}
\text{LSD}_{\alpha \to \infty} : ~~~~~ &{a_{6,2} - {5 \over 2} a_{6,1} \over a_{6,0}} =0 \,, ~~~~~ {a_{6,3} - {10 \over 3} a_{6,1} \over a_{6,0}} = 0 \,, \nn \\
0 &\leq {a_{6,1} \over a_{6,0}} \leq 6 \,, ~~~~~ 0 \leq {a_{6,2} \over a_{6,0}} \leq 15 \,, \nn \\
0 &\leq {a_{6,3} \over a_{6,0}} \leq 20 \,.
\ee
As for $k=4$ we can understood the result above by simply dropping higher-spin contributions in the dispersive representations for the couplings and keeping only the lowest-spin partial waves in each channel
\be
\label{eq:lowspindominancek6}
a_{6,0} &=\int_{m_{\rm gap}^2}^\infty {d m^2 \over \pi }  {\rho_{++0}(m^2) \over (m^2)^{11}} + \int_{m_{\rm gap}^2}^\infty {d m^2 \over \pi }  {\rho_{+-4}(m^2) \over (m^2)^{11}}  + {\rm higher \ spin} , \\
a_{6,1} &=6 \int_{m_{\rm gap}^2}^\infty {d m^2 \over \pi }  {\rho_{+-4}(m^2) \over (m^2)^{11}}  + {\rm higher \ spin} ,  \\
a_{6,2} &=15 \int_{m_{\rm gap}^2}^\infty {d m^2 \over \pi }  {\rho_{+-4}(m^2) \over (m^2)^{11}}  + {\rm higher \ spin} ,  \\
a_{6,3} &=20 \int_{m_{\rm gap}^2}^\infty {d m^2 \over \pi }  {\rho_{+-4}(m^2) \over (m^2)^{11}}  + {\rm higher \ spin} .
\ee
We emphasize again that while dropping infinitely many terms is not justified, the bound
\eqref{eq:LSDinftyk6} is rigorous. 

\begin{figure}[tbh]
\centering
\includegraphics[scale=.65]{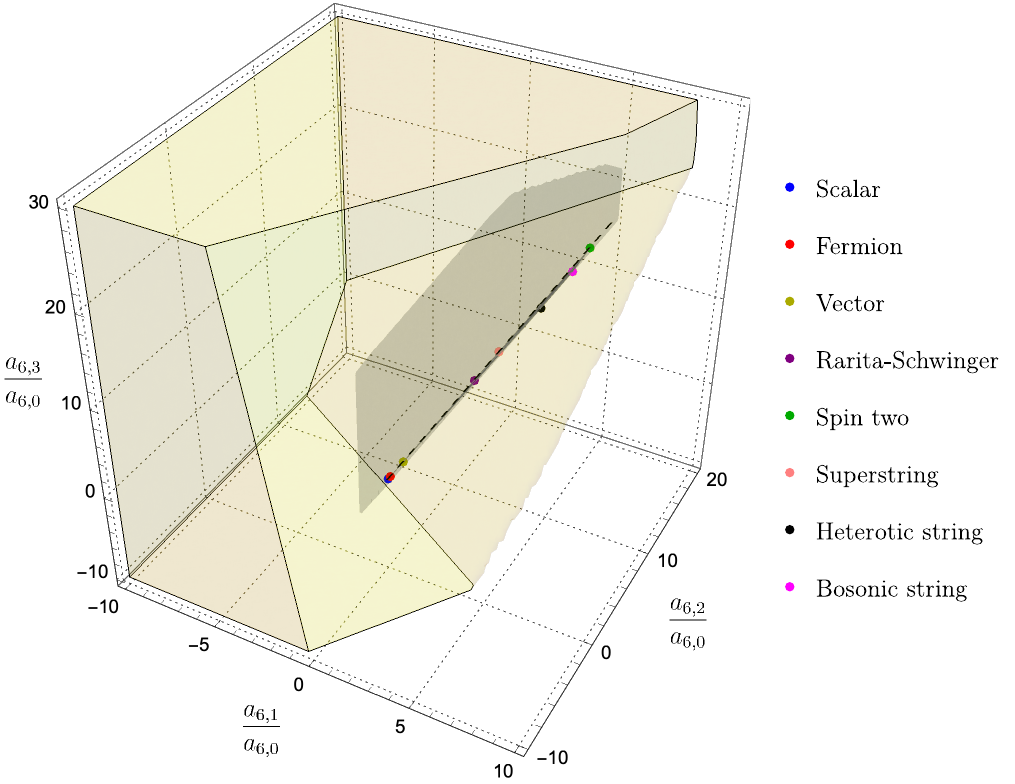}
\caption{\captionlineskip The scaled version of the theory island from \fig{fig:k6full}. The gray region is derived using the bounds \eqref{eq:extraboundsk6low} (and similar bounds for $a_{6,i}$ which we do not write down explicitly) and the LSD assumption \eqref{eq:lowspindomAlpha} with $\alpha = 10^{2}$. The dashed black line ${a_{6,2} \over a_{6,1}} = {5 \over 2}$ and ${a_{6,3} \over a_{6,1} }= {10 \over 3}$ corresponds to $\text{LSD}_{\alpha \to \infty }$ bounds given in \eqn{eq:LSDinftyk6}.
}
\label{fig:k6scaled}
\end{figure}

Assuming the strong version of the LSD and using \eqn{eq:extraboundsk6low} we again find the small region around the island occupied by the explicit examples, as shown in Fig.~\ref{fig:k6full} and Fig.~\ref{fig:k6scaled}.

${\bf k=8\!:}$

\begin{figure}[tbh]
  \hspace{-24pt}   
  \begin{tabular}{m{7cm}  m{7cm}}
\includegraphics[scale=.5]{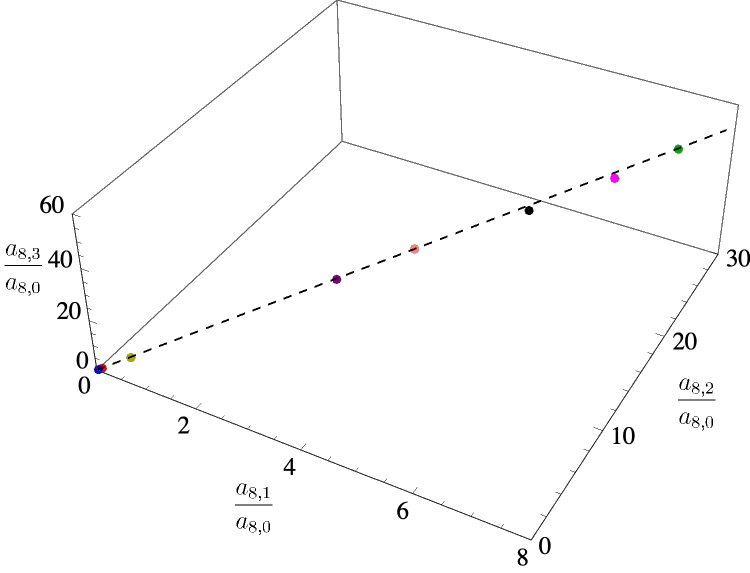} &
\includegraphics[scale=.5]{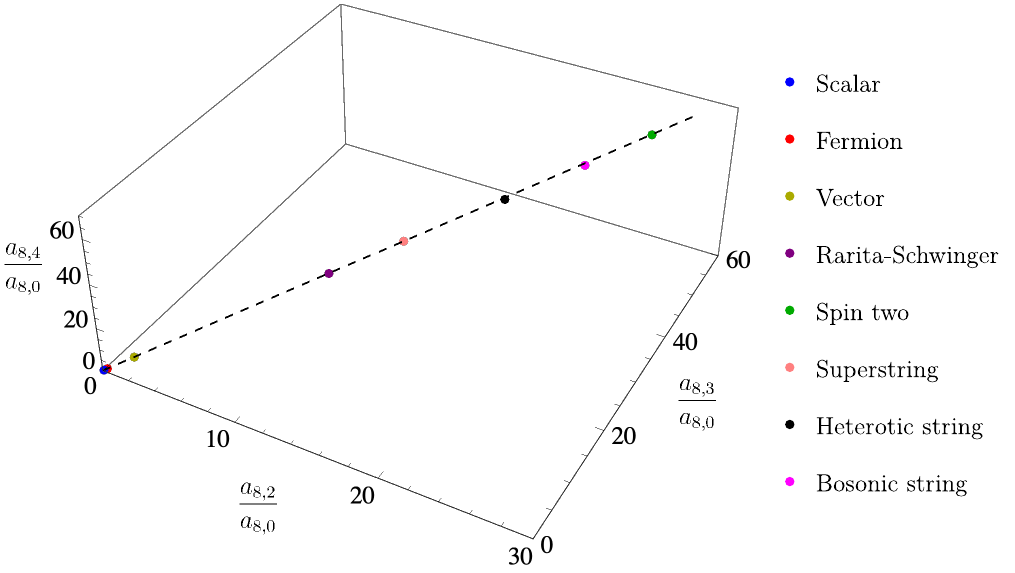}\\
\hskip 3.3 cm (a) & \hskip 3.3 cm (b)                               
 \end{tabular}
 \caption{\captionlineskip A plot for the data for $a_{8,j}$ in various theories. In panel a) we plot $({a_{8,1} \over a_{8,0}}, {a_{8,2} \over a_{8,0}} , {a_{8,3} \over a_{8,0}})$  and in panel b) $({a_{8,2} \over a_{8,0}}, {a_{8,3} \over a_{8,0}} , {a_{8,4} \over a_{8,0}})$. The dashed line corresponds to the low-spin dominant line defined by \eqn{eq:lowspindominancek8} upon neglecting the higher-spin contributions to the partial-wave expansion.
 }
  \label{fig:resultsk8}
\end{figure}

For $k=8$ we do not perform the analysis of finding the allowed region but simply report on the data in the specific theories. There are five independent couplings at this level which we choose to be $a_{8,0\leq j \leq 4}$. Other couplings $a_{8,5\leq j \leq 8}$ can be obtained by crossing. The dispersive representation of the couplings take the following form
\be
a_{8,0} &=\int_{m_{\rm gap}^2}^\infty {d m^2 \over \pi }  {\rho_{++0}(m^2) \over (m^2)^{13}} 
  + \int_{m_{\rm gap}^2}^\infty {d m^2 \over \pi }  {\rho_{+-4}(m^2) \over (m^2)^{13}}  + {\rm higher \ spin} \,, \nn\\
a_{8,1} &=8 \int_{m_{\rm gap}^2}^\infty {d m^2 \over \pi }  {\rho_{+-4}(m^2) \over (m^2)^{13}}  + {\rm higher \ spin} \,, \nn \\
a_{8,2} &=28 \int_{m_{\rm gap}^2}^\infty {d m^2 \over \pi }  {\rho_{+-4}(m^2) \over (m^2)^{13}}  + {\rm higher \ spin} \,, \nn \\
a_{8,3} &=56 \int_{m_{\rm gap}^2}^\infty {d m^2 \over \pi }  {\rho_{+-4}(m^2) \over (m^2)^{13}}  + {\rm higher \ spin} \,, \nn \\
a_{8,4} &=70 \int_{m_{\rm gap}^2}^\infty {d m^2 \over \pi }  {\rho_{+-4}(m^2) \over (m^2)^{13}}  + {\rm higher \ spin} \,,
\label{eq:lowspindominancek8}
\ee
where we keep the leading spin contributions in both channels. To visualize the data we consider two 3-dimensional slices of the space of couplings with coordinates $({a_{8,1} \over a_{8,0}}, {a_{8,2} \over a_{8,0}} , {a_{8,3} \over a_{8,0}})$ and $({a_{8,2} \over a_{8,0}}, {a_{8,3} \over a_{8,0}} , {a_{8,4} \over a_{8,0}})$. Formulas \eqref{eq:lowspindominancek8} define a line in this space upon neglecting the higher-spin partial wave contributions. We depict the result in \fig{fig:resultsk8} and it is again completely analogous to our observations for lower $k$'s.
It would be interesting to extend the analysis done for lower $k$ to this case as well.

\subsection{Crossing-symmetric dispersive representation of low-energy couplings}
\label{sec:susymmetric}

We can also analyze the same amplitude using dispersive representation with a different pair of channels. We start by recalling that
\be
\mathcal{M}_4  (1^+,2^-,3^-,4^+) = (\langle 23 \rangle [14])^4 \, f (s,u) \,.
\ee
We would like to use dispersion relations at fixed $t$ to derive the dispersive representation for the low-energy couplings. For this purpose it is natural to introduce the following parameterization of the low-energy expansion of $f(s,u)$
\be
\label{eq:ftildeexpans}
f(z-{t \over 2}, - z -{t \over 2}) &= \left( \frac{\kappa \null}{2} \right)^2 {1 \over t ({t^2 \over 4} - z^2)} +  |\beta_{R^3}|^2 { ({t^2 \over 4} - z^2) \over t}  - |\beta_\phi|^2  {1 \over t} \nn \\
& \hskip .3 cm 
+\sum_{k \geq q \geq 0}^\infty \tilde a_{k,q} z^{k-q} t^q, ~~~~~ z = s+{t \over 2} , ~~~~~ k-q \in 2 {\mathbb Z}_{\geq 0} \, .
\ee
Crossing symmetry $f(s,u) = f(u,s)$ acts as $z \to - z$ and therefore constrains $k-q$ to be even. 

To derive the bounds we can try to use dispersive representation for $\tilde a_{k,q}$, where we keep $t$ fixed and deform the $z$ integral
\be
\label{eq:tildea}
 \tilde a_{k,q}  &= {1 \over q!} \partial_t^q \oint {d z \over 2 \pi i} {1 \over  z^{k-q+1}} \nn \\
 & \Big[ f(z-{t \over 2}, - z -{t \over 2}) - \left( \frac{\kappa \null}{2} \right)^2 {1 \over t ({t^2 \over 4} - z^2)} -  |\beta_{R^3}|^2 { ({t^2 \over 4} - z^2) \over t} + |\beta_\phi|^2  {1 \over t}  \Big] \Big|_{t=0},
\ee
where the contour integral encircles the origin and we explicitly subtracted the contributions that are singular at $t=0$. As discussed in the previous section, unitarity constrains the form of the discontinuity of $f(z-{t \over 2}, - z -{t \over 2})$. 
Therefore, in evaluating \eqn{eq:tildea} we can open the contour and assuming we can drop the arcs at infinity to arrive at the following representation
\be
\label{eq:tildeadisprep}
\tilde a_{k,q}  = {2 \over q!} \partial_t^q  \int_{m_{\rm gap}^2}^\infty {d m^2 \over \pi} \sum_{J=4}^\infty  {\rho_{J}^{+-}(m^2) \over (m^2+{t \over 2})^{k-q+1}} {(-1)^J d_{4,-4}^J \left(1+{2t \over m^2} \right) \over t^4} \Bigr|_{t=0}, ~~~~~ k-q \in 2 {\mathbb Z}_{\geq 0}  \,. 
\ee
The factor of $2$ originates from the sum over the $s$-channel and the $u$-channel discontinuities. For odd $k-q$ they cancel each other and we get zero. An important factor ${1 / t^4}$ originates from the fact that unitarity constraints are formulated in terms of $\mathcal{M}_{+--+}$ which includes the prefactor $(\langle 23 \rangle [14])^4$---see \sect{sec:unitarity} for details. The factor $(-1)^J$ can be understood from the fact that with a given choice of helicity the discontinuity is positive for the forward limit $u=0$ (as opposed to $t=0$).\footnote{\footnotelineskip Consistency between \eqn{eq:dispersionrelationsf} and \eqn{eq:tildeadisprep}, namely matching of the $\rho_{J}^{+-}$ discontinuities, requires that
$
d_{4,4}^J(x) = (-1)^J d_{4,-4}^J(-x)
$,
which is indeed the case.} 
In using the representation (\ref{eq:tildeadisprep}) we should not forget that it was derived assuming that the Regge behavior of the amplitude is such that the arcs at infinity can be dropped. In particular, given that $ f(z-{t \over 2}, - z -{t \over 2})  \sim z^{J_0}$ for large $|z|$, and taking into account the subtractions of terms that are singular at $t=0$ the representation (\ref{eq:tildeadisprep}) is valid only for $k-q > {\rm max} \{ J_0,2 \}$.  As opposed to the previous section all $\tilde a_{k,q}$, whose dispersive representation is given in \eqn{eq:tildeadisprep}, are independent and there are no extra constraints coming from crossing.

The presence of $(-1)^J$ in the sum \eqref{eq:tildeadisprep} prevents us from deriving useful bounds from the representation \eqref{eq:tildeadisprep}.
We, however, present the data for $\tilde a_{k,q}$ obtained from the explicit amplitudes since it reveals an interesting aspect of the discussion in the previous 
section.

For example, consider $k=6$. The mapping between $a_{6,j}$ and $\tilde a_{6,j}$ takes the following form
\be
\tilde a_{6,0} &= 2 (a_{6,1} - a_{6,2}) + a_{6,3} \,,  \nn \\
\tilde a_{6,2} &= {1 \over 4} (-10 a_{6,1} + 10 a_{6,2} -3 a_{6,3} )  \,, \nn  \\
\tilde a_{6,4} &= {1 \over 16} (30 a_{6,1} - 14 a_{6,2} + 3 a_{6,3} ) \,, \label{eq:tildekvariables}
\ee
and we will not need $\tilde a_{6,6}$ for our purposes. The remarkable fact about $\tilde a_{6,j}$ is that the dashed LSD line from \fig{fig:k6scaled} maps to the point $({15 \over 4}, {15 \over 16})$ in the $({\tilde a_{6,2} \over \tilde a_{6.0}},{\tilde a_{6,4} \over \tilde a_{6.0}})$ plane, which can also be found by keeping the lowest-spin contribution in \eqn{eq:tildeadisprep}
\be
\label{eq:lowspin6}
\text{LSD}_{\alpha \to \infty}: ~~~\Big( {\tilde a_{6,2} \over \tilde a_{6,0}},{\tilde a_{6,4} \over \tilde a_{6,0}} \Big) = \Big({15 \over 4}, {15 \over 16} \Big) \simeq (3.75, 0.94) \,.
\ee
Therefore, the plane $({\tilde a_{6,2} \over \tilde a_{6,0}},{\tilde a_{6,4} \over \tilde a_{6,0}})$ happens to be precisely orthogonal to the line in \fig{fig:k6scaled} and it is  well-suited to study the fine structure of the distribution of points around the line.

\begin{figure}[tbh]
\centering
\includegraphics[scale=.7]{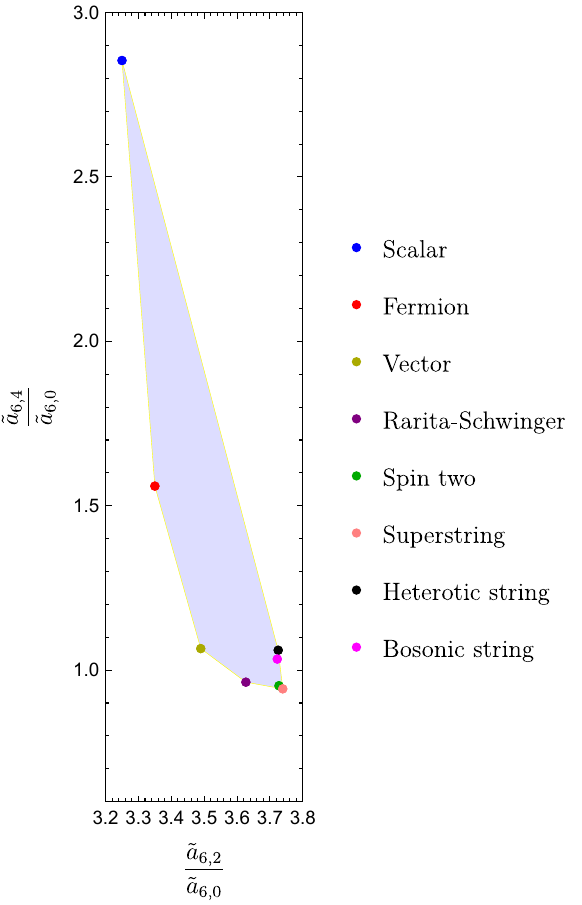}
\caption{\captionlineskip The transverse view of the theory island from \fig{fig:k6scaled}. We indicate various theories that form the vertices of the island. To reach a particular point inside the UV island we need to take a superposition of various amplitudes. 
}
\label{fig:z6scaled}
\end{figure}

Looking at the plot \ref{fig:k6scaled}, we see that both the scalar and the fermion lie pretty far from the naive LSD point. The reason we did not detect this on \fig{fig:k6scaled} is that for the scalar and fermion: $a_{6,0} \gg a_{6,1}, a_{6,2} , a_{6,3}$. In this way both points are very close to the LSD line in \fig{fig:k6scaled}, but for the trivial reason of being close to the origin. In \fig{fig:z6scaled} we resolve the origin by switching to variables \eqref{eq:tildekvariables} that do not depend on $a_{6,0}$. 

Working with $\tilde a_{k,q}$ emphasizes an important aspect of LSD. While assuming strong LSD leads to much tighter bounds on ratios ${a_{k,j} \over a_{k,0}}$ it does not lead to improved bounds for ${\tilde a_{k,q} \over \tilde a_{k,0}}$. The reason can be understood as follows: For any large, but finite $\alpha$, the admissible range for ${a_{k,j} \over a_{k,0}}$ includes $0$ and a small part of the negative axis. It is then easy to see from the definition \eqref{eq:tildekvariables} that arbitrarily small vicinity of the origin in the ${a_{k,j} \over a_{k,0}}$ includes all possible values for ${\tilde a_{k,q} \over \tilde a_{k,0}}$. This precludes using LSD to derive stronger bounds on ${\tilde a_{k,q} \over \tilde a_{k,0}}$. This simple fact highlights the point  that not all EFT coupling bases are equally illuminating for understanding the underlying structure. 

\begin{figure}[tbh]
  \hspace{-24pt}   
  \begin{tabular}{m{7cm}  m{4cm}}
\includegraphics[scale=.6]{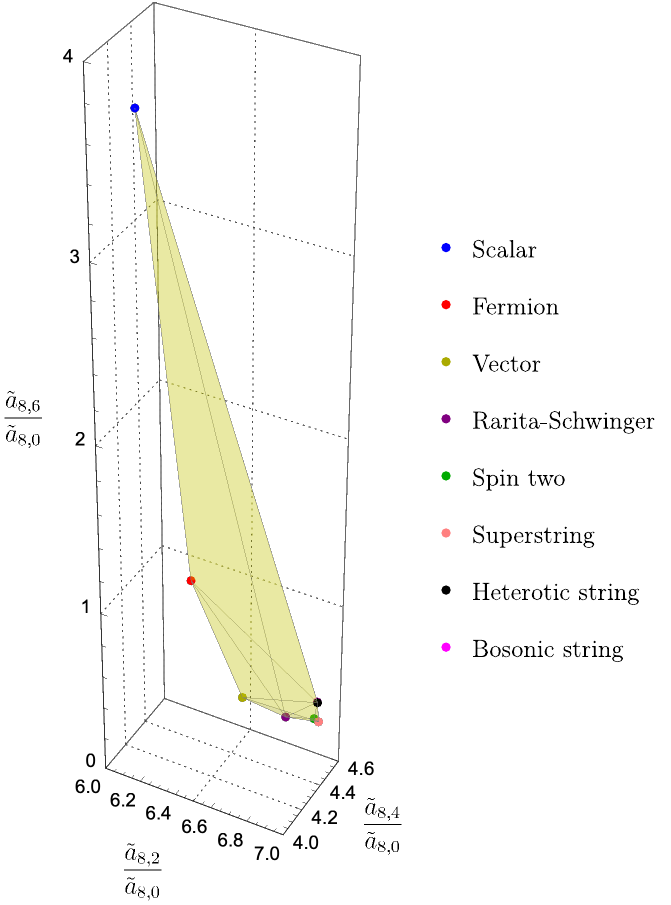} & 
\includegraphics[scale=.5]{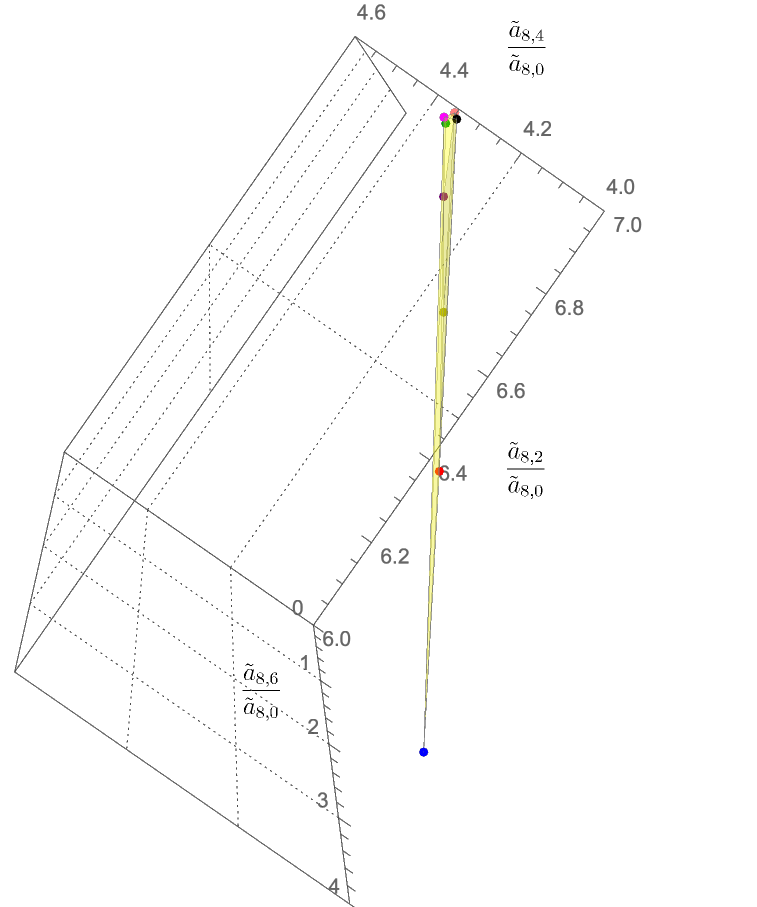}\\
\hskip 2 cm (a) & \hskip 1.4 cm (b)                               
 \end{tabular}
 \caption{\captionlineskip Various data points for $({\tilde a_{8,2} \over \tilde a_{8,0}} , {\tilde a_{8,4} \over \tilde a_{8,0}}, {\tilde a_{8,6} \over \tilde a_{8,0}})$. 
 Panels a) and b) present the same plot as seen from different vantage points. In particular, panel b) makes it clear that the points essentially lie on a plane.}
  \label{fig:k8tildetheories}
\end{figure}

A similar analysis can be performed for $k=8$. By switching to the $({\tilde a_{8,2} \over \tilde a_{8,0}} , {\tilde a_{8,4} \over \tilde a_{8,0}}, {\tilde a_{8,6} \over \tilde a_{8,0}})$ coordinates we note that the line from \fig{fig:resultsk8} maps to a single point. The same point can be obtained by keeping only $J=4$ contribution in (\ref{eq:tildeadisprep}) 
\be
\label{eq:lowspindomk8p}
\text{LSD}_{\alpha \to \infty}: ~~~ ({\tilde a_{8,2} \over \tilde a_{8,0}} , {\tilde a_{8,4} \over \tilde a_{8,0}}, {\tilde a_{8,6} \over \tilde a_{8,0}})  = (7, {35 \over 8}, {7 \over 16}) = (7 , 4.375 , 0.4375) \, .
\ee
Therefore by plotting the data in these coordinates we can resolve the points close to the origin in \fig{fig:resultsk8} which were located there due to the fact that $a_{8,0} \gg a_{8,i>0}$. One can check again that $({\tilde a_{8,2} \over \tilde a_{8,0}} , {\tilde a_{8,4} \over \tilde a_{8,0}}, {\tilde a_{8,6} \over \tilde a_{8,0}})$ is independent of $a_{8,0}$ and therefore this suppression does not take place anymore. The result is depicted in \fig{fig:k8tildetheories}.

\subsection{Spectral densities and low-spin dominance}
\label{sec:spectraldensities}

In the discussion above we considered various bounds that follow from the dispersive representation of the low-energy couplings and observed the phenomenon of LSD in the known physical theories. It manifests itself in the fact that the low-energy couplings 
occupy a very small region in the space of couplings allowed on general grounds. To see this phenomenon more clearly it is instructive to look at the spectral densities of the amplitudes directly.

Given a known expression for the amplitude it is not hard to compute various spectral densities. Indeed, it amounts to taking a discontinuity of the amplitude and integrating it against the proper Wigner d-function. These satisfy a familiar orthogonality relation
\be
\label{eq:orthogonality}
\int_{-1}^1 dx \, d_{\lambda, \lambda'}^J(x) d_{\lambda, \lambda'}^{\tilde J}(x) ={2 \over 2 J + 1} \delta_{J, \tilde J} \, .
\ee

For tree-level amplitudes the spectral density is a sum of delta-functions which correspond to the masses of exchanged particles. For the one-loop amplitudes the discontinuity has a continuous support above the two-particle threshold $4m^2$. In this case we perform the integration \eqref{eq:orthogonality} numerically. 
We focus on the first few spin spectral densities in both $\rho_{J}^{++}$ and $\rho_{J}^{+-}$ channels. To go from spectral densities to the coefficients in the low-energy expansion we need to compute the moments as in Eqs.~\eqref{eq:lowspink4}, \eqref{eq:lowspindominancek6}, and \eqref{eq:lowspindominancek8}.

\begin{figure}[tbh]
\centering
\includegraphics[scale=.65]{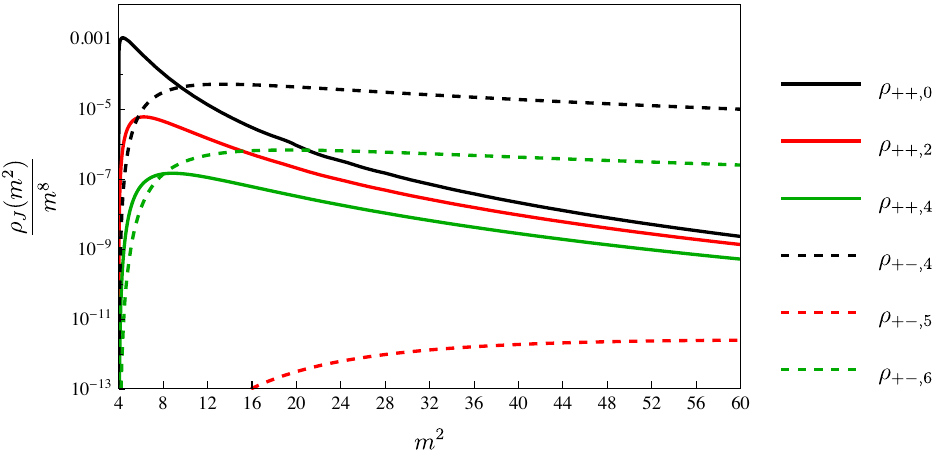}
\caption{\captionlineskip Spectral densities for the one-loop minimally-coupled scalar. We observe that $\rho_{0}^{++} \gg \rho_{J \geq 4}^{+-}, \rho_{J \geq 2}^{++}$ and that $\rho_{4}^{+-} \sim \rho_{2}^{++}$ close to the two-particle threshold. This is fully consistent with the features of the plots for various couplings in the previous section.
}
\label{fig:scalarspectrum}
\end{figure}

\begin{figure}[tbh]
\centering
\includegraphics[scale=.65]{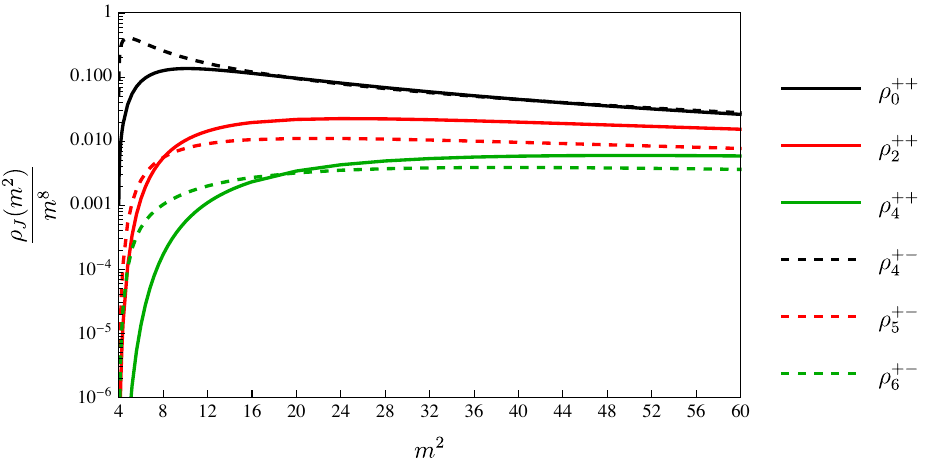}
\caption{\captionlineskip Spectral densities for the one-loop minimally coupled spin-2 particle. We observe that $\rho_{4}^{+-} \gg \rho_{J > 4}^{+-}, \rho_{J \geq 2}^{++}$ and that $\rho_{4}^{+-} \sim \rho_{0}^{++} $. Therefore in the space of couplings this amplitude is expected to lie on the low-spin dominance line.
}
\label{fig:spin2spectrum}
\end{figure}

\begin{figure}[tbh]
\centering
\includegraphics[scale=.65]{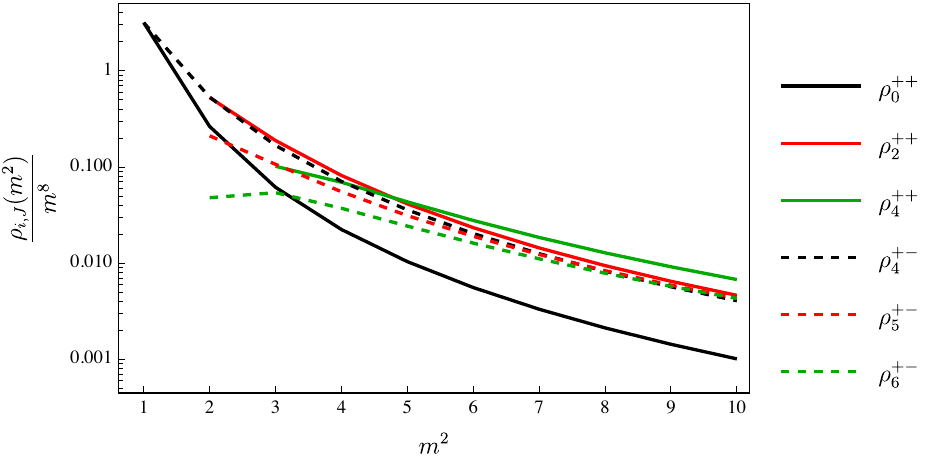}
\caption{\captionlineskip Spectral densities for the tree-level scattering of gravitons in the superstring theory. We observe that $\rho_{4}^{+-} \gg \rho_{J > 4}^{+-}, \rho_{J \geq 2}^{++}$ and that $\rho_{4}^{+-} \sim \rho_{0}^{++} $. Therefore in the space of couplings this amplitude is expected to lie on the low-spin dominance line.}
\label{fig:ssspectrum}
\end{figure}

\begin{figure}[tbh]
\centering
\includegraphics[scale=.65]{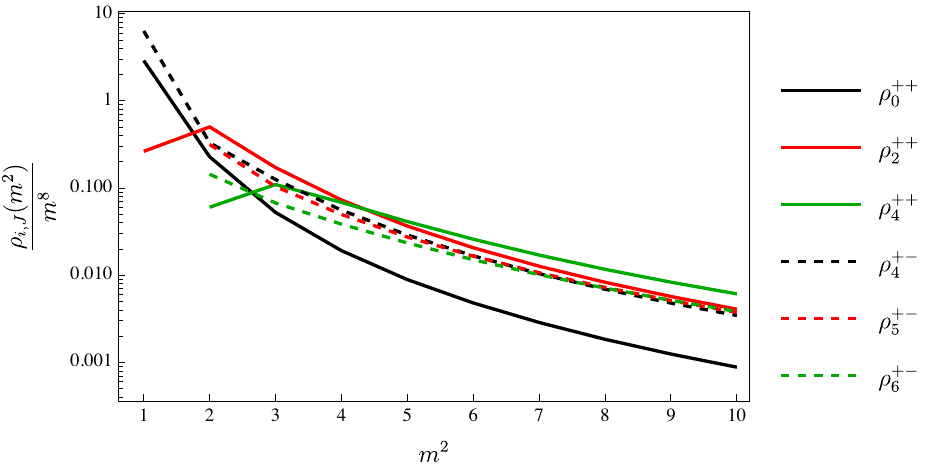}
\caption{\captionlineskip Spectral densities for the tree-level scattering of gravitons in the heterotic string theory. We observe that $\rho_{4}^{+-} \gg \rho_{J > 4}^{+-}, \rho_{J \geq 2}^{++}$ and that $\rho_{4}^{+-} \sim \rho_{0}^{++} $. Therefore in the space of couplings this amplitude is expected to lie on the LSD line.}
\label{fig:hsspectrum}
\end{figure}

\begin{figure}[tbh]
  \hspace{-24pt}   
  \begin{tabular}{m{7cm}  m{7cm}}
 \hskip 2.4 cm \text{Scalar loop} &  \hskip 2.4 cm \text{Spin 2 loop} \\
\includegraphics[scale=.5]{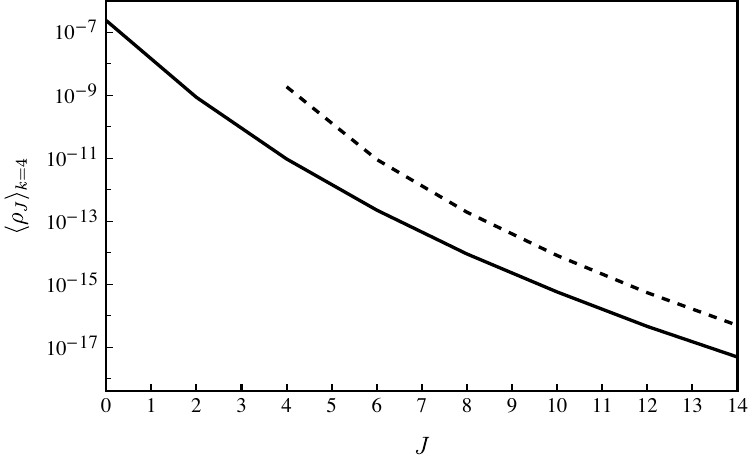} &
\includegraphics[scale=.5]{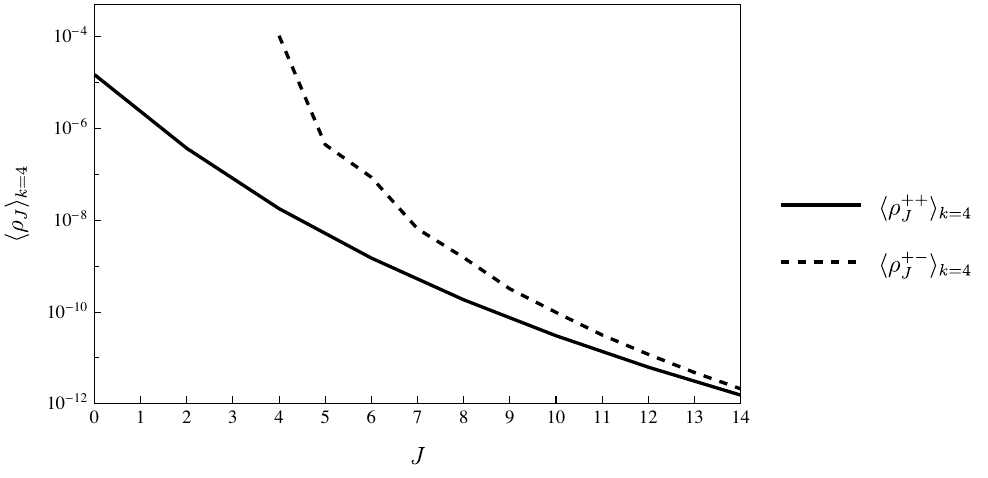}\\
\hskip 2.4 cm Superstring & \hskip 2.6 cm Heterotic     \\    
\includegraphics[scale=.5]{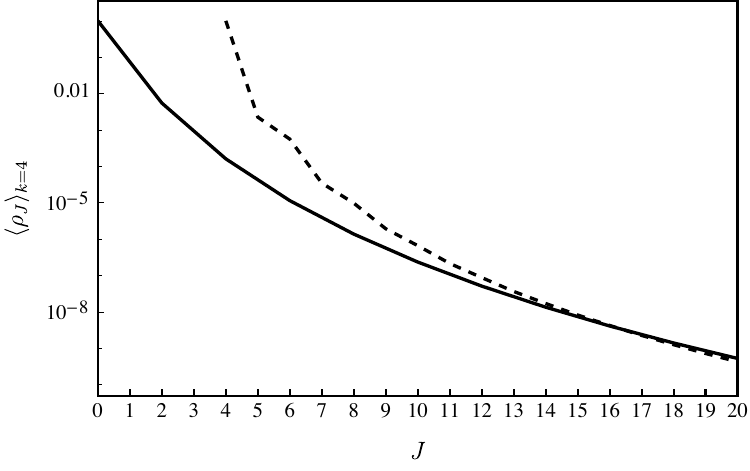} &
\includegraphics[scale=.5]{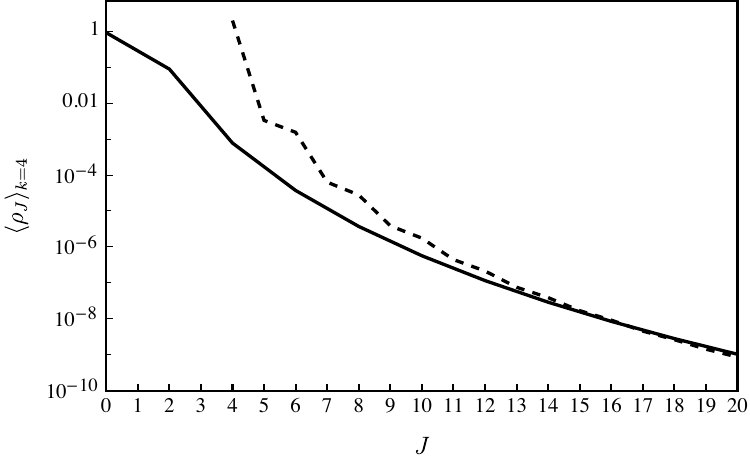} \\                      
 \end{tabular}
 \caption{\captionlineskip Moments of the spectral density $\langle \rho_{J} \rangle_4$ \eqref{eq:momentsrho} as a function of spin $J$ for various examples. For the scalar one loop result $\rho_J^{+-}$ 
 with $J$ being odd are negligible and are not presented in the figure. The line in the corresponding panel is obtained by connecting the even spin values of the spectral density moments.}
  \label{fig:lsdmoments}
\end{figure}

The results for a few selected cased are listed in \fig{fig:scalarspectrum}-\ref{fig:hsspectrum}. In all cases we see that the minimal-spin partial waves dominate in the corresponding channel. It is also instructive to plot the moments $\langle \rho_{J} \rangle_k$ which we present in Fig.~\ref{fig:lsdmoments}. The moments clearly satisfy LSD used in the previous section to derive stronger bounds.

In fact the moments of spectral density $\langle \rho_{J} \rangle_k$ in the examples we consider not only exhibit the dominance of the lowest-spin partial waves but also rapid decay at higher $J$. This latter feature is expected to be completely general. Indeed, the convergence of the sum rules \eqref{eq:dispcouplingrepresentation} requires that $\langle \rho_{J} \rangle_k$ decay faster than any polynomial at large $J$. This latter decay can be traced to locality of scattering in the impact-parameter space. 

Indeed, in the perturbative regime at large $J$ we have
\be
\label{eq:impactparameter}
\rho_J(s) \sim  {\rm Im} \, \delta (s, b) , ~~~~~ b = {2 J \over \sqrt{s}} \, ,
\ee
where $\delta(s,b)$ is the phase shift and $b$ is the impact parameter. For fixed $s$ and large impact parameters we
expect to have
\be
 {\rm Im} \delta (s, b) \sim e^{- m_{\rm gap} b} ,~~~~~ m_{\rm gap} b \gg 1 \,,
\ee
which controls decay of $\rho_J(s)$ at large $J$. In string theory the leading-order behavior is different and is controlled by the transverse spreading of strings $ {\rm Im} \, \delta (s, b)\sim e^{ - {b^2 \over 2 \alpha'  \log {s \alpha' \over 4}} } $---see Ref.~\cite{Brower:2006ea}.

A priori the large impact parameter discussion is not necessarily relevant for understanding the large-$J$ behavior of $\langle \rho_{J} \rangle_k$ which involves computing the moment \eqref{eq:momentsrho} over all energies (as opposed to keeping $s$ fixed as we take the large-$J$ limit). However, in analyzing $\langle \rho_{J} \rangle_k$ for the amplitudes considered in the present paper we experimentally observed that the integral over energies is peaked at energies
\be
\langle \rho_{J} \rangle_k: ~~~ {s_{*} \over m_{\rm gap}^2} \sim J .
\ee

Plugging this into the formula for the impact parameter \eqref{eq:impactparameter} we find that the dominant impact parameters are $m_{\rm gap} b_{*} \sim J^{1/2} \gg 1$ and therefore the large-$J$ behavior of the moments $\langle \rho_{J} \rangle$ is still controlled by large impact-parameter physics. We tested this picture against the data presented in Fig.~\ref{fig:lsdmoments} and found a qualitative agreement. It would be interesting to study the large-$J$ limit of  $\langle \rho_{J} \rangle_k$ more systematically. Of course, this discussion does not explain the fact that the hierarchical structure among partial waves continues all the way to the lowest spins in the examples we analyzed. It is this latter fact was crucial for the analysis in the previous section.

An important question is to understand how general is the picture that we observed in the tree-level string amplitudes and one-loop matter amplitudes. A priori these amplitudes look very different from each other, but at the level of the partial-wave analysis discussed in this section they exhibit remarkably similar behavior and strong version of LSD. This suggests that the hierarchical structures we observed in this paper could be a general property of consistent weakly-coupled gravitational $S$-matrices, but we do not have a proof yet.

\subsection{A hierarchy from unitarity}
\label{sec:hierarchyfromunitarity}

\begin{figure}[tb]
  \hspace{-24pt}   
  \begin{tabular}{m{7cm}  m{1cm}}
\includegraphics[scale=.7]{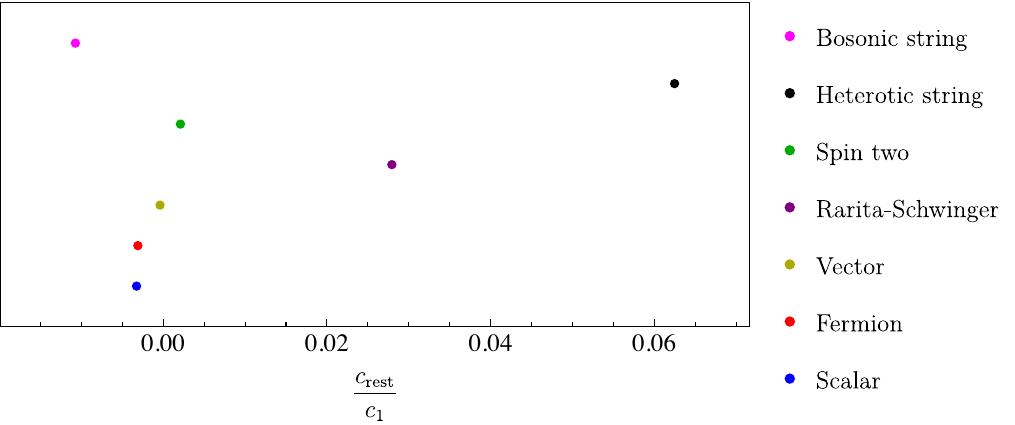} & $\null$ \hskip -2 cm 
\\[12pt]
\includegraphics[scale=.7]{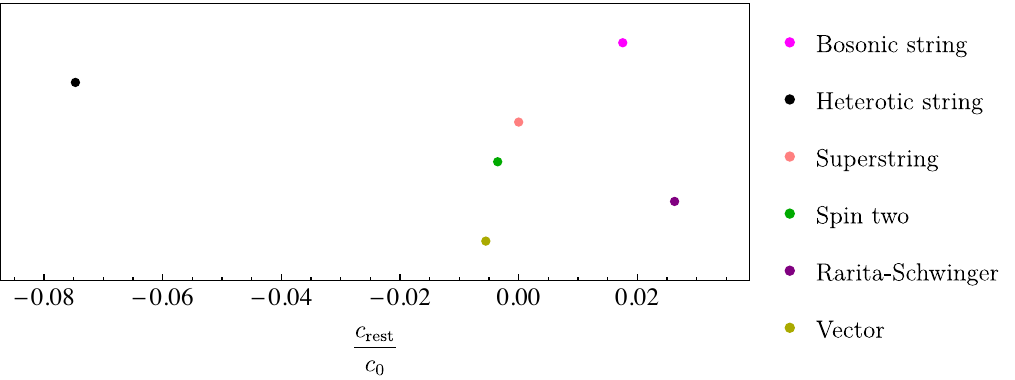} & $\null$ \hskip -2 cm                               
 \end{tabular}
\caption{\captionlineskip Examples of a hierarchy from unitarity. 
The horizontal axis shows illustrates that  the dominance of low-spin spectral densities naturally introduces a $10^{-2}$ hierarchy between various EFT coefficients in the absence of any symmetry. The vertical axis carries no meaning other than separating the points.  
}
\label{fig:hierarchy}
\end{figure}

The LSD discussion above illustrates an interesting phenomenon of emergent hierarchy between EFT coefficients in the absence of an underlying symmetry. 
Consider for example $D^8 R^4$ type corrections discussed in the $k=4$ section above. In writing down the relevant correction to the amplitude an EFT practitioner guided by the considerations of the supersymmetric decomposition \eqref{susyGR} can write the following ansatz
\begin{equation}
f_{D^8 R^4}(t,u) = c_{0}\, (s^2 + t^2 + u^2)^2 + c_1\, s^4 + c_{\rm {rest}}\, (t^2+u^2)^2 \,.
\end{equation}
The first term, which is proportional to $c_{0}$, is the completely
crossing-symmetric term of the type that appears in the new spin-2 part
of the amplitude given in \eqn{M4NewPieceExpand2}, corresponding to a massive
${\mathcal N}=8$ supersymmetric multiplet, obtained, for example,
by dimensional reduction from five dimensions.  In the ${\mathcal N}=8$
theory the full crossing symmetry of such terms is a consequence of
$\mathcal N = 8$ supersymmetric Ward identities~\cite{SWI}.  The
$c_{1}$ term corresponds to the additional terms needed in the
new spin-3/2 part of the amplitude \eqref{M4NewPieceExpand32} beyond the
fully crossing symmetric terms already appearing in the new spin-2 part
of the amplitude.  Finally, the third term containing $c_{\text{rest}}$
is the remaining independent term with $t$--$u$ crossing
symmetry. This term is distinguished from the $c_1$ term by its
differing behavior for $t \rightarrow \infty$ with $s$ fixed.

We then express the coefficient above in terms of more familiar
$a_{k,j}$ defined in \eqn{eq:fexpa} and used in the discussion of
various bounds. We get the following result
\begin{align}
c_{0} &= {1\over 4} (2 a_{4,1} - a_{4,2}) \,, \nn \\
c_{1} &= {1\over 4} (4 a_{4,0} -5 a_{4,1} +2 a_{4,2})\,, \nn \\
c_{\text{rest}} &= {1 \over 2} (a_{4,2} - {3 \over 2} a_{4,1}) \,.
\end{align}
Remarkably, we see that $c_{\text{rest}}$ vanishes along the low-spin dominance line that follows 
from \eqn{eq:lowspink4}. Plugging in
the values of $a_{k,j}$ for the theories discussed here we find a 
$10^{-2}$ hierarchy with $c_{\text{rest}} \ll c_{1}$ in the
absence of any symmetry, as shown in  \fig{fig:hierarchy}.  For spin $S=1,{3
  \over 2}, 2$ particle in the loop we also have $c_{\text{rest}} \ll
c_{0}$.

\section{Deriving bounds: multiple polarizations}
\label{Sec:BoundMultipleHelicity}

In this section we consider dispersion relations that combine
information from the different helicity configurations. More
precisely, we consider dispersive sum rules
\eqref{eq:superconvergence} which we apply to the matrices built out
of various scattering amplitudes \eqref{eq:matrixh},
\eqref{eq:matrixg}. We derive bounds on the inelastic amplitudes (single-minus and all-plus amplitudes)
in terms of the elastic double-minus amplitude.

\subsection{$s=0$}

The simplest bounds comes from setting $s=0$ in \eqn{eq:superconvergenceE}. The relevant equation takes the form
\be
\label{eq:momentssz}
 \oint_{t_0} {d t \over 2 \pi i} \mathcal{M}_h (s,t) {1 \over t} {1 \over (t (s+t))^{k} } \Big|_{s=0}  = \int_{m_{\rm gap}^2}^\infty {d t \over \pi} {2  \over t^{2k+1} }  {\rm Disc}_t \mathcal{M}_h (0,t)  \succeq 0\, .
\ee
By plugging the low-energy expansion of the amplitude in the LHS of the equation above and requiring that its eigenvalues are non-negative we get
\be
\label{eq:conditionfh}
2 f_{2k-4,0} \geq | h_{2k,0} | \,, ~~~~  k=2, 4, 6, \ldots  \, . 
\ee

The combination above for different $k$ form a set of moments as can be seen from the RHS of (\ref{eq:momentssz}). The difference compared to the recent analysis in Ref.~\cite{Bellazzini:2020cot} is that in our case we have a positive semi-definite spectral density matrix instead of a function. By contracting such matrix moments with an arbitrary polarization vector reduces the problem to the one considered in Ref.~\cite{Bellazzini:2020cot}. 

The situation simplifies in the parity preserving case when $h^*(s,u) = h(s,u)$. In this case the eigenvectors $\mathcal{M}_{h,\pm} (s,t)$ given in \eqn{eq:functionsparity} generate the low-energy expansion that satisfies the moment problem conditions considered in great detail in Ref.~\cite{Bellazzini:2020cot}. More precisely, we define a set of moments as $\mu_{{k -1}} \equiv 2 f_{2k-4,0} - h_{2k,0}$  and consider the Hankel matrix $H_{i j} = \mu_{i+j-1}$. 

Using the results in Appendix~\ref{HighOrdersAppendix} for the one-loop amplitude due to minimally-coupled scalar in the loop the first five moments take the following form
\be
H^{S=0} ={1 \over 140} \left( \frac{\kappa}{2} \right)^4 {1 \over (4 \pi)^2 } \left(
\begin{array}{ccc}
 \frac{1}{135 m^4} & \frac{1}{24024 m^8} & \frac{1}{1969110 m^{12}} \\
 \frac{1}{24024 m^8} & \frac{1}{1969110 m^{12}} & \frac{1}{109745064 m^{16}} \\
 \frac{1}{1969110 m^{12}} & \frac{1}{109745064 m^{16}} & \frac{1}{4833678850 m^{20}} \\
\end{array}
\right)  .
\ee
In agreement with the general prediction the one-loop moment matrix has nonnegative minors. 

\subsection{Away from $s=0$: first derivative}
\label{sec:R3bound}

We can use \eqref{eq:unitarityNew} to derive the bounds by taking the derivative of $B^+_k(s)$ with respect to $s$ before setting $s=0$. 
For example let us consider the first derivative with respect to $s$. In this way we get
\be
\partial_s \oint_{t_0} {d t \over 2 \pi i} & \mathcal{M}_{h,g}(s,t) {1 \over t} {1 \over (t (s+t))^{k} } \Bigr|_{s=0} \nn\\
&  = \int_{m_{\rm gap}^2}^\infty {d t \over \pi} {1  \over t^{2k+2} } \left( - (2k+1) {\rm Disc}_t \mathcal{M}_{h,g}(0,t) + 2 t \,\partial_s  {\rm Disc}_t \mathcal{M}_{h,g}(0,t) \right) .
\ee
The RHS is not positive semi-definite and therefore we cannot derive the bound on the LHS in a similar fashion. We can, however, consider the following linear combination 
\be
\label{eq:superconvergk2}
&{2 k+1 \over 2 m_{\rm gap}^2}  \oint_{t_0} {d t \over 2 \pi i} \mathcal{M}_{h,g}(s,t) {1 \over t} {1 \over (t (s+t))^{k} } + \partial_s \oint_{t_0} {d t \over 2 \pi i} \mathcal{M}_{h,g}(s,t) {1 \over t} {1 \over (t (s+t))^{k} } \Big|_{s=0} \nn \\
& \hskip .3 cm 
=\int_{m_{\rm gap}^2}^\infty {d t \over \pi} {1  \over t^{2k+2} } \left( \Bigl[{t \over m_{\rm gap}^2} - 1 \Bigr] (2k+1) {\rm Disc}_t \mathcal{M}_{h,g}(0,t) + 2 t \, \partial_s  {\rm Disc}_t \mathcal{M}_{h,g}(0,t) \right) \succeq 0 \, .
\ee

We now apply \eqn{eq:superconvergk2} to $\mathcal{M}_h$. By plugging the low-energy expansion (see \eqns{eq:lowenergyexp}{HelicityDef}) in the LHS of the formula above and imposing that the eigenvalues of the resulting matrix are nonnegative we can derive various bounds. Let us consider for example $k=2$. We get the following bound on the inelastic amplitude in terms of the elastic one
\be
\label{eq:boundthreepoint}
\boxed{
 \Big| h_{5,1} + {5 \over 2} {h_{4,0} \over m_{\rm gap}^2} \Big| \leq 5 \Bigl(f_{1,0} + {f_{0,0} \over m_{\rm gap}^2}\Bigr) - 2 | \beta_{R^3} |^2   \,. 
}
\ee
To apply the dispersive sum rule above to the bosonic string amplitude we first subtracted the contribution of the tachyon exchange to get $\delta f^{(\text{bs})}(s,u) =  f^{(\text{bs})}(s,u) + \left( \frac{\kappa}{2} \right)^2  {1 \over 1 + t}$ and $\delta\, h^{(\text{bs})}(s,u) = h^{(\text{bs})}(s,u)  + \left( \frac{\kappa}{2} \right)^2 \left({s^4 \over 1 + s} + {t^4 \over 1 + t} + {u^4 \over 1 + u} \right)$. The resulting functions satisfy all the properties needed to apply \eqref{eq:superconvergk2} for $k=2$. We then get, using the formulas from App. \ref{StringAppendix},
\be
\delta f^{(\text{bs})}_{0,0} &=\left( \frac{\kappa \null}{2} \right)^2 \Big(  3+2 \zeta(3) \Big), ~~~~~ \delta f^{(\text{bs})}_{1,0} = \left( \frac{\kappa \null}{2} \right)^2 (- 3) , \nn \\
\delta \, h^{(\text{bs})}_{4,0} &=\left( \frac{\kappa \null}{2} \right)^2  2, ~~~~~ \delta h^{(\text{bs})}_{5,1} =\left( \frac{\kappa \null}{2} \right)^2 (-3)  ,  ~~~ | \beta_{R^3}^{(\text{bs})} |^2  = \left( \frac{\kappa \null}{2} \right)^2  . 
\ee
which indeed satisfy \eqn{eq:boundthreepoint} where $m_{\rm gap}^2=1$ in the string case.

For the one-loop minimally coupled scalar $ | \beta_{R^3} |^2 \sim \kappa^6$ and appears only at two loops. For other coefficients that enter into \eqn{eq:boundthreepoint} we get, using Eq. \eqref{fExpansionShort} and Eq. \eqref{hExpansionShort},
\be
 f_{0,0}^{S=0} &=  \left( \frac{\kappa}{2} \right)^4 {1 \over (4 \pi)^2 } {1 \over 6300 m^4}, ~~~ f_{1,0}^{S=0} =  -\left( \frac{\kappa}{2} \right)^4 {1 \over (4 \pi)^2 } {1 \over 41580 m^6} \,, \nn \\
 h_{4,0}^{S=0} &= \left( \frac{\kappa}{2} \right)^4 {1 \over (4 \pi)^2 } {1 \over 3780 m^4}, ~~~ h_{5,1}^{S=0} = -  \left( \frac{\kappa}{2} \right)^4 {1 \over (4 \pi)^2 } {1 \over 7920 m^6} \,, 
\ee
so that, together with $m_{\rm gap}^2 = 4 m^2$, \eqref{eq:boundthreepoint} is again satisfied.

More generally, the formula above bounds the correction to the
three-point function of the graviton $ \beta_{R^3}$ from above in
terms of the EFT data. To make it more manifest we can rewrite the
above
\be
\label{eq:boundR3a}
2 | \beta_{R^3} |^2 \leq 5 \Bigl(f_{1,0} + {f_{0,0} \over m_{\rm gap}^2} \Bigr) 
 - \Big| h_{5,1} + {5 \over 2} {h_{4,0} \over m_{\rm gap}^2} \Big| \leq  {10 f_{0,0} \over m_{\rm gap}^2} \,,
\ee
where in the last inequality we used the fact that $|f_{1,0}| \leq {1
  \over m_{\rm gap}^2} f_{0,0}$ which readily follows from
\eqns{eq:firstfewcoefficient}{eq:dispcouplingrepresentation}.  We can
restate it more succinctly in terms of the Wilson coefficients of the
gravitational EFT
\be
\label{eq:boundR3A}
| \beta_{R^3} |^2 \leq 5 {\beta_{R^4}^+ \over m_{\rm gap}^2}  \,, 
\ee
where $m_{\rm gap}^2$ denotes the mass gap at which the massive degrees of freedom that induce the higher-derivative corrections appear. Recall that $f_{0,0} = \beta_{R^4}^+$ was defined in \eqn{eq:R4definition}. Bounds similar to \eqn{eq:boundR3a} can be derived by considering the superconvergence sum rules for $k>2$. We do not list them here.

In fact it is not difficult to strengthen the bound \eqref{eq:boundR3A} by considering the following unsubtracted dispersive sum rule (analogous to $k=0$ in \eqn{eq:superconvergence})
\be
\label{eq:relationB0}
\oint_{\infty} {d t \over 2 \pi i} {1 \over t} f (s,-s-t)   = 0 \,. 
\ee
The universal tree-level gravitational piece $\left( {\kappa \over 2} \right)^2 {1 \over s t u}$, which is singular at $s=0$, does not contribute to \eqn{eq:relationB0} therefore we can consider
\be
&\left( {1 \over m_{\rm gap}^2} + \partial_s \right) \eqref{eq:relationB0} \Big|_{s=0}: ~~~~ {\beta_{R^4}^+ \over m_{\rm gap}^2} - | \beta_{R^3} |^2 =  \nn \\
&  \hskip 2.5 cm 
\int_{m_{\rm gap}^2}^\infty {d t \over \pi} {1  \over t} \left( \Bigl[\partial_s + {1 \over m_{\rm gap}^2} \Bigr]{\rm Disc}_t f  + \Bigl[\partial_s + {t - m_{\rm gap}^2 \over m_{\rm gap}^2  t}\Bigr]{\rm Disc}_u f  \right) \Big|_{s=0} \geq 0 \,,
\ee
where nonnegativity of the RHS can be readily checked for each partial wave separately by plugging the discontinuities of $f(s,u)$, given in \eqn{eq:dispersionrelationsf}, into the formula above.\footnote{\footnotelineskip More precisely, it follows from $\partial_s^n \left( \frac{d_{4,4}^J(1+{2s \over m^2})}{(m^2+s)^4}  \right)\Big|_{s=0}  > 0$ for $n=0,1$ which can be readily checked using formulas from Appendix~\ref{app:wignerd}.} In this way we immediately get a bound
\be
\label{eq:boundR3}
\boxed{ | \beta_{R^3} |^2 \leq {\beta_{R^4}^+ \over m_{\rm gap}^2} \,.} 
\ee

The bound \eqref{eq:boundR3} is a step towards making the analysis of
Ref.~\cite{Camanho:2014apa} quantitatively precise. At least for the
$R^3$ correction to the graviton three-point coupling which is
considered here, it translates the problem to bounding the coefficient
$\beta_{R^4}^+$ in terms of $G / m_{\rm gap}^2$.\footnote{\footnotelineskip Recall
from \eqn{kappaDef}  that $\left( {\kappa /2} \right)^2 = 8 \pi G$.}  This problem was
beautifully solved recently in Ref.~\cite{Caron-Huot:2021rmr} for
$D=10$ maximal supergravity in a perturbative setting similar to
ours, and it was addressed nonperturbatively in
Ref.~\cite{Guerrieri:2021ivu}.  The method used in
Ref.~\cite{Caron-Huot:2021rmr} is not directly applicable in four
dimensions due to the IR divergences, but at least in $D\geq 5$ where
the $2 \to 2$ amplitude is nonperturbatively well-defined it is
natural to expect that one will be able to get a bound on
$\beta_{R^4}^+$ in terms of $ {G / m_{\rm gap}^2}$.  It would be
very interesting to demonstrate it explicitly.

Indeed, assuming the nonperturbative Regge bound \eqref{eq:QRG}, we can
consider the $(-2)$ subtracted dispersion relations for $f(s,u)$ which
expresses $- {8 \pi G /s}$ in terms of the contribution of heavy
states, cf. equation (3.37) in
Ref. \cite{Caron-Huot:2021rmr}. Existence of such a dispersion
relation in the absence of supersymmetry is crucially due to the fact
that we consider gravitons as external states.

\subsection{Away from $s=0$: second derivative}

Using exactly the same technique as above one can check that
\be
&{2 k^2 -1 \over 4} \oint_{t_0} {d t \over 2 \pi i} \mathcal{M}_{h,g}(s,t) {1 \over t} {1 \over (t (s+t))^{k} } +{2 k+1 \over 2 t_0} \partial_s \oint_{t_0} {d t \over 2 \pi i} \mathcal{M}_{h,g}(s,t) {1 \over t} {1 \over (t (s+t))^{k} } \nn \\
& \hskip 3 cm 
+ \partial_s^2 \oint_{t_0} {d t \over 2 \pi i} \mathcal{M}_{h,g}(s,t) {1 \over t} {1 \over (t (s+t))^{k} } \Big|_{s=0} \succeq 0 \,.
\ee

Consider $k=2$ and plug the low-energy expansion of $\mathcal{M}_{g}(s,t)$ in the formula above. We get a matrix whose eigenvalues should be non-negative. In this way we can bound the constant term in the $++-+$ amplitude
\be
\label{eq:boungg00}
\boxed{ | g_{0,0} | \leq {7 \over 4} {f_{0,0} \over (m_{\rm gap}^2)^2} + {5 \over 4} {5 f_{1,0} - 4 |\beta_{R^3} |^2 \over m_{\rm gap}^2} + {3 \over 2} \left( f_{2,0} + 3 f_{2,1} \right) \,. }
\ee

As in the previous section we can check \eqref{eq:boungg00} in the bosonic string theory. The extra data compared to the previous section takes the form
\be
\label{eq:newdatabs}
g_{0,0}^{(\text{bs})} &=\left( \frac{\kappa \null}{2} \right)^2 \Big(  2 \zeta_3 \Big), ~~~ \delta f_{2,0}^{(\text{bs})} =\left( \frac{\kappa \null}{2} \right)^2 \Big(  3 +2 \zeta(5) \Big), \nn \\
\delta f_{2,1}^{(\text{bs})} &=\left( \frac{\kappa \null}{2} \right)^2 \Big(2 +4 \zeta(3) + 2 \zeta(5) \Big) . 
\ee
Plugging \eqref{eq:newdatabs} into \eqn{eq:boungg00} we see that it is indeed satisfied.

For the one-loop minimally coupled scalar we get correspondingly
\be
g_{0,0}^{S=0} &= \left( \frac{\kappa}{2} \right)^4 {1 \over (4 \pi)^2 } {1 \over 6306300 m^8}\,, ~~~~~  f_{2,0}^{S=0}  =  \left( \frac{\kappa}{2} \right)^4 {1 \over (4 \pi)^2 } {3 \over 560560 m^8} , \nn \\
f_{2,1}^{S=0}  &= \left( \frac{\kappa}{2} \right)^4 {1 \over (4 \pi)^2 } {31 \over 9081072 m^8} \,. 
\ee
Again, we checked that \eqn{eq:boungg00} is satisfied, where we set $m_{\rm gap}^2 = 4 m^2$.

Using the bounds from the previous section and \eqref{eq:boungg00} we can bound $g_{0,0}$ in terms of $\beta^+_{R^4}$ defined in \eqn{eft} as follows
\be
\label{eq:boundsimpleg}
| g_{0,0} | \leq {815 \over 44} {\beta^+_{R^4} \over m_{\rm gap}^4} \, .
\ee 
In deriving \eqref{eq:boundsimpleg} we first expressed $f_{i,j}$,
defined in \eqn{eq:lowenergyexp}, in terms of $a_{k,j}$, defined in
\eqn{eq:fexpa}, and then used \eqns{eq:momentsak0}{eq:nonsymlev2bound}.
Assuming LSD we get a stronger bound 
\be
\label{eq:boundsimplegLSD}
\text{LSD}_{\alpha \to \infty}: ~~~~~ | g_{0,0} | \leq {25 \over 4} {\beta^+_{R^4} \over m_{\rm gap}^4} \, ,
\ee
where in deriving this we used the stronger LSD bound \eqref{eq:LSDinftyk2}. 

\section{Conclusions}
\label{ConclusionSection}

Our paper naturally consisted of two parts with the first part
providing theoretical data that we interpreted in the second part in
terms of bounds on coefficients of gravitational EFTs.  In the first
part, using amplitudes methods, we obtained the one-loop
four-graviton amplitude with a minimally-coupled massive particle of spin up to $S=2$ circulating in the loop.  By series expanding these
amplitude in large mass, we obtain theoretical data for Wilson
coefficients which we analyze in the second part.  Combining this data
with similar theoretical data obtained from tree-level string-theory amplitudes
we found that the Wilson coefficients fall on small islands compared to
the general bounds coming from consistency of $2\to 2$ scattering determined along the lines of
Refs.~\cite{Tolley:2020gtv,Caron-Huot:2020cmc,Arkani-Hamed:2020blm}.
It is quite striking that the EFT coefficients derived from 
both string-theory and one-loop-massive amplitudes land on the
same small islands.  Remarkably this can be explained as a
consequence of low-spin dominance in the partial-wave expansions.

\subsection{Obtaining one-loop amplitudes}  

In order to compute the one-loop amplitudes used to generate EFT data
we applied standard amplitudes methods, including spinor
helicity~\cite{SpinorHelicity}, generalized unitarity~\cite{Unitarity,
  BernMorgan}, the double copy~\cite{Kawai:1985xq, BCJ, BCJReview}
and integration by parts~\cite{IBP, Smirnov:2019qkx}.  Using
generalized unitarity we obtained all integral coefficients except for
a few whose integrals have no unitarity cuts in any channel.  We fixed
the coefficients of the latter integrals by using the known
ultraviolet properties of the amplitude.  To fully utilize this
information we made use of overcomplete integral basis that contains
higher-dimensional integrals, but whose coefficients do not depend on
the spacetime dimension or equivalently the dimensional-regularization parameter
$\epsilon$.  In addition, we also demonstrated that these coefficients
do not depend on the mass $m$ of the particles circulating in the
loops. The existence of such a basis imposes constraints
that makes it reasonably straightforward to determine any pieces not captured by
the $s$, $t$ and $u$ channel unitarity cuts.  In
this basis we used the fact that the coefficient of the potential
$1/\epsilon$ logarithmic ultraviolet divergence is
zero~\cite{tHooft:1972tcz} to completely fix the remaining integral
coefficients. A basis without $\epsilon$-dependent coefficients always
exists for one-loop problems~\cite{Bern:1995ix}, however the lack of
$m$-dependence appears to be special to our case of a closed massive loop
with external massless particles.

It would of course be interesting whether there is some way to
generalize our approach to more complicated situations with external
legs of differing masses.  To generalize our approach to the generic
case of a massive one-loop amplitude one could use information from
the higher-than-logarithmic divergences which are accessible in
dimensional regularization by shifting the dimension downwards as
discussed in \sect{MHVConstructionSection}. We showed that knowledge
of all these divergences is sufficient to fully constrain the
remaining ambiguities, something we expect to be true more
generally. We expect constraints from ultraviolet
divergences and from requiring proper decoupling in the large mass
limit to be sufficient to remove any ambiguities in terms that are not
fixed by the unitarity cuts, up to the usual ambiguities tied to scheme
choices and renormalization. This may provide an alternative method
for obtaining complete one-loop amplitudes using on-shell
techniques~\cite{BrittoMirabella}.

We exposed a useful supersymmetric decomposition for graviton
amplitudes with a massive particle in the loop. A similar
decomposition exists in gauge theory~\cite{Bern:1993tz}. This
decomposition expressed the amplitude with a particle of spin $S$ in
the loop in terms of amplitudes with lower-spin particles
and simpler to calculate pieces. These pieces correspond to amplitudes
with a massive BPS multiplet circulating in the loop.

Having constructed the one-loop four-graviton amplitudes with massive
particles in the loop, it was then straightforward to expand in large
mass, generating amplitudes matching those from a gravitational EFT.
Because the original amplitudes are sensible, satisfying appropriate
Regge behavior and unitarity, this EFT is also sensible and can be
used to test the constraints derived from dispersion relations.

Besides using these results in our study of bounds on EFT coefficients,
we also noted a simple relations between ultraviolet divergences
in higher dimensions and the coefficients in the $1/m$ expansion.
This relation requires a particular analytic continuation of the
amplitude to higher dimensions where we keep the number of physical
states at their four-dimensional values.  It would be interesting
to see if similar relations hold more generally, including higher loops,
and whether this connection can be exploited in studies of bounds 
on EFT coefficients.

\subsection{EFT bounds}

In order to derive bounds on the Wilson coefficients, we first reviewed
in \sect{Sec:GravAmplitudes} the basic properties of unitarity,
crossing, and bounds on the Regge limit in the context of perturbative
$2 \to 2$ scattering of gravitons in four dimensions. These allowed us
to study dispersion relations and derive bounds on the Wilson
coefficients.

In \sect{Sec:Bounds} we focused on the double-minus amplitude and
derived bounds on the Wilson coefficients along the lines of
Ref.~\cite{Tolley:2020gtv,Caron-Huot:2020cmc,Arkani-Hamed:2020blm}. We expressed the
low-energy expansion coefficients as dispersive integrals in
\eqn{eq:dispcouplingrepresentation}. We then identified the two-sided
bounds on the coefficients following the observation in
Ref.~\cite{Arkani-Hamed:2020blm} that the boundaries of the allowed
region inherit the cyclicity property from the one channel dispersion
relation. In this way, by explicitly extracting the mixed constraints
from the mixed $s$--$u$ channel partial waves we identified the new
boundaries of the allowed region. We do not rigorously prove that our
identification of the bounds is optimal, and we leave filling this
gap for future work. We then introduced the idea of LSD which
expresses a relationship between moments of the spectral functions of
various spins, and we used crossing symmetry or null constraints
\cite{Caron-Huot:2020cmc} to derive rigorous bounds in terms of a few
low-spin partial waves. We found that all the amplitudes considered in
this paper lie on small islands whose location and shape can be
determined using the {\it low-spin dominance} principle.

In Sect.~\ref{Sec:BoundMultipleHelicity} we studied dispersion
relations for the graviton amplitudes of various helicities. The key
observation is that we can apply dispersion relations to matrices  \eqref{Mmatrix}, composed of the different helicity amplitudes, whose
discontinuity is positive semi-definite. Applying dispersive sum
rules along the lines of Ref.~\cite{Caron-Huot:2020cmc} to these matrices
we derived constraints on the Wilson coefficients which appear in the
single-minus and all-plus gravitational scattering amplitudes. Notably
we placed a bound on the $R^3$ coefficient that corrects the graviton
three-point amplitude in terms of the $R^4$ coefficient, making a step
towards making the analysis of Ref.~\cite{Camanho:2014apa} quantitatively
precise.

The idea that the set of possibilities to UV complete gravity is much
sparser than one naively would have thought lies at the heart of the
swampland program~\cite{swampland}. Usually this sparseness is
associated with non-perturbative aspects of the UV completion. In
the present paper we analyzed the problem in a perturbative
setting. By minimally coupling low-spin matter to gravity we generated
$S$-matrices which we expect should satisfy the axioms of
causality and unitarity up to an arbitrary order in $G$ and in an
arbitrary $n$-point amplitude. We can ask therefore a weaker version
of the swampland question: what is the set of perturbatively
consistent weakly-coupled gravitational $S$-matrices? It is in the
framework of this question that the analysis of the present work can
be placed. Unexpectedly, we found that in known examples of
perturbatively consistent gravitational $S$-matrices the low-energy
couplings lie in regions much smaller than predicted on general
grounds of causality and positivity of $2 \to 2$ graviton scattering. Looking at the amplitudes for
tree-level string theory and minimally coupled one-loop matter it is
not a priori obvious at all that they should be ``close'' to each
other. Dispersive representation of the low-energy expansion of both
makes this similarity apparent. Indeed, as discussed in Sect.~\ref{Sec:Bounds} both amplitudes satisfy strong versions of LSD that
localizes their low-energy data to regions that are much smaller than
indicated by the general analysis of unitarity and crossing.

An obvious question is: how general is this universality? We motivated it by
pointing out that the examples considered in the present paper define consistent $S$-matrices
for any $n \to m$ scattering, whereas the bounds were derived by considering
$2 \to 2$ scattering only.
To answer this question we would need to better understand the landscape of
consistent gravitational amplitudes, including any unitary
perturbative or nonperturbative QFT coupled to gravity, as well as
amplitudes in theories with extra dimensions. In particular, 
it would be very interesting to understand if and how consistency of  $n \to m$
scattering can be used to rigorously establish stronger bounds on $2 \to 2$ scattering, potentially bringing us closer to the small theory
islands observed in the present work. It would also be very interesting to understand the 
example of large-$N$ QCD coupled to gravity~\cite{Kaplan:2020tdz}.

In this paper, we only considered the leading-order effect from
integrating out massive degrees of freedom. We did not address the
question of bounds that can be applied to IR-safe observables in four
dimensions, nor have we included the loops of massless particles which
generate more general logarithmic corrections in the low-energy
expansion of the amplitude. We leave these important questions to
future work. The logarithmic running was discussed in
Refs.~\cite{Bellazzini:2020cot,Trott:2020ebl,Arkani-Hamed:2020blm},
with the basic idea being that instead of expanding the amplitude
around $s,t=0$ one considers dispersive representations of the EFT
couplings defined at some scale. To deal with the IR divergences we
may consider dressed states (see e.g.
Ref.~\cite{Strominger:2017zoo,Arkani-Hamed:2020gyp} for a recent
discussion), for which the full implications of unitarity and crossing
are still to be fully understood. Another interesting problem is to
repeat the analysis of the present paper in the context of $\rm
AdS_4/CFT_3$ where the problem of IR divergences does not arise, but
dispersive techniques discussed in the present paper still
hold~\cite{CFTdispersion}.

In summary, motivated by low-energy theoretical data obtained from
one-loop field-theory and tree-level string-theory amplitudes, we put
forward the idea that EFTs that describe sensible weak gravitational theories
live on small islands that can be understood in terms of partial-wave low-spin
dominance.  It will be important to understand the extent to which this can
be extended to constrain gravitational theories.

\vskip .3 cm 
\noindent
{\bf Acknowledgements:}

We are grateful to Nima Arkani-Hamed, Thomas Dumitrescu, Kelian
H\"aring, Enrico Hermann, Tzu-Chen Huang, Yu-tin Huang, Julio
Para-Martinez, Marco Meineri, Irina Mocioiu, Baur Mukhametzhanov, Radu
Roiban, Eric Sawyer, Chia-Hsien Shen, and Piotr Tourkine for helpful
conversations.  This work was supported by the U.S. Department of
Energy (DOE) under award number DE-SC0009937.  This project has
received funding from the European Research Council (ERC) under the
European Union's Horizon 2020 research and innovation programme (grant
agreement number 949077).  D.~K. also thanks the Mani L. Bhaumik
Institute for Theoretical Physics for support.

\newpage
\appendix

\section{Minimal Coupling}
\label{app:MinimalCoupling}

\begin{figure}[t]
     \centering
     \includegraphics[scale=.65]{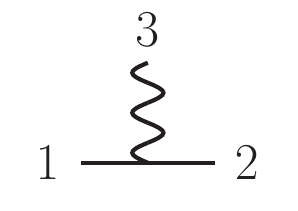}
     \caption{\captionlineskip The three-point amplitude necessary to study minimal coupling. The straight lines represent massive spinning particles, while the wiggly line denotes either a photon or a graviton. }
      \label{fig:appendix3pt}
\end{figure}

In this appendix we present in more detail our definition of minimal coupling of gravity to a massive spinning particle.
For massive spinning particles, the requirement of general-coordinate invariance and mass-dimension-4 operators leaves an ambiguity for integer spin-particles in defining the minimal coupling to gravity.  In this appendix we discuss this ambiguity and the choice we make in this paper. Our choice corresponds to the one that does not violate the unitarity and causality constraints of Refs.~\cite{treeLevelUnitarity,Bonifacio:2017nnt,Afkhami-Jeddi:2018own,Meltzer:2017rtf,Afkhami-Jeddi:2018apj}. 
We start by reviewing the case of electromagnetism (EM) where a similar situation exists, before we move on to consider the gravitational case.

This issue appears only for integer-spin particles. The source of the
ambiguity is the existence of a gauge-invariant operator that has the same mass
dimension as the kinetic term. For half-integer-spin particles, the
corresponding operator is of higher-mass dimension and hence would not count 
as minimal coupling. Therefore, the cases
that require attention are the coupling of a spin-1 or spin-2 particle
to gravity. 
Since the analysis of the coupling of a spin-1 particle to EM is
similar, we start by reviewing this case.

Here we fix the ambiguity by requiring that the interactions smoothly
match onto the massless limit. To explain our choice, it suffices to
evaluate the three-point amplitude between two massive spinning particles
and a photon or a graviton.\footnote{\footnotelineskip In calculating this amplitude one implicitly
  uses complex kinematics.} Specifically, we demand that by taking the
massless limit of these amplitudes and identifying the spin-1
particles as gauge bosons and the spin-2 particles as gravitons, we
recover the corresponding amplitudes in Yang-Mills and pure
gravity. The former is indeed realized for the coupling of the W boson
to the photon~\cite{PeskinSchroeder}.
The double-copy construction used in this paper is smooth in the massless limit, and hence selects the above prescription. 

For this discussion, we follow the formalism of
Refs.~\cite{Singh:1974qz}, which allows us to discuss particles of
arbitrary spin.\footnote{\footnotelineskip Descriptions of higher-spin
  particles date back to Pauli and Fierz~\cite{Pauli:1939xp}.}  After
obtaining the relevant three-point amplitudes, we specialize to the
cases of interest.  The part of the Lagrangian we give is just the one
necessary to obtain the three-point amplitudes relevant for this
discussion. We emphasize that these Lagrangians would need to be
modified with auxiliary-field terms in order to reproduce our one-loop
calculation.

For the case of a massive spinning particle $\phi$ coupled to EM we have
\begin{equation}
\mathcal{L} = -\frac{1}{4}F_{\mu \nu}F^{\mu \nu} + D_\mu^\dagger \bar \phi  D^\mu \phi - m^2 \bar \phi \phi  
+ e(g-1)F_{\mu \nu} \, \bar \phi M^{\mu \nu} \phi + \ldots \,,
\end{equation}
where $e$ is the charge of the particle, $g$ is the gyromagnetic ratio
and the ellipsis denote possible auxiliary-field terms.  For the
Lorentz generator $M$ in the representation of $\phi$ we may use
\begin{equation}
(M^{\mu\nu}){}_{c(s)}{}^{d(s)} = {2 {} i {} s}\delta^{[\mu}_{(c_1}\eta^{\nu](d_1}\delta^{d_2}_{c_2}\dots \delta^{d_s)}_{c_s)}\,,
\label{ExplicitLorentzGenerators}
\end{equation}
where the indices $c(s)$ and $d(s)$ stand for the symmetrized sets of
vector indices $\{c_1,\dots, c_s\}$ and $\{d_1,\dots, d_s\}$ and
symmetrizations include division by the number of terms.

We may now obtain the three-point amplitude between two massive spinning particles and a photon, depicted in Fig.~\ref{fig:appendix3pt}. Neglecting the overall normalization we find 
\begin{equation}
\mathcal{A}^{\text{EM}-s}_3 \sim \varepsilon_1 \cdot \varepsilon_2 \, \varepsilon_3 \cdot p_1 + i (g-1) \varepsilon_1 \cdot M[\varepsilon_3,p_3] \cdot \varepsilon_2 \,,
\end{equation}
where labels 1 and 2 denote the massive spinning particles, while particle 3 is a photon. We use the notation 
\begin{equation}
A \cdot M[p,q] \cdot B \equiv p_\mu q_\nu (M^{\mu\nu}){}_{c(s)}{}^{d(s)} A^{c(s)} B_{d(s)}\,. 
\end{equation} 
Plugging in the spin-1 representation for $M$ we find
\begin{equation}
\mathcal{A}^{\text{EM}-1}_3 \sim \varepsilon_1 \cdot \varepsilon_2 \, \varepsilon_3 \cdot p_1 + (g-1)\Big( \varepsilon_3 \cdot \varepsilon_1 \, \varepsilon_2 \cdot p_3 - \varepsilon_2 \cdot \varepsilon_3 \, \varepsilon_1 \cdot p_3 \Big) \,.
\end{equation}
Identifying all three particles as gauge bosons in the massless limit, i.e. demanding $\mathcal{A}^{\text{EM}-1}_3 = \mathcal{A}^\text{YM}_3$,  fixes $g=2$.  

The case of a massive spinning particle coupled to gravity is similar.  In this case, we have 
\begin{align}
{\cal L}&=-\frac{1}{16 \pi G} R + \frac{1}{2} g^{\mu\nu} \nabla_\mu\phi\nabla_\nu\phi 
- \frac{1}{2}m^2\phi\phi
+\frac{\rm H}{8}R_{\kappa \lambda \mu \nu} \, \phi M^{\kappa \lambda} M^{\mu \nu} \phi +\ldots \,,
\end{align}
where $\rm H$ is an arbitrary dimensionless coefficient and we again denote possible  auxiliary-field terms by the ellipsis.  Plugging in the spin-1 representation we see that the Riemann tensor in the term proportional to $\rm H$ contracts into the Ricci tensor or scalar which under the equations of motion is equivalent to a
$\phi^2$ term. Therefore, the $\rm H$ term may be replaced by using an appropriate field redefinition by a $\phi^4$-type term, which is not relevant for our discussion. Looking at the spin-2 representation and taking the massless limit, we recover the background-field-gauge gravitational Lagrangian of Ref.~\cite{Bern:1993wt} for $\rm H = 1$ upon identifying the spin-2 particle with the graviton.

We may alternatively reach the same conclusions by looking at the three-point amplitude between two massive spinning particles and a graviton. This amplitude, shown in Fig.~\ref{fig:appendix3pt}, is given by
\begin{equation}
\mathcal{M}^{\text{GR}-s}_3 \sim -\varepsilon_1 \cdot \varepsilon_2 (\varepsilon_3 \cdot p_1)^2 + i (\varepsilon_3 \cdot p_1) \varepsilon_1 \cdot M[\varepsilon_3,p_3] \cdot \varepsilon_2 + \frac{\rm H}{2} \varepsilon_1 \cdot M[\varepsilon_3,p_3] \cdot M[\varepsilon_3,p_3] \cdot \varepsilon_2 \,.
\end{equation}
For the spin-1 representation of $M$ we observe that the last term vanishes. For the spin-2 representation we have
\begin{align}
\mathcal{M}^{\text{GR}-2}_3 \sim -\varepsilon_1 \cdot \varepsilon_2 & (\varepsilon_3 \cdot p_1)^2 -2 \, \varepsilon_1^{\mu \kappa} \varepsilon_{2,\kappa}^{\nu} (\varepsilon_{3,\mu} p_{3,\nu} - \varepsilon_{3,\nu} p_{3,\mu})(\varepsilon_3 \cdot p_1) \nn \\
&-\text{H} \, \varepsilon_1^{\mu \nu} \varepsilon_2^{\alpha \beta} 
(\varepsilon_{3,\mu} p_{3,\nu} - \varepsilon_{3,\nu} p_{3,\mu}) 
(\varepsilon_{3,\alpha} p_{3,\beta} - \varepsilon_{3,\beta} p_{3,\alpha}) \,. 
\end{align}
Identifying the spin-2 particles with gravitons, i.e. setting $\mathcal{M}^{\text{GR}-2}_3$ equal to the three-point amplitude in pure gravity, we find $\rm H=1$.

With this choice our one-loop amplitudes do not violate unitarity or causality constraints~\cite{treeLevelUnitarity,Meltzer:2017rtf,Afkhami-Jeddi:2018apj,Camanho:2014apa}. Hence, they serve as toy models of causal  UV completions.


\section{Tree-level string amplitudes}
\label{StringAppendix}

Here we collect the relevant string four-graviton tree-level amplitudes \cite{Kawai:1985xq}
\be
\mathcal{M}_4^{(\text{ss})}(1^{+},2^{-},3^{-},4^{+}) &= - \left( \frac{\kappa}{2} \right)^2 (\langle 23 \rangle [14])^4 {\Gamma(-s) \Gamma(-t) \Gamma(-u) \over \Gamma(1+s) \Gamma(1+t)\Gamma(1+u)} \,,  \nn \\
\mathcal{M}_4^{(\text{hs})}(1^{+},2^{-},3^{-},4^{+}) &= - \left( \frac{\kappa}{2} \right)^2 (\langle 23 \rangle [14])^4 {\Gamma(-s) \Gamma(-t) \Gamma(-u) \over \Gamma(1+s) \Gamma(1+t)\Gamma(1+u)} \left( 1 - {s u \over t+1} \right) \,, \nn \\
\mathcal{M}_4^{(\text{bs})}(1^{+},2^{-},3^{-},4^{+}) &= - \left( \frac{\kappa}{2} \right)^2 (\langle 23 \rangle [14])^4 {\Gamma(-s) \Gamma(-t) \Gamma(-u) \over \Gamma(1+s) \Gamma(1+t)\Gamma(1+u)} \left( 1 - {s u \over t+1} \right)^2 \, ,
\label{eq:stringresults}
\ee
where we set the closed string tension $\alpha'=4$ and $(\text{ss})$, $(\text{hs})$, $(\text{bs})$ stand for superstring, heterotic string and bosonic string respectively. 
In the case of the bosonic string we have additional independent nonvanishing helicity configurations,
\begin{align}
\mathcal{M}_4^{(\text{bs})}(1^{+},2^{+},3^{-},4^{+})\, &=- \left( \frac{\kappa}{2} \right)^2 \left( [12] [14] \langle 13 \rangle \right)^4 {\Gamma(-s) \Gamma(-t) \Gamma(-u) \over \Gamma(1+s) \Gamma(1+t)\Gamma(1+u)} \,,  \nn \\
\mathcal{M}_4^{(\text{bs})}(1^{+},2^{+},3^{+},4^{+}) &=- \left( \frac{\kappa}{2} \right)^2 \left( \frac{[12][34]}{\langle 12 \rangle \langle 34 \rangle} \right)^2  {4 s^2 t^2 u^2 (1- {s t u \over 2})^2 \over (1+ s )^2 (1+ t )^2 (1+ u)^2} \nn \\
& \null \hskip 2 cm \times
 {\Gamma(-s) \Gamma(-t) \Gamma(-u) \over \Gamma(1+s) \Gamma(1+t)\Gamma(1+u)} \, .
\label{eq:bosonicextra}
\end{align}

In \sect{Sec:Bounds} we considered the dispersive representation of $f(s,u)$ and used it to derive various bounds on the Wilson coefficients. We focused on the polynomial expansion of the amplitudes at low energies around $s=t=0$ which are generated by exchanges of massive states above a certain gap $m_{\rm gap}^2$. To focus on such contributions let us write down explicitly the part of the amplitude due to the exchange by light states (in the case of strings these are tachyon, dilaton, graviton)
\be
f^{(\text{ss})}(s,u) &= - \left( \frac{\kappa \null}{2} \right)^2{\Gamma(-s) \Gamma(-t) \Gamma(-u) \over \Gamma(1+s) \Gamma(1+t)\Gamma(1+u)} 
           =\left( \frac{\kappa \null}{2} \right)^2 {1 \over s t u} + \delta f^{(\text{ss})}(s,u) \,,\nn \\
f^{(\text{hs})}(s,u) &= - \left( \frac{\kappa \null}{2} \right)^2 {\Gamma(-s) \Gamma(-t) \Gamma(-u) \over \Gamma(1+s) \Gamma(1+t)\Gamma(1+u)} \left( 1 - {s u \over t+1} \right) 
           = \left( \frac{\kappa \null}{2} \right)^2 \Big( {1 \over s t u} - {1 \over t} \Big) + \delta f^{(\text{hs})}(s,u) \,, \nn\\
f^{(\text{bs})}(s,u) &= - \left( \frac{\kappa \null}{2} \right)^2 {\Gamma(-s) \Gamma(-t) \Gamma(-u) \over \Gamma(1+s) \Gamma(1+t)\Gamma(1+u)} \left( 1 - {s u \over t+1} \right)^2 \nn \\
&=\left( \frac{\kappa \null}{2} \right)^2 \Big(  {1 \over s t u} - {1 \over t+1} + {s u - 2\over t} \Big) + \delta f^{(\text{bs})}(s,u) \, ,
\label{eq:explicitform}
\ee
where ${\rm Disc} \ \delta f(s,u)$ is nonzero only for $s,t,u \geq m_{\rm gap}^2 = 1$ and $\delta f(s,u)$ admits the low-energy expansion \eqref{eq:lowenergyexp} with the dispersive representation of the couplings \eqref{eq:dispcouplingrepresentation}.

To write the dispersive representations for string amplitudes let us recall their Regge limit behavior.
In the $t$-channel we have
\be
\lim_{|t| \to \infty} | f^{(\text{ss}, \text{hs}, \text{bs})}(s,u) | &\leq {\text{const} \over |t|^2}, ~~~~~ s \leq 0 \,.  
\ee
Therefore, we can write dispersion representation for $f(s,u)$ without subtractions and $\delta f(s,u)$ will take the form \eqref{eq:dispersionrelationsf}
and the corresponding bounds of \sect{sec:nonsusymmetric} apply to $\delta f(s,u)$.

The situation is different in the $s$-channel relevant for \sect{sec:susymmetric}. In this case we have
\be
\lim_{|s| \to \infty} | f^{(\text{ss})}(s,u) | &\leq {\text{const} \over |s|^2}, ~~~~~ t \leq 0 \,,  \\
\lim_{|s| \to \infty} | f^{(\text{hs})}(s,u) | &\leq \text{const}, ~~~~~ t \leq 0 \,, \\
\lim_{|s| \to \infty} | f^{(\text{bs})}(s,u) | &\leq \text{const} |s|^2 , ~~~~~ t \leq 0\, . 
\ee
Therefore, for the heterotic and bosonic strings we would need to consider dispersion relations with subtractions. In  \sect{sec:susymmetric} instead 
we directly considered the dispersive representation of the relevant couplings \eqref{eq:tildea} in the expansion of $\delta f(s,u)$. As can be seen from 
\eqref{eq:explicitform} the $s$-channel Regge limit of $\delta f(s,u)$ coincides with the ones of the corresponding $f(s,u)$. In particular, the set of couplings that can admit a dispersive representation in the $s$-channel in the three cases correspond to $J_0^{(\text{ss})} = -2$, $J_0^{(\text{hs})} = 0$, $J_0^{(\text{bs})} = 2$, where $J_0$ is the Regge intercept see the discussion after \eqref{eq:tildeadisprep}. 

In \sect{Sec:BoundMultipleHelicity} we consider superconvergence relations applied to various helicity amplitudes. The relevant Regge limits for $g^{(\text{bs})}(s,u)$ and $h^{(\text{bs})}(s,u)$ can be easily read off from \eqn{eq:bosonicextra} and take the following form
\be
\lim_{|t| \to \infty} | g^{(\text{bs})}(s,u) | &\leq {\text{const} \over |t|^2} , ~~~~~ s \leq 0 \,, \\
\lim_{|t| \to \infty} | h^{(\text{bs})}(s,u) | &\leq \text{const} |t|^2, ~~~~~ s \leq 0 \, .
\ee
All the Regge bounds discussed in this section are in agreement with the general tree-level Regge bound \eqref{eq:CRG}.

\section{Bounding the coupling space at $k=6$}
\label{app:fullk6analysis}

In the main body of the paper we did not list the complete set of linear constraints that characterizes the space of admissible couplings
in the case of $k=6$. For completeness we list full set of constraints that we did not present in the main text. To reduce cluttering we avoid writing the subindices that we used in main text and specify which channel should be used in evaluating the determinant explicitly.
The constraints are:

$\< suuuuu\>$:
\be
& \bigl\{ \< 2,4,5,6,7,9 \>,\, \< 4,4,5,7,8,9 \>,\, - \< 6,4,5,6,7,8 \>,\, - \< 6,4,5,6,8,9 \>,\, - \< 6,4,6,7,8,9 \>,\, \nn \\
&-\< 2,5,6,7,k,k+1 \>_{k \geq 9},\, -\< 2,5,j,j+1, k , k+1 \>_{k>j \geq 9},\, -\< 4,5,7, 8 , k , k+1 \>_{k\geq 9} ,\, \\
&-\< 4, 5 , 8 , 9 , k , k+1 \>_{k\geq 10} \bigr\} \geq 0 \, . \nn
\ee

$\< ssuuuu\>$:
\be
& \bigl\{ -\< 2,4,5,7, k, k+1 \>_{k \geq 9} ,\, -\< 2,4,4,5,6,7 \>,\, -\< 2,4,4,5,7,9 \>,\, -\< 2,6,4,5,8,9 \>,\, 
 \nn \\
& -\< 4,6,4,5,6,7 \>,\,  - \< 4,6,4,5,7,8\>,\, - \< 4,6,5,6,7,8 \>,\, \< 2,4,4,5,8,9 \>,\, \< 2,6,4,5,6,9 \>,\, 
\nn \\
& \< 2,6,4,6,7,9 \>,\,  \< 4,6,4,7,8,9 \>,\,  \< 4,8,5,6,7,8 \>,\, \< 2, 4, 5, 9, k, k + 1 \>_{k \geq 10},\, \< 2, 6, 6, 7, k, k + 1 \>_{k \geq 9 } ,\, 
 \nn \\
& \< 2, 6, 5, 6, k, k + 1 \>_{k \geq 9} ,\,  \< 6, 8, 5, 6, k, k + 1 \>_{k \geq 8},\, \< 2, 4, j, j + 1, k, k + 1 \>_{k>j \geq 9},\, 
 \nn \\
& \< 4, 6, j, j + 1, k, k + 1 \>_{k>j \geq 7 },\,
 \<  6, 8, j, j + 1, k, k + 1\>_{k>j\geq 6},\, \nn \\
& \< i, i + 2, j, j + 1, k, k + 1 \>_{i \geq 8; k>j\geq 5} \bigr\} \geq 0 \, .
\ee

$\< sssuuu\>$:
\be
&\bigl\{ -\< 2,4,6,4,5,6 \>,\, - \< 2,4,6,4,6,7 \>,\, - \< 2,4,6,4,7,9\>,\, \< 2, 4, 6, 4, 5, 8 \>,\, \< 2, 4, 6, 4, 8, 9 \>,\,
\nn \\
& \< 2, 4, 8, 5, 6, 7 \>,\,   \< 4, 6, 8, 5, 6, 8 \>,\, \< 4, 6, 8, 6, 7, 8 \>,\, \< 2, 4, 6, 7, k, k + 1 \>_{k \geq 9} ,\, 
\nn \\
& \< 2, i, i + 2, 5, j, j + 1 \>_{i \geq 8 ; j \geq 6},\, 
\< 4, i, i + 2, 5, 7, 8 \>_{i \geq 8} \bigr\} \geq 0 \, . 
\ee

$\< ssssuu\>$:
\be
&\bigl\{ \< 2, 4, 6, 8, 5, 6 \>,\, -\< 2,4,j,j+2,5,8 \>_{6 \leq j \leq 14} ,\,    \< 2, 4, j, j + 2, 5, 7 \>_{j \geq 8} ,  - \< 2, 4, j, j + 2, 8, 9 \>_{6 \leq j \leq 14} , \nn \\
&   - \< 2, 4, j, j + 2, 6, 7 \>_{8 \leq j \leq 24} ,\,   - \< 2, 6, j, j + 2, 5, 6 \>_{j \geq 8} ,\,   - \< 2, 4, j, j + 2, k, k + 1 \>_{j \geq 6, k \geq 9} ,
\nn \\
&  -\< 4, 6, j, j + 2, k, k + 1 \>_{j \geq 8 , k \geq 6} ,\, 
-\< i, i + 2, j, j + 2, k, k + 1 \>_{j>i\geq 6, k \geq 5} \bigr\} \geq 0 \, . 
\ee

$\< sssssu\>$:
\be
& \bigl\{ -\< 2, 4, 6, j, j + 2, 6 \>_{j \geq 8} , - \< 2, i, i + 2, j, j + 2, 5 \>_{j>i \geq 6} \bigr\} \geq 0 \,.
\ee

\noindent
In the above formulas, $s$ indicates that in evaluating the determinant one should use $i_s$ vectors in the corresponding position, whereas $u$ means that the corresponding vector should
be $j_u^{(6)}$. Note also that $i_s$ take only even integer values. To summarize, the precise meaning of $\< sssuuu\>:  \< 2, 4, 6, 7, k, k + 1 \>_{k \geq 9} \geq 0$ is
\be
\< {\bf a}_6 , 2_s, 4_s , 6_s , 7_u^{(6)}, k_u^{(6)},  k_u^{(6)}+1 \>_{k \geq 9} \geq 0 \,.
\ee
On the crossing-symmetric slice we found that the strongest constraints arise from the $\< ssuuuu\>$ case. To generate Fig.~\ref{fig:k6full} the maximal spin in the $\< ssuuuu\>$ bounds above is set to $i_{\rm max}, j_{\rm max}, k_{\rm max}=20$. We used dispersive representations for the couplings truncated to $J_{\rm max} = 100$ to check that all the determinants are non-negative. We found that the final allowed region for the couplings is not sensitive to the precise value of $k_{\rm max}$.

\section{Amplitude with an accumulation point in the spectrum}
\label{app:Mstumodel}

In this appendix we analyze a toy model that violates the LSD property.  Nevertheless, 
we find that the model still winds up on the LSD island. Consider the amplitude function
\be
\label{eq:Mstumodel}
f(t,u) &= -{1 \over (t-m_1^2)(u-m_1^2)(s - m_2^2)}  \\
&= {1 \over t-m_1^2} {1 \over m_1^2 + m_2^2 + t} \left( {1 \over s-m_2^2} + {1 \over u - m_1^2} \right) , \nn
\ee
where in the second line we rewrote the amplitude in the dispersive representation using partial fractions. For scattering of external scalars, a similar model with $m_{1}=m_2$ was considered in \cite{Caron-Huot:2020cmc}. This amplitude saturates the tree-level Regge bound and it has an accumulation point in its spectrum, by which we mean that the residue of the amplitude at either $m_1$ or $m_2$ involves infinitely many particles of all spins in the partial-wave expansion.  Such models should not be considered physical, but nevertheless it is useful to illustrate
features when LSD is violated.

Expanding the $s$- and $u$-channel residues in the corresponding partial waves we find that the amplitude is unitary for 
\be
m_2 \geq m_1 . 
\ee
Let us therefore set $m_1 = 1$ and study the model as a function of $m_2 \geq 1$.

It is easy to find $\rho_J^{++}(m_2^2)$ explicitly with the following result
\begin{equation}
\label{eq:stuspectraldensity}
 -{1 \over t-1} {1 \over 1 + m_2^2 + t} \Big|_{t = - {m_2^2 \over 2} (1-x)} = \sum_{J=0}^\infty \rho_J^{++}(m_2^2) \, d^{J}_{0,0}(x) \,, 
\end{equation}
where
\begin{equation}
\rho_J^{++}(m_2^2) = {4(2J+1) \over m_2^2 (2+m_2^2) } Q_J \left( {2+m_2^2 \over m_2^2} \right) ,
\end{equation}
is determined by projecting the left-hand-side of
\eqn{eq:stuspectraldensity} onto the Legendre polynomials.
The function $Q_J(z)$ is
the four-dimensional Legendre Q-function\footnote{\footnotelineskip Note that the Mathematica LegendreQ function is defined somewhat differently. The precise relation is $Q_J(z) = \text{LegendreQ}[J,z+ i 0] - i {\pi \over 2} \text{LegendreP}[J,z]$ for $z>1$.} that can be found in
Eq.~(2.44) of 
Ref.~\cite{Correia:2020xtr}. By increasing $m_2$ we can 
make the spectral densities of non-minimal spin dominant. 
One way to understand this is by noting that the LHS of
\eqref{eq:stuspectraldensity} develops a singularity at $x = \pm 1$
when $m_2 = \infty$ which translates into enhancement of higher-spin
contributions.
Therefore,
\eqref{eq:Mstumodel} is an explicit example where LSD
does not hold. 

In the other channel we have
\be
-{1 \over t-1} {1 \over 1 + m_2^2 + t} \Big|_{t = - {1 \over 2} (1-x)} = \sum_{J=4}^{\infty} {\rho_J^{+-}(1) \over ({1+x \over 2})^4} d^{J}_{4,4}(x) .
\ee
It is then straightforward to check that $\rho_J^{+-}(1)$ satisfies LSD with $\alpha \simeq 10$ for any value of $m_2$. 

Curiously, if we now consider the values of the couplings $a_{k,j}$ for various values of $m_2$ they all end up being located at the LSD islands. For $m_2$ close to $1$ it is manifest in the properties of the spectral densities described above.  At large values of $m_2$ when the $s$-channel spectral density $\rho_J^{++}(m_2^2)$ violates the LSD property the reason is that its contribution to the low-energy couplings $a_{k,j}$ is suppressed by an extra factor of $\left( {1 \over m_2^2} \right)^{k-j}$ coming from the expansion of ${1 \over s-m_2^2}$ at small $s$, compared to the $u$-channel contribution that does satisfy $\alpha \simeq 10$ LSD. In other words, for $m_2 \gg 1$ we have $\langle \rho_J^{+-} \rangle_k \gg \langle \rho_J^{++} \rangle_k $.

To summarize, while \eqref{eq:Mstumodel} with an accumulation point in the spectrum does provide an example of an amplitude function which violates the LSD assumption at ${m_2 \over m_1} \gg 1$, from the low-energy couplings point of view it still ends up in the LSD island region because the $\rho_J^{++}(m_2^2)$ spectral function that violates the LSD property ends up being irrelevant. 
While this model should not be considered physical, it does illustrate
the idea that potential violations of the LSD principle do not affect the location of the island if there is
a separation of scales between the lowest-mass state and higher-mass
physics that sources the violation.

\section{Wigner d-matrices}
\label{app:wignerd}

Here we list convenient formulas for Wigner d-matrices used in bulk of the paper
\be
d_{4,4}^J(x) &= 2^{-J} (x+1)^J \, _2F_1\left(-J-4,4-J;1;\frac{x-1}{x+1}\right) , \\
d_{4,-4}^J(x) &=\frac{2^{-J-7} }{315} {\Gamma(J+5) \over \Gamma(J-3)} (1-x)^4 (x+1)^{J-4} \, _2F_1\left(4-J,4-J;9;\frac{x-1}{x+1}\right) , \\
d_{4,0}^J(x) &=\frac{2^{-J-3}}{3} \sqrt{{\Gamma(J+5) \over \Gamma(J-3)}} (1-x)^2 (x+1)^{J-2} \, _2F_1\left(4-J,-J;5;\frac{x-1}{x+1}\right) , \\
  d_{0,0}^J(x) &= \, _2F_1\left(-J,J+1;1;\frac{1-x}{2}\right) . 
\ee
By expanding the formulas above around $x=1$ it is easy to check formulas \eqref{eq:legendrepos}, \eqref{eq:legendrepos2}, and \eqref{eq:legendrepos3} for the first few $n$'s.  In Mathematica notation the $d_{\lambda_1,\lambda_2}^J(x)$ functions are given by ${\rm WignerD}[\{J, \lambda_2, \lambda_1\}, {\rm ArcCos}[x]]$. For a detailed derivation of the partial-wave expansion for spinning external particles see Ref. \cite{Hebbar:2020ukp}. 

\section{Explicit values of one-loop four-graviton amplitudes}
\label{AmplitudesResultsAppendix}

In this appendix we collect the final results for the integrated one-loop
amplitudes. For the double-minus configuration, we give the results for the
$\mathcal{M}_4^{\newPiece{\it S}}$ defined in \eqn{susyGR}.  These results are also collected in
a Mathematica ancillary file~\cite{AttachedFile}. 
The supersymmetric
decomposition \eqref{susyGR}
directly gives the amplitude for any massive particle up to spin 2
circulating in the loop. For the single-minus and all-plus helicity
configurations, we present the amplitude with a scalar particle
circulating in the loop. The amplitudes with a higher-spin particle
circulating in the loop are all proportional to this one as shown in
\eqn{allPlusSUSY}.  In this appendix we give the results for the
amplitudes in terms of scalar integral functions whose 
values we cite in Appendix~\ref{IntegralValuesAppendix}.

\subsection{The double-minus configuration}

We first give our results formally to all orders in the
dimensional-regularization parameter $\epsilon$.  We write them in terms of an overcomplete basis that
contains higher-dimensional integrals as we find this form to be the
most concise.  This basis is chosen so that the coefficients of the
integrals are free of both $\eps$ and $m$.  As we explained in
Appendix~\ref{IBPsec}, using dimension-shifting relations the
higher-dimension integrals are directly expressible in terms of
standard ($4-2\epsilon$)-dimensional ones.  We collect the explicit
values of the ($4-2\epsilon$)-dimensional integrals to leading orders
in $\epsilon$~\cite{Integrals} in Appendix~\ref{IntegralsAppendix}.

We manifest Bose symmetry under a $2 \leftrightarrow 3$
relabeling (which implies $s \leftrightarrow u$) by writing the
amplitudes as
\begin{equation}
\mathcal{M}_4^{\newPiece{\it S}}(1^{+},2^{-},3^{-},4^{+}) = -\frac{1}{(4\pi)^{2-\epsilon}} 
\left( \frac{\kappa}{2} \right)^2 \mathcal{M}^\text{tree}_4 
 (F_1^{\newPiece{\it S}}(s,u) + F_2^{\newPiece{\it S}}(s,u) + F_2^{\newPiece{\it S}}(u,s) ) \,,
\label{fullAns}
\end{equation}
where $F_1^{\newPiece{\it S}}(s,u)=F_1^{\newPiece{\it S}}(u,s)$ with kinematics in the Euclidean region, and
the tree-level amplitude $\mathcal{M}^\text{tree}_4$ is given in
Eq.~(\ref{mtree}).

\noindent
For the $\mathcal{M}_4^{\newPiece{0}}$  pieces we have
\begin{align}
F_1^{\newPiece{0}}(s,u) = \null & 
\frac{13s^4 + 52s^3 u + 75 s^2 u^2 +52 s u^3 + 13u^4}{96 t^4} \, I_1  \nn\\[3pt]
&
- \frac{8s^4+40s^3 u +55 s^2 u^2 + 40s u^3 + 8 u^4}{8s u t^3} \, I_1^{6-2\epsilon} \nn\\[3pt]
&
 - \frac{(s^2-s u+u^2)^2 t }{64 s^2 u^2} \, I_2{(t)} 
+ \frac{1}{32} \Bigl( 16 - \frac{7 s}{u} - \frac{ 7 u}{s} \Bigr) \, I_2^{6-2\epsilon}{(t)} 
- \frac{45}{16t} \, I_2^{8-2\epsilon}{(t)} \nn \\[3pt]
&
+ \frac{s^8 + s^7 u + s u^7 + u^8}{128 s^3 u^3} \, I_3{(t)} 
- \frac{5(s^5 + u^5)}{64 s^2 u^2}\, I_3^{6-2\epsilon}{(t)} 
+ \frac{25(s^2-s u+u^2)}{32s u}\, I_3^{8-2\epsilon}{(t)} \nn\\[3pt]
&
+ \frac{105}{16t} \, I_3^{10-2\epsilon}{(t)} 
- \frac{s^5 u^5}{256 t^7} \, I_4(s,u) 
+ \frac{7 s^4 u^4}{32 t^6} \, I_4^{6-2\epsilon}(s,u)
- \frac{105 s^3 u^3}{32 t^5} \, I_4^{8-2\epsilon}(s,u)  \nn \\[3pt]
&
+ \frac{105 s^2 u^2 }{8 t^4} \, I_4^{10-2\epsilon}(s,u)
- \frac{105 s u }{16 t^3}\, I_4^{12-2\epsilon}(s,u)  
\, , \hskip .2 cm \label{F1su0}
\end{align}
\begin{align}
F_2^{\newPiece{0}}(s,u) = \null &
 -\frac{s^3(s^2 + 2s u +2 u^2)(s^4 +4 s^3 u +5 s^2 u^2 +2s u^3 +u^4)}{64u^2 t^6} \, I_2{(s)} \nn \\[3pt]
&- \frac{s^2 (7s^4 + 30s^3 u + 50 s^2 u^2 + 40s u^3 - 12 u^4)}{32u t^5} \, I_2^{6-2\epsilon}{(s)} \nn \\[3pt]
&- \frac{45s^4 + 118s^3 u + 294 s^2 u^2 + 96s u^3 + 16 u^4}{16s t^4}\, I_2^{8-2\epsilon}{(s)} \nn \\[3pt]
&+ \frac{s^6 ( s^6 + 7s^5 u + 21 s^4 u^2 + 35 s^3 u^3 + 35 s^2 u^4 + 21 s u^5 + 7 u^6)}{128 u^3 t^7} \, I_3{(s)} \nn \\[3pt]
&+ \frac{s^4(5s^5 + 25 s^4 u + 50 s^3 u^2 + 50 s^2 u^3 + 25 s u^4 + 32 u^5)}{64 u^2 t^6} \, I_3^{6-2\epsilon}{(s)} \nn \\[3pt]
&+ \frac{5 s^3 (5 s^3 + 15s^2 u + 15 s u^2 - 32 u^3)}{32u t^5} \, I_3^{8-2\epsilon}{(s)}
+ \frac{3 s^2(35s + 128 u)}{16t^4} \, I_3^{10-2\epsilon}{(s)} \nn \\[3pt]
&- \frac{s^5 t}{256 u^3}\, I_4(s,t)
- \frac{s^4}{32 u^2} \, I_4^{6-2\epsilon}(s,t)
- \frac{9 s^3}{32t u} \, I_4^{8-2\epsilon}(s,t)
- \frac{15s^2}{8t^2} \, I_4^{10-2\epsilon}(s,t) \hskip 2 cm \nn \\[3pt]
&- \frac{105s u }{16t^3}\, I_4^{12-2\epsilon}(s,t)
\,. \label{F2su0}
\end{align}
\noindent 
For the $\mathcal{M}_4^{\newPiece{1/2}}$ pieces we have
\begin{align}
F_1^{\newPiece{1/2}}(s,u) = \null &
 \frac{4s^2 + 7s u +4u^2}{8t^2} \, I_1 
+ \frac{(s-u)^2 t}{16s u} \, I_2{(t)} 
+ \frac{3}{4} \, I_2^{6-2\epsilon}{(t)} 
+ \frac{(s^5+u^5)t}{32s^2u^2} \, I_3{(t)}\nn \\[3pt]
& + \frac{s^3+u^3}{4s u} \, I_3^{6-2\epsilon}{(t)}
- \frac{15}{8} \, I_3^{8-2\epsilon}{(t)} 
- \frac{s^4 u^4}{64t^5} \, I_4(s,u)
+ \frac{15s^3u^3}{32t^4} \, I_4^{6-2\epsilon}(s,u) \nn \hskip 2 cm \\[3pt]
&
- \frac{45s^2u^2}{16t^3} \, I_4^{8-2\epsilon}(s,u) 
+ \frac{15s u}{8t^2} \, I_4^{10-2\epsilon}(s,u) 
\,,   \label{F1su1o2}
\end{align}
\begin{align}
F_2^{\newPiece{1/2}}(s,u) = \null &
 \frac{s^3(s+2u)(s^2+2s u+2u^2)}{16u t^4} \, I_2{(s)}
+\frac{3s^3+7s^2u +12s u^2 +2u^3}{4t^3} \, I_2^{6-2\epsilon}{(s)} \nn \\[3pt]
&- \frac{s^4(s+2u)(s^4+3s^3u+4s^2u^2+2s u^3+u^4)}{32u^2t^5} \, I_3{(s)}\nn\\[3pt]
&- \frac{s^3(2s^3+6s^2u+6s u^2-5u^3)}{8u t^4} \, I_3^{6-2\epsilon}{(s)} 
- \frac{3s^2(5s+16u)}{8t^3} \, I_3^{8-2\epsilon}{(s)} \hskip 4 cm\nn  \\[3pt]
&+ \frac{s^4t}{64u^2} \, I_4(s,t)
+ \frac{3s^3}{32u} \, I_4^{6-2\epsilon}(s,t)
+ \frac{9s^2}{16t} \, I_4^{8-2\epsilon}(s,t)
+ \frac{15s u}{8t^2} \, I_4^{10-2\epsilon}(s,t) 
\, .  \label{F2su1o2}
\end{align}
\noindent 
For the $\mathcal{M}_4^{\newPiece{1}}$ pieces we have
\begin{align}
F_{1}^{\newPiece{1}}(s,u) = \null & 
- \frac{t}{4} \, I_2{(t)}
+ \frac{s^4+s^3 u +s u^3 +u^4}{8s u} \, I_3{(t)}
+\frac{3t}{4} \, I_3^{6-2\epsilon}{(t)}
- \frac{s^3 u^3 }{16t^3} \, I_4(s,u) \nn \hskip 2 cm  \\[3pt]
&
+ \frac{3 s^2 u^2 }{4t^2} \, I_4^{6-2\epsilon}(s,u)
- \frac{3s u}{4t} \, I_4^{8-2\epsilon}(s,u) 
\,, \label{F1su1}
\end{align}
%
\begin{align}
F_{2}^{\newPiece{1}}(s,u) = \null &
 -\frac{s^{} (s^2+2s u + 2u^2)}{4t^2} \, I_2{(s)}
+ \frac{s^4(s^2+3s u +3u^2)}{8u t^3} \, I_3{(s)} 
+ \frac{s^2(3s+8u)}{4t^2} \, I_3^{6-2\epsilon}{(s)} \hskip .5 cm  \nn  \\[3pt]
&
- \frac{s^3 t}{16u} \, I_4(s,t)
-\frac{s^2}{4} \, I_4^{6-2\epsilon}(s,t)
- \frac{3s u}{4t} \, I_4^{8-2\epsilon}(s,t)
\,.   \label{F2su1}
\end{align}
%
\noindent
For the $\mathcal{M}_4^{\newPiece{3/2}}$ pieces we have
\begin{align}
F_{1}^{\newPiece{3/2}}(s,u) = \null &
 -\frac{t^2}{2} \, I_3{(t)}
- \frac{s^2 u^2}{4t} \, I_4(s,u)
+\frac{s u}{2} \, I_4^{6-2\epsilon}(s,u) 
\,, \hskip 4.2 cm  \label{F1su3o2} \\
F_{2}^{\newPiece{3/2}}(s,u) = \null &
 -\frac{s^2(s+2u)}{2t} \, I_3{(s)}
+ \frac{s^2 t}{4} \, I_4(s,t)
+\frac{s u}{2} \, I_4^{6-2\epsilon}(s,t) 
\, . \hskip 3.4 cm \label{F2su3o2} 
\end{align}
%
%
%
%
\noindent
Finally, for the $\mathcal{M}_4^{\newPiece{2}}$ pieces we have
\begin{align}
F_{1}^{\newPiece{2}}(s,u) = \null &
-s t u \, I_4(s,u)
\, , \hskip 9.2 cm \label{F1su2}
 \\
F_{2}^{\newPiece{2}}(s,u) = \null &
-s t u  \, I_4(s,t)
\,. \label{F2su2}
\end{align}
We note that the progression of the new pieces from more complicated
contributions to simpler ones as the spin increases is a direct
consequence of the supersymmetric decomposition \eqref{susyGR}.  As the spin
increases the new pieces have lower and lower power counts
corresponding to increasing supersymmetry.  The final pieces
\eqref{F1su2} and \eqref{F2su2} correspond to $D=(4-2 \eps)$ scalar
box integrals with no powers of loop momentum in the numerator.  This
may be compared to the $\mathcal{M}_4^{\newPiece{0}}$ contribution
which has 8 powers of loop momentum in the numerator. This high power
count results in, for example, the $D=(12-2\eps)$ box integrals
appearing in \eqns{F1su0}{F2su0}.

Next we expand the above results to leading order in the
dimensional-regularization parameter $\epsilon$. Using
Eq.~(\ref{dimShift}) we express the higher-dimensional integrals in
terms of ($4-2\epsilon$)-dimensional ones whose explicit values
through $\mathcal{O}(\epsilon^0)$ are collected in
Appendix~\ref{IntegralsAppendix}. In the following expressions, the
integrals are understood as truncated to this order.

Both $F_1^{\newPiece{\it S}}$ and $F_2^{\newPiece{\it S}}$ are
ultraviolet divergent. However, when put together in
Eq.~(\ref{fullAns}), the ultraviolet divergence cancels as
expected~\cite{tHooft:1974toh}. We expose this cancellation by
separating the bubble integral in \eqn{BubbleIntegral} into a
divergent part $I_2(0)$ and a finite part defined in
\eqn{BubbleFinite}.  After canceling the $1/\epsilon$-pole we write
the amplitude as
\begin{equation}
\mathcal{M}_4^{\newPiece{\it S}}(1^{+},2^{-},3^{-},4^{+}) = -\frac{1}{(4\pi)^{2}} 
\left( \frac{\kappa}{2} \right)^2 \mathcal{M}^\text{tree}_4 ( f_1^{\newPiece{\it S}}(s,u) + f_2^{\newPiece{\it S}}(s,u) + f_2^{\newPiece{\it S}}(u,s) ) \,,
\label{fullAnsExpanded}
\end{equation}
where $f_1^{\newPiece{\it S}}(s,u) = f_1^{\newPiece{\it S}}(u,s)$ and
$f_{1,2}^{\newPiece{\it S}}$ are ultraviolet finite.

\noindent
For the $\mathcal{M}_4^{\newPiece{0}}$ pieces corresponding to \eqns{F1su0}{F2su0} we have
\begin{align}
f_1^{\newPiece{0}}(s,u) = \null & 
  -  \frac{1}{360t^5} \Big( 540m^4 s u t^2 + s u {} (2s^4 + 23 s^3 u + 222 s^2 u^2 + 23 s u^3 + 2 u^4) \nn \\[3pt]
& \hskip 1 cm 
 - 2 m^2 t {} (8s^4 + 5s^3 u - 366 s^2 u^2 + 5 s u^3 + 8 u^4) \Big)  \nn \\[3pt] 
&- \frac{s u}{2 t^7}\Big( s^4 u^4 + 8m^2 s^3 u^3 t + 20 m^4 s^2 t^2 u^2 + 16 m^6 s u t^3 +2m^8 t^4 \Big) I_4(s,u) 
\,, \hskip .1 cm  
\end{align}
%
\begin{align}
f_2^{\newPiece{0}}(s,u) = \null & 
 \frac{u \null}{60 s t^6} \Big( 2 m^4 t^2 ( 73s^3 - 147s^2 u -48s u^2 -8u^3) \nn \\[3pt]
& \null 
+ 2m^2 s t {}(9s^4 + 78s^3 u -105 s^2 u^2 -28s u^3 - 4u^4) \nn \\[3pt]
&\null 
+ s^2 (s-u) (s^4 + 9s^3 u +46s^2 u^2 + 9 s u^3 + u^4) \Big) I_2^{\rm fin}{(s)} \nn  \\[3pt]
&\null
- \frac{s^2 u}{t^7} (s u+2m^2 t)(s^2u^2 + 4m^2 s t u +2 m^4 t^2) \, I_3{(s)}  - \frac{m^8 s u}{t^3} \, I_4(s,t)
\,, \hskip .5 cm 
\end{align}
where  the integrals are defined in Eqs.~\eqref{BubbleIntegral}, \eqref{triangles} and \eqref{boxes}.
%
%
%
\noindent
For the  $\mathcal{M}_4^{\newPiece{1/2}}$ pieces corresponding to \eqns{F1su1o2}{F2su1o2} we have
\begin{align}
f_1^{\newPiece{1/2}}(s,u) = \null &
 -\frac{(s^2 + 14 s u+u^2+ 24 m^2 t) s u}{24t^3} \nn \\[3pt]
&- \frac{(s u+2m^2t)(s^2 u^2+4m^2 s t u+m^4t^2)s u }{2t^5} \, I_4(s,u)
\, , \hskip 3 cm \\[6pt]
%
f_2^{\newPiece{1/2}}(s,u) = \null & 
    \frac{\Big((s-u)(s^2+8s u+u^2)s +4m^2(-4s^3+2s^2u+7s u^2+u^3) \Big)u}{12t^4} \, I_2^{\rm fin}{(s)} \nn \\[3pt]
& - \frac{(s u+3m^2t)(s u+m^2t) s^2 u}{t^5} \, I_3{(s)}
- \frac{m^6 s u}{t^2} \, I_4(s,t)
\,. 
\end{align}
%
%
%
\noindent
For the  $\mathcal{M}_4^{\newPiece{1}}$ pieces corresponding to \eqns{F1su1}{F2su1} we have
\begin{align}
f_1^{\newPiece{1}}(s,u)  = \null &
 -\frac{s u}{2t}
-\frac{ (s^2 u^2+4 m^2 s t u +2 m^4 t^2) s u }{2t^3} \, I_4(s,u)
\,, \hskip 4 cm \\[6pt]
f_2^{\newPiece{1}}(s,u)  = \null & 
\frac{ (s-u) s u}{2 t^2} \, I_2^{\rm fin}{(s)}
-\frac{(s u + 2m^2 t) s^2 u }{t^3} \, I_3{(s)}
-\frac{m^4 s u}{t} \, I_4(s,t)
\,.
\end{align}
%
%
%
For the  $\mathcal{M}_4^{\newPiece{3/2}}$ pieces corresponding to \eqns{F1su3o2}{F2su3o2} we have
\begin{align}
f_1^{\newPiece{3/2}}(s,u) = \null &
 -\frac{(s u + 2 m^2t) s u}{2 t}\, I_4(s,u)
\,, \hskip 7 cm \\
f_2^{\newPiece{3/2}}(s,u) = \null &
 - \frac{s^2 u}{t} \, I_3{(s)} 
 - m^2 s u \, I_4(s,t)
\,.
\end{align}
%
Finally, for the  $\mathcal{M}_4^{\newPiece{2}}$ pieces corresponding to \eqns{F1su2}{F2su2} we have
\begin{align}
f_1^{\newPiece{2}}(s,u) = \null &
-s t u \, I_4(s,u)
\, , \hskip 9.2 cm \\
f_2^{\newPiece{2}}(s,u) = \null &
 -s t u \, I_4(s,t)
\,.
\end{align}

\subsection{The all-plus configuration}

As noted in \eqn{allPlusSUSY}, the result for particles of any spin $0 \le S \le 2$ circulating in the loop is proportional 
to the $S=0$ case.
The all-orders-in-$\epsilon$ form of this amplitude using the higher-dimensional integral basis is
\begin{equation}
\mathcal{M}_4^{S=0}(1^{+},2^{+},3^{+},4^{+}) = \frac{1}{(4\pi)^{2-\epsilon}} \left( \frac{\kappa}{2} \right)^4 \left( \frac{[12][34]}{\langle 12 \rangle \langle 34 \rangle} \right)^2 
\frac{1}{2}(F_3(s,t,u) + F_4(s,t) + F_4(t,u) + F_4(u,s)) \,,
\end{equation}
where
\begin{align}
F_3(s,t,u) =  & \null 
    \frac{(s^2 + t^2 + u^2)^2}{64 s t u} \, I_1
- \frac{15}{4} \, I_1^{6-2\epsilon} 
\,, \\[6pt]
F_4(s,t) = \null & \null 
    \frac{u^2 (s^3+t^3)^2}{32 s^3 t^3} \, I_2{(u)}
+ \frac{u^3 (7s^2 - 16s t + 7t^2)}{16 s^2 t^2} \, I_2^{6-2\epsilon}{(u)} 
+ \frac{45u^2}{8 s t} \, I_2^{8-2\epsilon}{(u)} \nn \\[3pt]
&\null 
+ \frac{u^4 (s^7+t^7)}{64 s^4 t^4} \, I_3{(u)}
+ \frac{5 u^3 (s^5+t^5)}{32 s^3 t^3} \, I_3^{6-2\epsilon}{(u)}
+ \frac{25 u^2 (s^3+t^3)}{16 s^2 t^2} \, I_3^{8-2\epsilon}{(u)} \nn \\[3pt]
& \null
- \frac{105 u^2}{8s t} \, I_3^{10-2\epsilon}{(u)} 
+ \frac{s^4 t^4}{128 u^4} \, I_4(s,t)
+ \frac{s^3 t^3}{16 u^3} \, I_4^{6-2\epsilon}(s,t)
+ \frac{9 s^2 t^2}{16u^2} \, I_4^{8-2\epsilon}(s,t) \nn \\[3pt]
& \null
+ \frac{15 s t}{4u} \, I_4^{10-2\epsilon}(s,t)
+ \frac{105}{8} \, I_4^{12-2\epsilon}(s,t)
\,.
\end{align}
There is no corresponding tree-level amplitude for the all-plus
helicity. Instead, we choose the above spinor-helicity combination to
be completely Bose symmetric. Given this choice, the combination in
the parenthesis also has this property. Furthermore, we arrange our
functions such that $F_3$ is completely Bose symmetric, while
$F_4(s,t) = F_4(t,s)$. 

The expression simplifies significantly if we expand in $\epsilon$ and drop the $\mathcal O (\epsilon)$ pieces. We have
\begin{align}
\mathcal{M}_4^{S=0}(1^{+},2^{+},3^{+},4^{+}) = &
    \frac{1}{(4\pi)^{2}} \left( \frac{\kappa}{2} \right)^4 \left( \frac{[12][34]}{\langle 12 \rangle \langle 34 \rangle} \right)^2 
\frac{1}{2} \Big( 
-\frac{1}{120}(120 m^4 + s^2 + t^2 + u^2 ) \nn \\[3pt]
& \null 
\hskip 2.2 cm 
+ 2m^8 ( I_4(s,t) + I_4(t,u) + I_4(u,s) )  
+ \mathcal{O}(\epsilon) \Big) \,.
\label{allPlusLargeMass}
\end{align}
Because the corresponding tree-level amplitude vanishes this amplitude is infrared finite.  As for the other 
helicities it is ultraviolet finite because of the lack of a viable counterterm~\cite{tHooft:1974toh}. Another interesting 
property is that for $m \rightarrow 0$ it has no logarithms. The all-minus amplitude follows from parity and is given by 
swapping angle and square brackets.

\subsection{The single-minus configuration}

As for the all-plus case, for the single-minus configuration we only need the $S=0$ case.
The single-minus amplitude in a form valid to all orders in $\epsilon$ is
\begin{equation}
\mathcal{M}_4^{S=0}(1^{+},2^{+},3^{-},4^{+}) = \frac{1}{(4\pi)^{2-\epsilon}} \left( \frac{\kappa}{2} \right)^4
\left( [12] \langle 13 \rangle [14] \right)^4 \frac{1}{2}(F_5(s,t,u) + F_6(s,t) + F_6(t,u) + F_6(u,s)).
\end{equation}
We choose the little group combination to have complete Bose symmetry. The function $F_5$ is completely Bose symmetric, while $F_6(s,u)=F_6(u,s)$. We find 
\begin{align}
F_5(s,t,u) &= 
\frac{(s^2+t^2+u^2)^2}{64 (s t u)^3} I_1 +
\frac{246 (s t u)^2 -9 (s^2+t^2+u^2)^3}{4 (s t u)^4} I_1^{6-2\epsilon}
\,,
\end{align}
\begin{align}
F_6(s,u) = &  
\frac{s^6 + u^6}{32 s^5 u^5} \, I_2{(t)} 
-\frac{s^4+2s^3u-2s^2u^2+2s u^3 +u^4}{16s^4 u^4 t} \, I_2^{6-2\epsilon}{(t)} \nn\\[3pt]
& \null
-\frac{3(9s^4 + 22s^3u + 42s^2u^2 + 22s u^3 + 9u^4)}{8s^3 u^3 t^4} \, I_2^{8-2\epsilon}{(t)}\nn\\[3pt]
& \null
+\frac{s^9+2s^8u+s^7u^2+s^2u^7+2s u^8 + u^9 }{64s^6u^6} \, I_3{(t)} \nn\\[3pt]
& \null
+ \frac{3(s^6+s^5u+s u^5 +u^6)}{32s^5 u^5} \, I_3^{6-2\epsilon}{(t)} \nn\\[3pt]
& \null
+ \frac{31 s^6 + 93 s^5 u + 93 s^4 u^2 + 126 s^3 u^3 + 93 s^2 u^4 + 93 s u^5 + 31 u^6} {16 s^4 u^4 t^3}
    \, I_3^{8-2\epsilon}{(t)} \nn\\[3pt] & \null
+\frac{3(29s^4 + 52s^3u -18s^2 u^2 +52 s u^3 +29 u^4)}{8s^3 u^3 t^4} \, I_3^{10-2\epsilon}{(t)}
+ \frac{s^2 u^2}{128 t^6} \, I_4(s,u) \nn\\[3pt]
& \null
- \frac{s u}{16 t^5} \, I_4^{6-2\epsilon}(s,u) 
-\frac{15}{16t^4} \, I_4^{8-2\epsilon}(s,u)
- \frac{15}{4s u t^3} \, I_4^{10-2\epsilon}(s,u)
+\frac{105}{8s^2 t^2 u^2} \, I_4^{12-2\epsilon}(s,u)
\,.
\end{align}

Next, we expand the above amplitude to leading order in $\epsilon$. We write
\begin{equation}
\mathcal{M}_4^{S=0}(1^{+},2^{+},3^{-},4^{+}) = \frac{1}{(4\pi)^2} \left( \frac{\kappa}{2} \right)^4
\left( [12] \langle 13 \rangle [14] \right)^4 \frac{1}{2}(f_5(s, t,u) + f_6(s,t)+f_6(t,u) + f_6(u,s)).
\end{equation}
As in the double-minus configuration, we extract the ultraviolet poles from the
bubble integrals in order to manifest the ultraviolet-divergence
cancellation. In this way we may express the amplitude in terms of the
ultraviolet-finite functions $f_{5}$ and $f_{6}$. We choose these functions such that
$f_5$ is completely Bose symmetric and $f_6(s,u) = f_6(u,s)$. We have
\begin{align}
f_5(s,t,u) = &
\frac{1}{360 (s t u)^4} \Bigg( (s^2+t^2+u^2) (s t u)^2 -15 m^2 (s t u) (s^2+t^2+u^2)^2 
	\nn\\[3pt] & \null + m^4 \Big( 90 (s^2+t^2+u^2)^3 - 2520 (s t u)^2 \Big) \Bigg)
\,,
\label{singleMinusLargeMass1}
\end{align}
and
\begin{align}
f_6(s,u) = & 
 \frac{m^4 }{s^3u^3t^4}  \Bigl(2s^4+5s^3 u + 5s u^3 + 2u^4\Bigr) \, I_2^{\rm fin}{(t)} \nn\\[3pt]
 & \null 
+ \frac{2m^4}{s^4t^4u^4} \Big( s^7 + 4 s^6 u + 6 s^5 u^2 + 4 s^4 u^3 + 4 s^3 u^4 + 
	6 s^2 u^5 + 4 s u^6 + u^7 
\nn\\[3pt] & \null \hskip 1.7 cm
 	- m^2 ( 2 s^5 u + 6 s^4 u^2 + 6 s^3 u^3 + 6 s^2 u^4 + 
	2 s u^5 )
 \Big) \, I_3{(t)} \nn\\[3pt]
& \null
+ \frac{m^4}{s^2u^2t^4} \Big(s^2u^2 +4m^2 (s t u) +2m^4 t^2 \Big) \, I_4(s,u)
\,.
\label{singleMinusLargeMass2}
\end{align}
As for the all-plus amplitude amplitude, because there is no corresponding 
tree-level amplitude there are no infrared singularities and again in 
the $m \rightarrow 0$ limit the expression is free of logarithms.
The single-plus helicity configuration follows from parity.

\subsection{Pure gravity}
\label{sec:puregravity}

For completeness, we also give the corresponding one-loop amplitude
with a massless graviton circulating in the loop.  
We obtain this result by
taking the massless limit of the amplitude with a massive spin-2
particle circulating in the loop, after accounting for the additional
states (see Eq.~(\ref{masslessLimit})). Our results match the ones
previously obtained in Ref.~\cite{Dunbar:1994bn}. 

Referring to this amplitude as $\mathcal{M}_4^\text{GR}$, for the
all-plus configuration we have
\def\allPlusLittleGroup{\left( \frac{[12][34]}{\langle 12 \rangle \langle 34 \rangle} \right)^2}
\begin{equation}
\mathcal{M}_4^{\text{GR}}(1^{+}, 2^{+},3^{+},\,4^{+}) = 
\frac{-1}{(4\pi)^{2}} \left( \frac{\kappa \null}{2} \right)^4 \allPlusLittleGroup
\frac{s^2 + t^2 + u^2}{120} 
\,,
\end{equation}
while for the single-minus configuration we find~\cite{Bern:1993wt}
\def\singleMinusLittleGroup{\left( [12] \langle 13 \rangle [14] \right)^4}
\begin{equation}
\mathcal{M}_4^{\text{GR}}(1^{+}, 2^{+},3^{-},\,4^{+}) = 
\frac{1}{(4\pi)^{2}} \left( \frac{\kappa \null}{2} \right)^4
\singleMinusLittleGroup
\frac{s^2 + t^2 + u^2}{360(s t u)^2} 
\,.
\end{equation}
For the double-minus configuration the amplitude takes the form
\begin{align}
\mathcal{M}_4^{\text{GR}}(1^{+}, 2^{-},&\,3^{-},4^{+}) = 
\frac{r_\Gamma}{(4\pi)^{2-\epsilon}} 
\left( \frac{\kappa \null}{2} \right)^2 s t u \mathcal{M}^\text{tree}_4 \Bigg[ \nn \\[3pt]
 & \frac{\null 2 \null}{\null \epsilon} \frac{1}{s t u}  {} \left(s \log\Bigl(\frac{-s}{\mu^2}\Bigr) 
         + t \log\Bigl(\frac{-t}{\mu^2} \Bigr) + u \log\Big(\frac{-u}{\mu^2}\Bigr) \right) \nn \\[3pt] 
& +  \frac{2}{\null s t u} {} \left(u \log\Bigl(\frac{-t}{\mu^2} \Bigr)\log\Bigl(\frac{-s}{\mu^2} \Bigr) 
    + t \log\Bigl(\frac{-s}{\mu^2}\Bigr) \log\Bigl(\frac{-u}{\mu^2} \Bigr) 
    + s \log\Bigl(\frac{-t}{\mu^2}\Bigr) \log\Bigl(\frac{-u}{\mu^2}\Bigr) \right) \nn \\[3pt]
& + \frac{4s^6 +14s^5u +28s^4u^2 +35s^3u^3 +28s^2u^4 + 14s u^5 +4u^6}{t^8} \left( \log^2\left(\frac{\null s}{u}\right) +\pi^2 \right) \nn \\[3pt]
& + \frac{(s-u)(261 s^4+809 s^3 u+1126 s^2 u^2 + 809 s u^3 +261u^4)}{30 t^7} \log\Bigl(\frac{\null s}{u}\Bigr) \nn \\[3pt] 
& + \frac{1682 s^4 + 5303 s^3 u +7422 s^2 u^2 + 5303 s u^3+1682 u^4}{180 t^6}
\Bigg]
\,,
\label{eq:doubleminusGR1loop}
\end{align}
where
\begin{equation}
r_\Gamma = \frac{\Gamma(1+\epsilon)\Gamma^2(1-\epsilon)}{\Gamma(1-2\epsilon)} \,.
\end{equation}
Here $\mu$ is an infrared dimensional regularization scale. 
Since massless gravitons circulate in the loop there is an infrared
divergence.  On the other hand there is no ultraviolet divergence
because there is no available counterterm~\cite{tHooft:1974toh}.  In
this expression we use the four-dimensional helicity
scheme~\cite{FDHScheme}.   For the massless case the analytic 
continuation from the Euclidean region to the physical one is 
simple and accomplished by taking $\log(-s) \rightarrow \log(s) - i \pi$.
We have explicitly verified that our results for the graviton 
in the loop match the ones calculated in Ref.~\cite{Dunbar:1994bn}, up to the opposite
sign for the  $\mathcal{M}_4^{\newPiece{1/2}}$ piece already noted in Ref.~\cite{Abreu:2020lyk}.


\section{Values of one-loop integrals}
\label{IntegralValuesAppendix}

In this appendix we give the values of the integrals appearing in the
amplitudes collected in Appendix~\ref{AmplitudesResultsAppendix}.  We
first present the $(4-2 \eps)$-dimensional integrals and then discuss
the higher-dimensional integrals. Furthermore, we provide an algorithmic procedure for
obtaining an expression for the amplitudes with no $\eps$ or mass
dependence in the integral coefficients.

\subsection{Explicit values of one-loop integrals}
\label{IntegralsAppendix}

We now collect the values for the integrals appearing in the above amplitudes~\cite{Integrals}.
We present the integrals in the unphysical Euclidean region where $s,t,u <0$
and then discuss the analytic continuation to the physical region.
We define a generic $D$-dimensional $n$-point integral by
\begin{equation}
I_n^D = i (-1)^{n+1} (4 \pi)^{D/2} \int \frac{d^D p}{(2 \pi)^D} \frac{1}{(p^2 -m^2)((p - p_{1})^2 - m^2) \cdot ((p - p_{n-1})^2 - m^2)} \,,
\label{eq:genericIntegralDef}
\end{equation}
where the $p_i$'s are linear combinations of the external
momenta.  The integral $I_n^D$ is also labeled by the specific choice of the
$p_i$'s. For example, we use $I_2^D(s)$ for a $D$-dimensional bubble
integral that has an invariant mass square of $s=(k_1+k_2)^2$ flowing
through its external legs. Similarly, we use $I_3^D(s)$ and
$I_4^D(s,t)$ for a $D$-dimensional triangle and box respectively,
where we use all scales that may appear in the integral as arguments
of the corresponding function. When dealing with a
($4-2\epsilon$)-dimensional integral we suppress the superscript
writing $I_n^{4-2\epsilon} \equiv I_n$. 

For the purposes of this paper it is sufficient to present the explicit
expressions for the $(4-2\epsilon)$-dimensional, one- through
four-point integrals up to $\mathcal{O}(\epsilon^0)$.
The tadpole (one-point) integral takes the form
\begin{equation}
I_1 = m^{2-2\epsilon} \frac{\Gamma(1+\epsilon)}{\epsilon(\epsilon-1)} \,.
\label{TadpoleIntegral}
\end{equation}
The bubble (two-point) integral with a kinematic invariant $s$ is given by 
\begin{equation}
I_2(s) = I_2(0) + I_2^{\rm fin}(s) + \mathcal{O}(\epsilon) \,,
\label{BubbleIntegral}
\end{equation}
where 
\begin{equation}
I_2(0) = m^{-2\epsilon} \frac{\Gamma(1+\epsilon)}{\epsilon} 
  = \frac{1}{\epsilon} + \mathcal{O}(\epsilon^0) \,,
\label{Bubble0Integral}
\end{equation}
and 
\begin{equation}
I_2^{\rm fin}(s) = 2 + x^{(s)} \log \left( \frac{x^{(s)}-1}{x^{(s)}+1} \right) \,,
\label{BubbleFinite}
\end{equation}
with $x^{(s)} \equiv \sqrt{1 - 4m^2/s}$.
The triangle (three-point)  integral is
\begin{equation}
I_3(s) = -\frac{1}{2s} \log^2 \left( \frac{x^{(s)}+1}{x^{(s)}-1} \right) + \mathcal{O}(\epsilon) \,.
\label{triangles}
\end{equation}
Finally, the box (four-point) integral is given by
\begin{align}
I_4(s,t) &= \frac{2}{s t x^{(s\,t)}} \Bigg[
	2 \log^2 \left( \frac{x^{(s\,t)}  + x^{(s)} }{x^{(s\,t)}  + x^{(t)} } \right) + 
	\log \left( \frac{x^{(s\,t)}  -x^{(s)} }{x^{(s\,t)}  + x^{(s)} } \right) \log \left( \frac{ x^{(s\,t)}  - x^{(t)} }{x^{(s\,t)}  + x^{(t)} } \right) - \frac{\pi^2}{2} \nn \\
	&+ \sum_{i=s,\,t} \left( 2 \text{Li}_2 \left( \frac{x^{(i)} - 1}{x^{(s\,t)}  + x^{(i)}} \right) - 
	2 \text{Li}_2 \left( -\frac{x^{(s\,t)}  - x^{(i)}}{x^{(i)} +1} \right) - 
	\log^2 \left( \frac{x^{(i)} + 1}{x^{(s\,t)}  + x^{(i)}} \right) \right) \Bigg] \,,
\label{boxes}
\end{align}
where $x^{(s\,t)} \equiv \sqrt{1-4m^2/s-4m^2/t}$.
To evaluate the expressions in physical regions, e.g. $s>0$, $t,u <0$, we need to account for the $i \epsilon$ prescription 
which for all our integrals is obtained by shifting  the mass by $m^2 \rightarrow m^2 - i \epsilon$.

In order to match to the low-energy EFT, we expand the above integrals
in the large-mass limit.  It is straightforward to expand the tadpole,
bubble and triangle integrals in this limit. For the box integral we
use
\begin{align}
I_4(s,t)  = \null &  
\frac{1}{6m^4} + \frac{s+t}{60m^6} + \frac{2 s^2 + s t + 2 t^2}{840 m^8} + \frac{(s+t)(3 s^2 -2 s t + 3 t^2)}{7560m^{10}} \nn \\
&+ \frac{12 s^4 + 3 s^3 t + 2 s^2 t^2 + 3 s t^3 + 12 t^4}{166320 m^{12}} + \frac{(s+t)(10 s^4 - 8 s^3 t + 9 s^2 t^2 - 8 s t^3 +10 t^4)}{720720m^{14}} \nn \\
&+ \frac{60 s^6 + 10 s^5 t + 4 s^4 t^2 + 3 s^3 t^3 + 4 s^2 t^4 + 10 s t^5 + 60 t^6}{21621600 m^{16}} \nn \\
&+ \frac{105 s^7 + 15 s^6 t + 5 s^5 t^2 + 3 s^4 t^3 + 3 s^3 t^4 + 5 s^2 t^5 + 15 s t^6 + 105 t^7}{183783600 m^{18}} \nn \\
&+ \frac{280 s^8 + 35 s^7 t + 10 s^6 t^2 + 5 s^5 t^3 + 4 s^4 t^4 + 5 s^3 t^5 + 10 s^2 t^6 + 35 s t^7 + 280 t^8}{2327925600 m^{20}} \nn \\
&+ \frac{252 s^9 + 28 s^8 t + 7 s^7 t^2 + 3 s^6 t^3 + 2 s^5 t^4 + 2 s^4 t^5 + 3 s^3 t^6 + 7 s^2 t^7 + 28 s t^8 + 252 t^9}{9777287520 m^{22}} \nn \\
&+  \frac{1}{449755225920 m^{24}} \Bigl(2520 s^{10} + 252 s^9 t + 56 s^8 t^2 + 21 s^7 t^3 + 12 s^6 t^4 + 10 s^5 t^5  \nn \\ 
& \hskip 2 cm  \null  + 12 s^4 t^6 + 21 s^3 t^7  + 56 s^2 t^8 + 252 s t^9 + 2520 t^{10} \Bigr) \nn \\
&+ \frac{1}{1873980108000 m^{26}} \Bigl( 2310 s^{11} + 210 s^{10} t + 42 s^9 t^2 + 14 s^8 t^3 + 7 s^7 t^4 + 5 s^6 t^5 \nn \\ 
&  \hskip 2 cm  \null + 5 s^5 t^6 + 7 s^4 t^7 + 14 s^3 t^8 + 42 s^2 t^9 + 210 s t^{10} + 2310 t^{11} \Bigr) \nn \\
&+  \frac{1}{101194925832000 m^{28}} \Bigl(27720 s^{12} + 2310 s^{11} t + 420 s^{10} t^2 + 126 s^9 t^3 + 56 s^8 t^4 + 35 s^7 t^5  \nn \\ 
& \hskip 2 cm  \null + 30 s^6 t^6 + 35 s^5 t^7 + 56 s^4 t^8 + 126 s^3 t^9 + 420 s^2 t^{10} + 2310 s t^{11} + 27720 t^{12} \Bigr) 
\nn \\
&+ \mathcal{O}(m^{-30}) \,.
\end{align}
This expansion is included in the ancillary files~\cite{AttachedFile}.

\subsection{Higher-dimension integrals}
\label{IBPsec}

In constructing the amplitudes we used an overcomplete basis of
integrals containing both ($4-2\epsilon$)- and higher-dimensional
integrals, which as we noted has the advantage of removing $\eps$ and
$m$ dependence from the integral coefficients.  We now explain the
construction of this form of the amplitudes and how one returns to the
usual integral basis containing only the ($4-2\epsilon$)-dimensional
scalar integrals introduced in Eq.~(\ref{ReductionTarget}).  In the
($4-2\epsilon$)-dimension form the coefficients of the integrals have explicit
$\epsilon$ and $m$ dependence.  As we discussed in \sect{uvSec}, the
basis including higher-dimension integrals is useful for exploiting
known properties of the amplitude in order to fix the coefficients
$a_0$ and $b_0$ in Eq.~(\ref{ReductionTarget}), which we cannot obtain
from the generalized-unitarity cuts.

Higher-dimension integrals occur naturally in the course of evaluating the loop integrands.
Integrals with powers of the higher-dimensional components of loop momentum $\mu$ (defined in \eqn{muDef}) in the numerator 
may be expressed directly in terms of higher-dimensional integrals. Following Ref.~\cite{BernMorgan} we have, 
\begin{equation}
\int \frac{d^4 \ell}{(2 \pi)^4} \frac{d^{-2\epsilon} \mu}{(2 \pi)^{-2\epsilon}}  \frac{(\mu^2)^r}{D_{abcd}}  
= \mathcal{P}(\epsilon,r) (4\pi)^r \int \frac{d^{4+2r-2\epsilon} L}{(2 \pi)^{4+2r-2\epsilon}}
\frac{1}{{D}_{abcd}}\, ,
\label{EqMu2epsilon}
\end{equation}
where the loop momentum on the right-hand side is integrated over a $(4+2r-2\epsilon)$-dimensional space, $D_{abcd}$ is defined in Eq.~(\ref{denGen}), and
\begin{equation}
{\mathcal P}(\epsilon,0) = 1 \,, \hskip 2 cm 
{\mathcal P}(\epsilon,r) = -\epsilon(1-\epsilon)(2-\epsilon)\ldots(r-1-\epsilon)\,, \quad r>0 \,.
\end{equation}

\begin{figure}
\centering
\includegraphics[scale=.65]{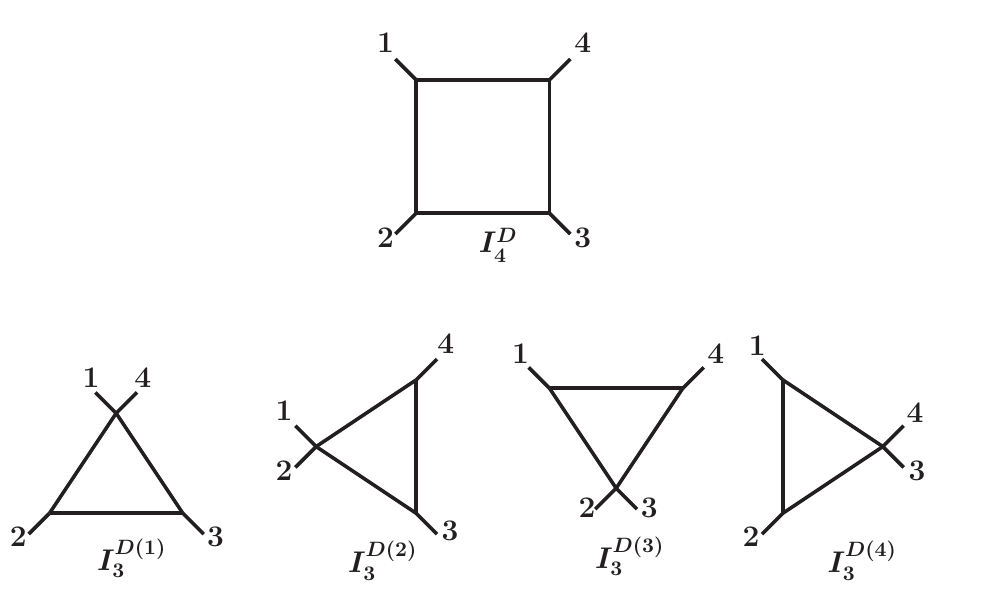}
\caption{\captionlineskip The $D$-dimensional box integral and its four triangle-integral daughters. }
\label{6dbox}
\end{figure}

The resulting higher-dimensional integrals may be expressed in terms the $(4-2\epsilon)$-dimensional ones using 
the dimension-shifting formula \cite{Bern:1992em}:
\begin{equation}
I_n^{D+2} = \frac{1}{(n-D-1)c_0} \Bigl[ 2 I_n^D - \sum_{i=1}^n c_i I_{n-1}^{D(i)} \Bigr] \,,
\label{dimShift}
\end{equation}
which holds for any spacetime dimension $D$ and $n \leq 5$. $I_n^D$ refers to an
$n$-gon integral in $D$ dimensions, defined in Eq.~(\ref{eq:genericIntegralDef}). We use $I_{n-1}^{D(i)}$
for the integral obtained by $I_n^D$ by removing the propagator
between legs $(i-1)$ and $i$.
The $c_i$   are combinations of kinematic factors given by
\begin{equation}
c_i = \sum_{j=1}^n S_{ij}^{-1}, \hskip 1 cm c_0 = \sum_{i=1}^n c_i \,,
\end{equation} 
where the matrix $S$ for the cases of interest to us is given by 
\begin{equation}
S_{ij} = m^2 -\frac{1}{2}p_{ij}^2, \qquad \text{with} \qquad p_{ij} = p_{i-1} - p_{j-1} \,.
\end{equation}
For example, a ($6-2\epsilon$)-dimensional box integral is expressed
in terms of a ($4-2\epsilon$)-dimensional box and four
($4-2\epsilon$)-dimensional triangles, illustrated in
Fig.~\ref{6dbox}.

Using Eq.~(\ref{dimShift}), we can reduce an expression that contains higher-dimensional integrals to one that does not.  For the reverse process, i.e. in order to eliminate all $\epsilon$ and $m$ dependence in a one-loop amplitude $\mathcal{M}_4^\text{1-loop}$,  we adopt the following strategy: We start by inspecting the coefficients of the boxes, which take the following schematic form
\begin{align}
\begin{split}
\mathcal{M}_4^\text{1-loop} 
&= \frac{P(\epsilon)Q(s,t,m)}{(4-(4-2\epsilon)-1)(4-(6-2\epsilon)-1)(4-(8-2\epsilon)-1)(4-(10-2\epsilon)-1)}I_4 \\
&+ \ldots
\end{split}
\label{BoxStart}
\end{align}
where $P(\epsilon)$ is at most an order-four polynomial in $\epsilon$
that does not cancel any of the poles of the expression and
$Q(s,t,m)$ is some rational function.  The maximum
degree in $\epsilon$ is directly tied to the maximal power of loop
momentum that can appear in the numerator for minimally-coupled
gravity.  Here we assume the highest possible power of loop momentum, which corresponds to a spin-0 particle circulating in the loop.  $I_4$ in \eqn{BoxStart} stands for a box integral in our amplitude, whose arguments we
do not specify since we are being schematic. This discussion applies
to all box integrals in our amplitude. The ellipsis contains other
master integrals and their coefficients. Looking at
Eq.~(\ref{dimShift}), we may identify this term as coming from the
$I_4^{12-2\epsilon}$ integral,
\begin{equation}
\mathcal{M}_4^\text{1-loop} = P(\epsilon)Q'(s,t,m)I_4^{12-2\epsilon} + \ldots \,,
\label{Dpol}
\end{equation}
where $Q'(s,t,m)$ is some new rational function. Note that the ellipsis also changes as dictated by Eq.~(\ref{dimShift}). 
Now we may set $D=4-2\epsilon$ in Eq.~(\ref{dimShift}) and rewrite it as follows,
\begin{equation}
\epsilon I_n^{6-2\epsilon} = \frac{1}{2 c_0}[ 2 I_n - \sum_{i=1}^n c_i I_{n-1}^{(i)} - c_0 (n-5)I_n^{6-2\epsilon} ] \,,
\end{equation}
where $I_n^{4-2\epsilon} \equiv I_n$. Similarly, setting $D=6-2\epsilon$ in Eq.~(\ref{dimShift}) we get
\begin{equation}
\epsilon I_n^{8-2\epsilon} = \frac{1}{2 c_0}[ 2 I_n^{6-2\epsilon} - \sum_{i=1}^n c_i I_{n-1}^{6-2\epsilon(i)} - c_0 (n-7)I_n^{8-2\epsilon} ] \,,
\end{equation}
etc. Combining the two we may trade for example $\epsilon^2 I_n^{8-2\epsilon}$ for an expression containing $I_n^{8-2\epsilon}$, $I_n^{6-2\epsilon}$, $I_n$ and lower-point integrals with $\epsilon$ dependence only in the coefficients of the lower-point integrals. In this fashion we turn Eq.~(\ref{Dpol}) into
\begin{equation}
\mathcal{M}_4^\text{1-loop} = \sum_{r=0}^4 Q_r(s,t)I_4^{4+2r-2\epsilon} + \ldots
\end{equation}
with some different rational functions $ Q_r(s,t)$. In this way we
eliminate all explicit $\epsilon$ and $m$ dependence in the
coefficients of the boxes (we discuss the $m$ dependence
momentarily). Only after this step is completed, the poles in
$\epsilon$ in the coefficients of the triangles have a similar
interpretation, i.e. as coming from higher-dimensional triangles. This
is because there is a feed down from the coefficients of the boxes to
those of the triangles due to Eq.~(\ref{dimShift}). The fact that the
poles of the triangles align correctly is a nontrivial check of our
calculation. We repeat this process sequentially for all lower-point
integrals to completely remove the explicit $\epsilon$ and $m$
dependence in the coefficients.

This procedure always succeeds. The reason is that there exists an
alternative process of reducing the integrals to master integrals that
does not introduce any explicit $\epsilon$ dependence, but instead
introduces these higher-dimensional integrals (for an extensive
discussion we refer the reader to Appendix~I of
Ref.~\cite{Bern:1995ix}). The existence of such a process guarantees
the success of a procedure like the one outlined above.  In addition,
we may understand why there is no $m$ dependence in the coefficients
of the master integrals in this basis as follows: We imagine
performing the calculation in a covariant gauge, in which all
propagators in the integrals have the canonical form $(p^2 - m^2)$,
and there are no other poles in the loop momentum or the mass. We use
\begin{equation}
m^2 = L^2 - (L^2-m^2) \,, \quad \quad \mu^2 = -L^2 + \ell^2
\end{equation}
to trade all $m$ and $\mu$ dependence in the coefficients for tensor integrals and lower-point integrals. Once all $m$ and $\mu$ dependence has been eliminated in this way, we may reduce the tensor integrals using the IBP reduction procedure described in Appendix~I of Ref.~\cite{Bern:1995ix}. The resulting expression contains higher-dimensional integrals without $\epsilon$ or $m$ dependence.

\section{High-order expansion of the one-loop four-graviton amplitudes in the large-mass limit}
\label{HighOrdersAppendix}

In this appendix, we present the large-mass expansion of our one-loop four-graviton amplitudes through
$\mathcal O (m^{-20})$. The same results are collected in a Mathematica ancillary file~\cite{AttachedFile}. 
We give the amplitudes in terms of loop integrals in
Appendix~\ref{AmplitudesResultsAppendix}.  We obtain the results of this appendix by expanding the ones of Appendix~\ref{AmplitudesResultsAppendix} in the large-mass limit. We give the values and expansion of the integrals in Appendix~\ref{IntegralsAppendix}.
The present representation corresponds to low-energy effective description of the gravitational theories under consideration.

The data contained here should be useful for systematic
investigations at higher orders in the $1/m$ expansion than carried
out in this paper.  The examples in
Sects.~\ref{Sec:Bounds}-\ref{Sec:BoundMultipleHelicity} based on low
orders in the expansion suggest that Wilson coefficients in physical
theories lie on small islands in the allowed parameter space.  Rather
strikingly, these islands are one dimensional to a good approximation,
which is due to the LSD property. Rather remarkably the string-theory
data we use also satisfy similar properties, so the  Wilson
coefficients obtained from string theory also populate these
islands. We hope that the data presented here will facilitate further
investigations of these features. 

Using the amplitudes expanded in the large-mass limit one may obtain
the low-energy effective description of the theory, along the lines of
Sect.~\ref{EFTMatchingSection}. Besides the operators present in the
action of Eq.~(\ref{eft}), one should include operators of the
schematic form $D^{2k} R^4$.  The operators $D^{2k} R^4$ correspond to
the terms of $\mathcal{O}(m^{-(4+2k)})$.

We organize the amplitudes in a supersymmetric decomposition \eqref{susyGR} in
terms of the new contributions for a given spin,
$\mathcal{M}_4^{\newPiece{\it S}}$. We may then
assemble these pieces into the contributions for a particle of a given
spin circulating in the loop using \eqn{susyGR}.  Regarding the
double-minus configuration, starting with the spin-0 contribution and
moving to the additional new pieces through spin 2, we have,
\begin{align}
f^{\newPiece{0}}&(s,u) 
= \mathcal K 
 \Bigg( \frac{1}{6300 m^4} + \frac{t}{41580 m^6} \nn \\[3pt]
&+ \frac{81(s^2+u^2) + 155 s u}{15135120 m^8}
+ \frac{t \big(161 (s^2+u^2) + 324 s u \big) }{151351200 m^{10}}	\nn \\[3pt]
&+ \frac{3556 (s^4 + u^4) + 14035 (s^3 u + s u^3 ) + 21030 s^2 u^2}{15437822400 m^{12}} \nn \\[3pt]
&+ \frac{t \big( 2052 (s^4 + u^4) + 8218 (s^3u + s u^3) + 12287 s^2 u^2  \big)}{41902660800 m^{14}} \nn \\[3pt]
&+ \frac{4634 (s^6 + u^6) + 27650 (s^5 u + s u^5) + 69026 (s^4 u^2 + s^2 u^4) +91987 s^3 u^3}{430200650880 m^{16}} \nn \\[3pt]
&+\frac{t \big( 87780(s^6 + u^6) + 526770(s^5 u + s u^5) + 1314684 (s^4 u^2 + s^2 u^4) + 1752653 s^3 u^3 \big)}{37104806138400 m^{18}} \nn \\[3pt]
&+ \frac{2551824(s^8 + u^8) + 20357964(s^7 u + s u^7) + 71183961(s^6 u^2 + s^2 u^6) }{4823624797992000 m^{20}} \nn \\[3pt]
&+ \frac{142285437(s^5 u^3 + s^3 u^5) + 177823240 s^4 u^4}{4823624797992000 m^{20}} \Bigg) 
\,,
\label{M4NewPieceExpand0}
\end{align}
\begin{align}
f^{\newPiece{1/2}}&(s,u) 
 = \mathcal K 
\Bigg( \frac{1}{1120 m^4} + \frac{t}{8400 m^6} \nn \\[3pt]
&+ \frac{15(s^2+u^2) + 28 s u}{554400 m^8}
+ \frac{t \big(153 (s^2+u^2) + 313 s u \big) }{30270240 m^{10}}	\nn \\[3pt]
&+ \frac{665 (s^4 + u^4) + 2596 (s^3 u + s u^3 ) + 3890 s^2 u^2}{605404800 m^{12}} \nn \\[3pt]
&+ \frac{t \big( 581 (s^4 + u^4) + 2345 (s^3u + s u^3) + 3495 s^2 u^2  \big)}{2572970400 m^{14}} \nn \\[3pt]
&+ \frac{29106 (s^6 + u^6) + 172676 (s^5 u + s u^5) + 430955 (s^4 u^2 + s^2 u^4) + 574230 s^3 u^3}{586637251200 m^{16}} \nn \\[3pt]
&+\frac{t \big( 34440(s^6 + u^6) + 207564(s^5 u + s u^5) + 517436 (s^4 u^2 + s^2 u^4) + 690109 s^3 u^3 \big)}{3226504881600 m^{18}} \nn \\[3pt]
&+ \frac{39270(s^8 + u^8) + 312240(s^7 u + s u^7) + 1091604(s^6 u^2 + s^2 u^6) }{16491024950400 m^{20}} \nn \\[3pt]
&+ \frac{2181716(s^5 u^3 + s^3 u^5) + 2726549 s^4 u^4}{16491024950400 m^{20}} \Bigg) 
\,,
\end{align}
\begin{align}
f^{\newPiece{1}}&(s,u) 
= \mathcal K 
\Bigg( \frac{1}{180 m^4} + \frac{t}{1680 m^6} \nn \\[3pt]
&+ \frac{22(s^2+u^2) + 39 s u}{151200 m^8}
+ \frac{t \big(20 (s^2+u^2) + 43 s u \big) }{831600 m^{10}}	\nn \\[3pt]
&+ \frac{825 (s^4 + u^4) + 3125 (s^3 u + s u^3 ) + 4684 s^2 u^2}{151351200 m^{12}} \nn \\[3pt]
&+ \frac{t \big( 315 (s^4 + u^4) + 1308 (s^3u + s u^3) + 1930 s^2 u^2  \big)}{302702400 m^{14}} \nn \\[3pt]
&+ \frac{1036 (s^6 + u^6) + 6027 (s^5 u + s u^5) + 15036 (s^4 u^2 + s^2 u^4) + 20030 s^3 u^3}{4410806400 m^{16}} \nn \\[3pt]
&+\frac{t \big( 7056(s^6 + u^6) + 43316(s^5 u + s u^5) + 107555 (s^4 u^2 + s^2 u^4) + 143715 s^3 u^3 \big)}{146659312800 m^{18}} \nn \\[3pt]
&+ \frac{11760(s^8 + u^8) + 92232(s^7 u + s u^7) + 322372(s^6 u^2 + s^2 u^6) }{1075501627200 m^{20}} \nn \\[3pt]
&+ \frac{644205(s^5 u^3 + s^3 u^5) + 805050 s^4 u^4}{1075501627200 m^{20}} \Bigg)
\,,
\end{align}
\begin{align}
f^{\newPiece{3/2}}&(s,u) 
 = \mathcal K 
\Bigg( \frac{1}{24 m^4} + \frac{t}{360 m^6} \nn \\[3pt]
&+ \frac{9(s^2+u^2) + 14 s u}{10080 m^8}
+ \frac{t \big(8 (s^2+u^2) + 21 s u \big) }{75600 m^{10}}	\nn \\[3pt]
&+ \frac{10 (s^4 + u^4) + 34 (s^3 u + s u^3 ) + 51 s^2 u^2}{332640 m^{12}} \nn \\[3pt]
&+ \frac{t \big( 225 (s^4 + u^4) + 1075 (s^3u + s u^3) + 1518 s^2 u^2  \big)}{50450400 m^{14}} \nn \\[3pt]
&+ \frac{105 (s^6 + u^6) + 558 (s^5 u + s u^5) + 1391 (s^4 u^2 + s^2 u^4) + 1852 s^3 u^3}{86486400 m^{16}} \nn \\[3pt]
&+\frac{t \big( 224(s^6 + u^6) + 1533(s^5 u + s u^5) + 3729 (s^4 u^2 + s^2 u^4) + 5035 s^3 u^3 \big)}{1102701600 m^{18}} \nn \\[3pt]
&+ \frac{5292(s^8 + u^8) + 38416(s^7 u + s u^7) + 134211(s^6 u^2 + s^2 u^6) }{97772875200 m^{20}} \nn \\[3pt]
&+ \frac{268122(s^5 u^3 + s^3 u^5) + 335050 s^4 u^4}{97772875200 m^{20}} \Bigg) 
\,,
\label{M4NewPieceExpand32}
\end{align}
\begin{align}
f^{\newPiece{2}}&(s,u) 
 = \mathcal K 
 \Bigg( \frac{1}{2 m^4} + \frac{s^2+s u+u^2}{120 m^8} \nn \\[3pt]
&+ \frac{ s t u }{504 m^{10}} 
+ \frac{(s^2+s u+u^2)^2}{3780 m^{12}}
+ \frac{(s^2+s u+u^2) s t u}{7920 m^{14}} \nn \\[3pt]
&+ \frac{75 (s^6 + u^6) + 225 (s^5 u + s u^5) + 559 (s^4 u^2 + s^2 u^4) + 743 s^3 u^3}{7207200 m^{16}} \nn \\[3pt]
&+\frac{ 3 (s^2 + s u + u^2)^2 s t u}{400400 m^{18}} \nn \\[3pt]
&+ \frac{(s^2+s u+u^2)(56(s^6+u^6) + 168(s^5 u + s u^5) + 557(s^4 u^2 + s^2 u^4) + 834 s^3 u^3)}{122522400 m^{20}}  \Bigg) 
\,.
\label{M4NewPieceExpand2}
\end{align}
We give the relation between $\mathcal{M}_4(1^{+},2^{-},3^{-},4^{+})$ and $f(s,u)$ in \eqn{HelicityDef} and define $\mathcal K$ in \eqn{eq:KDef}.

For the all-plus and single-minus configurations it suffices to give
the result for the spin-0 contribution since we obtain the remaining
amplitudes via Eq.~(\ref{allPlusSUSY}). For the all-plus configuration
we have
\begin{align}
h^{S=0}&(s,u)  = 
   \mathcal K 
\Bigg( 
    \frac{s t u}{504 m^2} + \frac{(s^2 + s u + u^2)^2}{3780 m^4} \nn \\
    &+ 
    \frac{(s^2 + s u + u^2) s t u}{7920 m^6} 
    + \frac{75(s^6+u^6) + 225 (s^5 u + s u^5) + 559 ( s^4 u^2 + s^2 u^4) + 743 s^3 u^3}{7207200 m^8}    \nn \\
    &+ \frac{3 (s^2 + s u + u^2)^2 s t u}{400400 m^{10}} \nn \\
    &+ \frac{(s^2 + s u + u^2)\big( 56(s^6+u^6) + 168 (s^5 u + s u^5) + 557 (s^4 u^2 + s^2 u^4) + 834 s^3 u^3 \big)}{122522400 m^{12}}
    \nn \\
    &+ \frac{\big( 392 (s^6 + u^6) + 1176(s^5 u + s u^5) + 2481 (s^4 u^2 + s^2 u^4) + 3002 s^3 u^3 \big) s t u}{888844320 m^{14}} \nn \\
    &+ \frac{(s^2 + s u + u^2)^2 \big( 105(s^6+u^6) + 315(s^5 u + s u^5) + 1412 (s^4 u^2 + s^2 u^4) + 2299 s^3 u^3 \big)}{4888643760 m^{16}} \nn \\
    &+ \frac{(s^2 + s u + u^2)\big( 150(s^6+u^6) + 450(s^5 u + s u^5) + 1049 (s^4 u^2 + s^2 u^4) + 1348 s^3 u^3 \big) s t u}{5766092640 m^{18}}  \nn \\
    &+ \frac{1}{33731641944000 m^{20}} \Big( 35640(s^{12}+u^{12})  \nn \\
    & \hskip 2 cm  \null + 213840(s^{11} u + s u^{11}) + 1174365 (s^{10} u^2 + s^2 u^{10}) + 3911625 (s^9 u^3 + s^3 u^9) \nn \\
    & \hskip 2 cm  \null + 8797526 (s^8 u^4 + s^4 u^8)  + 14072594 (s^7 u^5 + s^5 u^7) + 16416696 s^6 u^6 \Big) \Bigg)
\,,
\end{align}
while for the single-minus configuration we find
\begin{align}
g^{S=0}&(s,u)  = \mathcal K 
\Bigg(
	\frac{1}{5040 m^2 s t u} + \frac{1}{6306300 m^8} \nn \\
    &+ \frac{(s^2 + s u + u^2)}{441080640 m^{12}} + 
    \frac{s t u}{2715913200 m^{14}} + 
    \frac{(s^2 + s u + u^2)^2}{22406283900 m^{16}} 
    + \frac{ (s^2 + s u + u^2)  s t u}{64250746560 m^{18}} \nn \\
    &+
    \frac{\big( 27( s^6 + u^6 ) + 81 ( s^5 u + s u^5 ) + 197 ( s^4 u^2 + s^2 u^4) + 259 s^3 u^3 \big)}{25057791158400 m^{20}}  
\Bigg)
\,.  
\end{align}
The relation of $\mathcal{M}_4(1^{+},2^{+},3^{+},4^{+})$ and $\mathcal{M}_4(1^{+},2^{+},3^{-},4^{+})$ to $h(s,u)$ and $g(s,u)$ is found in \eqn{HelicityDef}.


\end{document}